\documentclass{aa}
\usepackage{epsf}
\def \la {\mathrel{\vcenter
     {\offinterlineskip \hbox{$<$}\hbox{$\sim$}}}}
\def \ga {\mathrel{\vcenter
     {\offinterlineskip \hbox{$>$}\hbox{$\sim$}}}}
\def\eck#1{\left\lbrack #1 \right\rbrack}
\def\rund#1{\left( #1 \right)}
\def\ave#1{\langle #1 \rangle}
\begin{document}
\thesaurus{06
           (08.19.4;
            02.05.1; 
            02.08.1; 
            02.01.2)}
 
\title{Conditions for shock revival by neutrino heating~\\
in core-collapse supernovae}

 
\author{H.-Thomas Janka} 
 
\offprints{H.-Th.~Janka\\(thj@mpa-garching.mpg.de)}
 
\institute{Max-Planck-Institut f\"ur Astrophysik,
Karl-Schwarzschild-Stra\ss e 1, D-85741 Garching, Germany}
 

 
\maketitle
 
\begin{abstract}
Energy deposition by neutrinos can rejuvenate the stalled bounce shock 
and can provide the energy for the supernova explosion of a massive
star. This neutrino-heating mechanism, though investigated by numerical
simulations and analytic studies, is not finally accepted or proven as
the trigger of the explosion. Part of the problem is that different groups
have obtained seemingly discrepant results, and the complexity of the 
hydrodynamic models often hampers a clear and simple interpretation
of the results. This demands a deeper theoretical understanding of the
requirements of a successful shock revival.\\
A toy model is developed here for discussing the neutrino heating 
phase analytically. The neutron star atmosphere 
between the neutrinosphere and the supernova shock can well be considered
to be in hydrostatic equilibrium, with a layer of net neutrino 
cooling below the gain radius and a layer of net neutrino heating above.
Since the mass infall rate to the shock is in general different from the rate
at which gas is advected into the neutron star, the mass in the gain layer
varies with time. Moreover, the gain layer receives additional 
energy input by neutrinos emitted from the neutrinosphere and the cooling
layer. Therefore the determination of the shock 
evolution requires a time-dependent treatment. To this end the 
hydrodynamical equations of continuity and energy are integrated
over the volume of the gain layer to obtain conservation laws for the 
total mass and energy in this layer.
The radius and velocity of the supernova shock can then be calculated from
global properties of the gain layer as solutions of an initial value 
problem, which expresses the fact that the behavior of the shock is controlled
by the cumulative effects of neutrino heating and mass accumulation in the 
gain layer.
The described toy model produces steady-state accretion and mass outflow
from the nascent neutron star as special cases.\\
The approach is useful to illuminate the conditions that can lead to 
delayed explosions and in this sense supplements detailed numerical 
simulations. On grounds of the model developed
here, a criterion is derived for the requirements of 
shock revival. It confirms the existence of a minimum neutrino luminosity 
that is needed for shock expansion, but also demonstrates the
importance of a sufficiently large mass infall rate to the shock. 
If the neutrinospheric luminosity or accretion rate by the shock are too
low, the shock is weakened because the gain layer loses more mass than 
is resupplied by inflow.
On the other hand, very high infall rates damp the shock expansion and 
above some threshold, the development of positive total energy in the 
neutrino-heating layer is prevented.
Time-dependent solutions for the evolution of the gain layer show that
the total specific energy transferred to nucleons by neutrinos is
limited by about $10^{52}$~erg$\,$M$_{\odot}^{-1}$ ($\sim 5$~MeV per
nucleon). This excludes the possibility of very energetic explosions
by the neutrino-heating mechanism, because
the typical mass in the gain layer is about 0.1~M$_{\odot}$
and does not exceed a few tenths of a solar 
mass. The toy model also allows for a crude discussion of the global effects of
convective energy transport in the neutrino-heating layer. Transfer of energy 
from the region of maximum heating to radii closer behind the shock mainly 
reduces the loss of energy by the inward flow
of neutrino-heated matter through the gain radius. 
\end{abstract}
 

\section{Introduction}
\label{sec:intro}

Neutrinos dominate the energetics of core-collapse supernovae.
Only about one percent or $\sim 10^{51}\,$erg of the gravitational binding
energy released in the formation process of the compact remnant,
usually a neutron star, end up as kinetic energy of the expanding ejecta,
whereas 99\% of this energy are radiated away in neutrinos. Electron 
captures on protons and nuclei trigger the gravitational instability
of the iron core of an evolved massive star, because the electron number and 
thus the pressure are reduced
by the escape of electron neutrinos (see, e.g., Bruenn 1986a). Later the loss of
energy by the diffusion of neutrinos and antineutrinos of all flavors drives the
evolution of the nascent neutron star from a hot, inflated configuration to
the compact and very dense final state (Burrows \& Lattimer 1986).

Colgate \& White (1966) were the first to suggest that neutrinos may
also play a crucial role for the explosion by taking up the gravitational
binding energy of the collapsing core and depositing it in the rest of the
star. Subsequent improvements and more realistic treatments of the 
microphysics, like equation of state (EoS) and neutrino transport, have
changed our modern picture of stellar core collapse dramatically compared
to the pioneering simulations by Colgate \& White (1966).
Because of the discovery of weak neutral currents and the corresponding
importance of neutrino scattering off nucleons and nuclei, the forming
neutron star was recognized to be highly opaque to neutrinos. Therefore the 
neutrino luminosities turned out to be too low, and the energy transfer 
rate by neutrinos not large enough to invert the infall
of the surrounding gas into an explosion. For many years, hopes and
efforts therefore concentrated on the prompt bounce-shock mechanism: The
energy given to the hydrodynamical shock wave in the moment of core bounce 
was thought to lead directly to the ejection of the stellar mantle and envelope.
Detailed models, however, showed that the shock experiences such severe
energy losses by photodisintegration of iron nuclei and additional neutrino 
emission, that its outward propagation stops still well inside the iron core
(e.g., Bruenn 1985, 1989a,b, 1993; Baron \& Cooperstein 1990;
Hillebrandt 1987; Myra et al. 1987, 1989). 

Wilson (1985), however, discovered that neutrinos can indeed
cause an explosion on a timescale much longer than previously thought.
More than 100 milliseconds after core bounce the conditions for neutrino
energy deposition have significantly improved (Bethe \& Wilson 1985), and 
the mass infall rate and thus the ram pressure of the shock have decreased,
making an explosion at later times easier than right after bounce 
(Burrows \& Goshy 1993, Bethe 1995). 
Although Wilson et al. (1986) obtained such
``delayed'' explosions via the neutrino-heating mechanism, their
simulations gave rather low explosion energies, and their successes
could not be confirmed by independent models with supposedly superior
treatment of the neutrino physics and EoS (Bruenn 1986b, 1989a,b).
Later simulations by  Wilson \& Mayle (1988, 1993) and Mayle \& Wilson (1988)
included neutron-finger convection in the nascent neutron star, which 
boosts the neutrino luminosities and thus increases the neutrino heating and
the explosion energy. But whether neutron-finger convection actually occurs in 
the hot neutron star, or Ledoux-type convection (Burrows 1987, Keil et al. 1996,
Pons et al. 1999), or none (Bruenn et al. 1995, Mezzacappa et al. 1998a) seems
to depend on the properties of the nuclear EoS and possibly also
on the treatment of the neutrino physics. 

More recently, multi-dimensional simulations showed that
convective overturn in the region of net neutrino heating between shock and
gain radius (that is the position outside the neutrinosphere where neutrino 
cooling is balanced by neutrino heating; Bethe \& Wilson 1985) 
can aid the explosion (Herant et al. 1994;
Janka \& M\"uller 1995, 1996; Burrows et al. 1995) and can produce successes
even when spherically symmetric models fail. This ``convective engine'' 
(Herant et al. 1994) or ``boiling'' (Burrows et al. 1995) transports cool gas 
into the region of strongest heating while at the same time hot gas rises towards 
the shock. Both effects increase the efficiency of neutrino energy transfer, 
reduce the energy loss by the reemission of neutrinos from the heated gas, and 
raise the postshock pressure, thus leading to more favorable conditions for shock
expansion. While the existence and importance of postshock convection is not 
questioned, simulations with the most advanced treatment of the neutrino 
transport applied to multi-dimensional supernova calculations so far 
(Mezzacappa et al. 1998b, Lichtenstadt et al. 1999) nourished doubts whether the 
effects of convection are sufficiently strong to cause explosions.

Therefore scepticism about the viability of the delayed explosion 
mechanism by neutrino heating still remains (Thompson 2000), and seems justified 
even more because of recent observations which indicate a possible connection 
between gamma-ray bursts and at least some supernovae (e.g., Galama et al. 1998,
Bloom et al. 1999).
If confirmed, this discovery would require to consider large energies and/or 
asphericities of the explosions (Iwamoto et al.~1998,
Woosley et al.~1999, H\"oflich et al.~1999)
which might be hard to explain by the neutrino-driven mechanism. Therefore, despite 
the fact that the observations are still far from being conclusive, theorists feel
tempted to speculate about alternative ways to power stellar explosions, e.g., 
by invoking magnetically driven jets (Wang \& Wheeler 1998, Khokhlov et al. 1999).
However, while we know about the crucial role of neutrinos,
we have no observational evidence or convincing theoretical argument in support
of a dynamically important strength of magnetic fields in combination with a 
significant degree of rotation in the iron cores of all massive stars.
Rather than in ordinary core-collapse supernovae, jets and a magnetohydrodynamic
mechanism may be at work in cases where the neutrino-driven mechanism
definitely fails, e.g., for progenitor main sequence masses above about
25$\,$M$_{\odot}$ (Fryer 1999) and when a black hole forms at the center of a 
rapidly spinning massive star (MacFadyen \& Woosley 1999, MacFadyen et al.\ 1999).

When judging about the viability of the neutrino-driven mechanism,
one must, however, keep in mind the enormous complexity of the problem. Because
of this complexity a number of approximations and simplifications had to be made
in even the currently most refined hydrodynamical calculations. Some of these
deficiencies have probably disadvantageous consequences for the efficiency
of neutrino energy deposition in the postshock layers. Until very recently, all
published hydrodynamical models employed, for example, a
still unsatisfactory treatment of the neutrino transport. 
Instead of solving the Boltzmann transport equation, they used flux-limited 
diffusion schemes, a fact which underestimates the neutrino 
heating above the gain radius and overestimates the energy loss by neutrino
emission below it (Janka 1991a, 1992; Messer et al. 1998; Yamada et al. 1999). 
Moreover, multidimensional supernova simulations have so far
not been able to resolve the convective processes inside the nascent
neutron star, although cooling models of neutron stars show their 
potential importance (Burrows 1987, Keil et al. 1996, Pons et al. 1999).
Even more, recent investigations (e.g., Raffelt \& Seckel 1995; Janka et al. 1996;
Burrows \& Sawyer 1998, 1999; Reddy et al. 1998, 1999;
Yamada 2000; Yamada \& Toki 2000, and references therein)
suggest that neutrino interaction rates in hot nuclear matter are 
suppressed compared to the standard description used in the numerical codes.
Both the latter effects imply that the neutrino luminosities from the
post-collapse core are most likely underestimated in current supernova
models.

The neutrino-driven mechanism is by its nature sensitive to the 
neutrino-matter coupling in the heating region, which depends on the
properties, i.e., spectra and luminosities, of the neutrino emission from the
neutrinosphere and on the angular distribution of the neutrinos exterior to the
neutrinosphere (Messer et al.\ 1998, Yamada et al. 1999, Burrows et al.\ 2000).
These issues require not only the best possible technical treatment of the 
neutrino transport (cf.\ Mezzacappa et al.\ 2000, Liebend\"orfer et al.\ 2000,
Rampp \& Janka 2000) and of the description of the neutrino opacities, but
they can vary with the structure of the progenitor star, with general
relativity, and with the nuclear EoS and therefore the compactness of the
nascent neutron star. Differences of the simulations by different groups may
be associated with one or more of these issues. Unfortunately,
a detailed analysis and direct comparison is essentially impossible because 
of largely different numerical approaches and a complicated interdependence
of effects.

In this unclear and extremely unsatisfactory situation a better fundamental
understanding of the conditions and requirements for shock revival by
neutrino heating is highly desirable. Several attempts were made for a 
discussion by analytic means (Bruenn 1993; Bethe 1993, 1995, 1997; Shigeyama 1995;
Thompson 2000) or on grounds of simplified numerical analysis (Burrows \& Goshy 
1993). While each of them contains interesting aspects and can shed light
on certain results of simulations, they have led to contradictory conclusions,
and none is general enough to be finally convincing.
For example, assuming steady-state conditions (Burrows \& Goshy 1993) 
cannot explain how accretion is reversed into expansion,
and why an accretion shock should contract again after moving outward for 
some while, a possibility which was in fact observed in many
hydrodynamical simulations. The beginning of the reexpansion of the
stalled shock and the phase when most of the explosion energy is deposited can
also not be described by a stationary neutrino-driven 
baryonic wind (Qian \& Woosley 1996).
Bethe (1990, 1993, 1995, 1997) gave a very useful and detailed discussion
of the physics of neutrino heating, the structure and composition of the heating 
region, and the shock energetics and nucleosynthesis, using observational constraints
from Supernova~1987A and numerical results provided mainly by Jim Wilson. 
Although addressing the question of the start of the shock, his analysis 
does not really reveal the requirements for a successful shock revival. Moreover,
aspects were disregarded which have been recognized to be important for the outcome
of simulations, for example the fact that rapid neutrino losses in the cooling 
region can weaken or even prevent an explosion (Woosley \& Weaver 1994, 
Janka \& M\"uller 1996, Messer et al. 1998). Bethe arrived at the conclusion that 
the explosion energy is delivered by neutrinos, whereas Bruenn (1993) and Thompson 
(2000) argued that neutrino heating is insufficient to cause an explosion because
the advection timescale of the gas between shock and gain radius is too short
for large energy deposition. Shigeyama (1995), on the other hand, performed a
quasi-stationary analysis by expanding the physical variables in a power series
of a small parameter, but his approach obscures the essential physics of shock
revival rather than illuminating them.

The work presented here is a new approach for an analytic discussion of
the conditions which can lead to the reexpansion of the supernova shock.
The analysis is based on a simplified model for the post-bounce structure
of the collapsed stellar core and generalizes the treatment of neutron
star accretion by 
Chevalier (1989; see also Brown \& Weingartner 1994, Fryer et al. 1996).
It is not meant to yield quantitative results or to be able to compete with 
detailed hydrodynamical simulations, but it should allow one to reproduce the basic
features of the shock stagnation, accretion, and shock revival phases. It is therefore
a supplementary tool which helps one getting a qualitative understanding of the
processes that determine the post-bounce evolution of the collapsed stellar core.
In particular, the relative strength of competing effects that play a role in the
neutrino-heating mechanism and their influence
on the behavior of the supernova shock, i.e., its radial position and velocity 
as a function of time, can be estimated. This should help explaining why some 
models fail to produce explosions while others succeed.

The paper is organized in the following way. In Sect.~\ref{sec:physics} the 
physics of the post-bounce accretion phase will be described, in 
Sect.~\ref{sec:basic} the basic equations and corresponding assumptions used
in the simplified analytic model will be introduced, in Sect.~\ref{sec:radii}
the characteristic radii of the problem and their properties will be formally 
defined, in Sect.~\ref{sec:atmosphere} the structure of the collapsed stellar
core behind the stalled supernova shock will be discussed, in Sect.~\ref{sec:heatcool}
expressions for the neutrino heating and cooling will be derived, in 
Sect.~\ref{sec:massacc} the mass accretion rate of the nascent neutron star will
be estimated, and in Sect.~\ref{sec:meconservation} the equations of mass 
and energy conservation will be applied to the neutrino heating layer, which 
leads to a criterion for the revival of a stalled
supernova shock in Sect.~\ref{sec:application}. The equations
derived in this paper will then be combined to an analytic toy model which 
allows one to integrate the shock position, shock radius, and properties of the 
gain layer as functions of time by solving an initial value problem.
A summary and conclusions will follow in Sect.~\ref{sec:summary}.


\section{Physical picture}
\label{sec:physics}

Right after core bounce the hydrodynamic shock propagates outward in mass
as well as in radius, being strongly damped by energy losses due to the 
photodisintegration of iron-group nuclei and neutrinos. The neutrino
emission rises significantly when the shock breaks out into the
neutrino-transparent regime. As a consequence, the pressure behind the
shock is reduced and the velocities of the shock and of the fluid
behind the shock, both of which were positive initially, decrease. Finally,
the outward expansion of the shock stagnates, and the shock transforms 
into a standing accretion shock with negative gas velocity in the postshock
region. The gas of the progenitor star, which 
continues to fall into the shock at a velocity near free fall, is 
decelerated abruptly within the shock. Below the shock it moves 
much more slowly towards the center, where it settles onto the surface
of the nascent neutron star. 

\begin{figure}
\epsfxsize=0.99\hsize
\epsffile{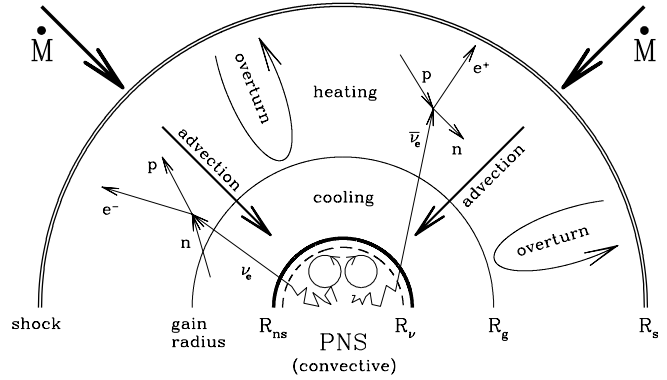}
\caption{
Sketch which summarizes the processes that determine the evolution of the
stalled supernova shock after core bounce. Stellar matter falls into the shock
at radius $R_{\mathrm{s}}$
with a mass accretion rate $\dot M$ and a velocity near free fall. After
deceleration in the shock, the gas is much more slowly advected towards the nascent 
neutron star through the
regions of net neutrino heating and cooling, respectively. The radius $R_{\mathrm{ns}}$
of the neutron star is defined by a steep decline of
the density over several orders of magnitude outside the neutrinosphere at $R_{\nu}$.
Heating balances cooling at the gain radius $R_{\mathrm{g}}$. 
The dominant processes of energy deposition and loss are
absorption of electron neutrinos onto neutrons and electron antineutrinos onto
protons as indicated in the figure. Convective overturn mixes the layer between 
gain radius and shock, and convection inside the neutron star helps the explosion by 
boosting the neutrino luminosities.
}
\label{fig:1}
\end{figure}

Figure~\ref{fig:1} displays the most important physical elements which
determine this evolutionary stage. Around the neutrinosphere at radius
$R_{\nu}$, which is close to the radius $R_{\mathrm{ns}}$ of the proto-neutron star 
(PNS), the hot and comparatively dense gas loses energy by radiating
neutrinos. If this energy sink were absent, the gas that is accreted 
through the shock at a rate $\dot M$ would pile up in a growing, 
high-entropy atmosphere on top of the compact remnant
(Colgate et al. 1993, Colgate \& Fryer 1995, Fryer et al. 1996).
But since neutrinos are emitted efficiently at the thermodynamical conditions
around the neutrinosphere, the entropy of the gas is reduced so
that the gas can be absorbed into the surface of the neutron star.
The mass flow through the neutrinospheric region is therefore triggered by the
neutrino energy loss and allows more gas to be advected inward from larger radii.
In case of stationary accretion the temperature at the base of the atmosphere
ensures that the emitted neutrinos carry away the gravitational binding energy
of the matter which is added to the neutron star at a given accretion rate.
In fact, this requirement closes the set of equations that determines the 
steady state of the accretion system and allows one to determine the radius
$R_{\mathrm{s}}$ of the accretion shock
(see, e.g., Chevalier 1989, Brown \& Weingartner 1994, Fryer et al. 1996).
 
At the so-called gain radius $R_{\mathrm{g}}$ (Bethe \& Wilson 1985) between
neutrinosphere $R_{\nu}$ and shock position $R_{\mathrm{s}}$, the temperature of the 
atmosphere becomes so low that the absorption of high-energy electron neutrinos 
and antineutrinos starts to exceed the neutrino emission. This radius
therefore separates the region of net neutrino cooling below from a layer of net
heating above. Since the neutrino heating is strongest just outside the gain radius
and the propagation of the shock has weakened before stagnation, a negative
entropy gradient is built up in the postshock region. This leads to convective 
overturn roughly between $R_{\mathrm{g}}$ and $R_{\mathrm{s}}$, which transports hot matter
outward in rising high-entropy bubbles. At the same time cooler material is mixed 
inward in narrow, low-entropy downflows (Herant et al. 1994,
Burrows et al. 1995, Janka \& M\"uller 1996). Inside the nascent neutron star,
below the neutrinosphere, convective motions can enhance the neutrino emission
by carrying energy faster to the surface than neutrino diffusion does
(Keil et al. 1996).

\begin{figure}
\epsfxsize=0.99\hsize
\epsffile{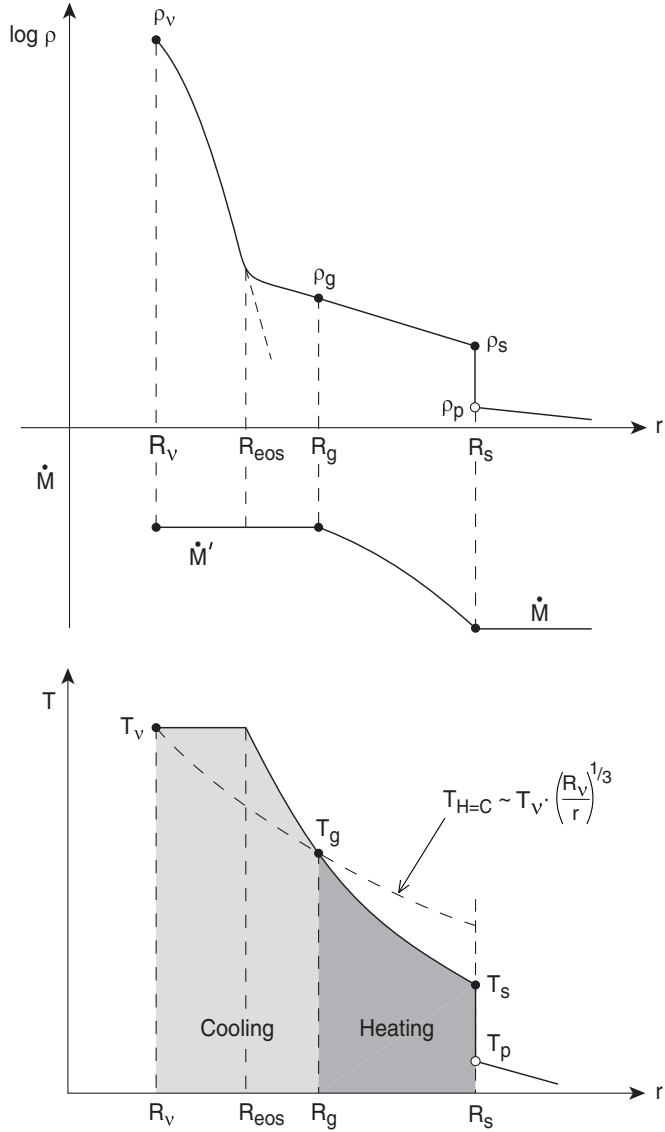}
\caption{
Schematic profiles of density, temperature, and mass accretion rate 
between neutrinosphere at radius $R_{\nu}$ and shock at $R_{\mathrm{s}}$ some
time after core bounce. $R_{\mathrm{g}}$ denotes the position of the gain radius.
At the shock, $\rho$ and $T$ jump discontinuously from 
their preshock values $\rho_{\mathrm{p}}$ and $T_{\mathrm{p}}$ to the postshock values
$\rho_{\mathrm{s}}$ and $T_{\mathrm{s}}$, respectively. 
For $r < R_{\mathrm{eos}}$ the density declines steeply because the 
pressure is mainly caused by the nonrelativistic
Boltzmann gases of free neutrons and protons. Outside of $R_{\mathrm{eos}}$ the gas
is radiation dominated and the density decrease much flatter. In general, 
some of the gas falling into the shock at rate $\dot M$ may stay in the region of
neutrino heating while another part (rate $\dot M'$) is advected into the nascent 
neutron star. Note that $\dot M(r)$ is continuous at the shock in the rest frame 
of the star only in case of a stalled shock front.
Between $R_{\nu}$ and $R_{\mathrm{eos}}$ the temperature can
be considered roughly as constant, whereas its negative gradient in the 
radiation dominated region ensures hydrostatic equilibrium. There is net energy
loss between $R_{\nu}$ and $R_{\mathrm{g}}$ where $T(r)$
exceeds the temperature $T_{\mathrm{H=C}} \sim T_{\nu}(R_{\nu}/r)^{1/3}$,
for which neutrino heating equals cooling. Net energy
deposition occurs between $R_{\mathrm{g}}$ and $R_{\mathrm{s}}$. 
}
\label{fig:2}
\end{figure}

Between neutrinosphere and the supernova shock a number of 
approximations apply to a high degree of accuracy, which help one developing
a simple analytic understanding of the effects that influence the evolution
of the supernova shock. Figure~\ref{fig:2} shows schematically the profiles
of density, temperature and mass accretion rate in that region. 
A formal discussion follows in the subsequent sections. Outside
the neutrinosphere (typically at about $10^{11}\,$g/cm$^3$) the temperature
drops slowly compared to the density decline, which is steep. When
nonrelativistic nucleons dominate the pressure, the decrease of the density yields
the pressure gradient which ensures hydrostatic equilibrium in the gravitational 
field of the neutron star. Assuming a temperature equal to the neutrinospheric
temperature in this region is a reasonably good approximation for the 
following reasons. On the one hand,
the cooling rate depends sensitively both on density and temperature, 
and the density drops rapidly. Therefore the total energy loss is determined in
the immediate vicinity of the neutrinosphere and the details of the 
temperature profile do not matter very much. On the other hand, efficient
neutrino heating prevents that the temperature can drop much below the 
neutrinospheric value. If, instead, the temperature would rise significantly above
this latter value, the matter would become optically thick to the energetic  
neutrinos produced in the hot gas (the opacity increases roughly with the square 
of the neutrino energy) and the neutrinosphere would move farther out to a lower
density (and thus typically a lower temperature).
 
Below a density between $10^9\,$g/cm$^3$ and $10^{10}\,$g/cm$^3$,
relativistic electron-positron pairs and radiation determine the pressure, 
provided the temperature is sufficiently 
high, typically around 1~MeV or more (see Woosley et al.~1986). 
Exterior to the corresponding radius $R_{\mathrm{eos}}$, where this
transition from the baryon-dominated to the radiation-dominated regime takes 
place, the temperature must therefore decrease so that the negative
temperature gradient can yield the force which balances gravity.

The gain radius $R_{\mathrm{g}}$ is located at the radial position where the 
temperature profile $T(r)$ intersects with the curve of temperature values,
$T_{\mathrm{H=C}}(r)$, for which heating is equal to cooling by neutrinos,
roughly given by
%
\begin{equation}
T_{\mathrm{H=C}}(r)\,\sim\,T_{\nu}\cdot \left( {R_{\nu}\over r}\right)^{\!
{1 \over 3}}
\label{eq:1}
\end{equation}
(Bethe \& Wilson 1985). In Eq.~(\ref{eq:1}) $T_{\nu}$ means the temperature
at the radius $R_{\nu}$ of the neutrinosphere. The shock at $R_{\mathrm{s}}$
is taken to be infinitesimally thin compared to the scales considered.
Within the shock the density and temperature therefore jump 
from their preshock values $\rho_{\mathrm{p}}$ and $T_{\mathrm{p}}$,
to the postshock values $\rho_{\mathrm{s}}$ and $T_{\mathrm{s}}$,
respectively. A part of the gas which falls into the shock with a mass 
accretion rate $\dot M$ can stay in the region of neutrino heating, whereas 
another part is advected with rate $\dot M'$ through the cooling region to
be added to the neutron star inside $R_{\nu}$.

The approach to the problem of shock revival taken in this paper is 
considerably different from the discussion of steady-state accretion or winds.
Steady-state assumptions, for example, were also used by Burrows and Goshy (1993) 
in their theoretical analysis of the explosion mechanism. Having realized the fact,
however, that the mass and energy in the gain layer vary because of different rates
of mass flow through the boundaries and additional neutrino heating, one is forced to
the following conclusions. Firstly, the discussion has to be time-dependent, which 
means that the time derivatives in the continuity and energy equations cannot be
ignored. (Dropping the total time-derivative in the momentum equation by assuming 
hydrostatic equilibrium is less problematic and yields a reasonably good approximation.)
Secondly, the properties of the shock and of the gain layer must be determined as 
solutions of an initial value problem rather than from a steady-state picture.
This reflects essential physics, namely that the shock behavior is controlled by
the cumulative effects of neutrino heating and mass accumulation in the gain layer.
For these reasons conservation laws for the total mass and energy in the gain layer
will be derived by integrating the hydrodynamic equations of continuity and energy,
including the terms with time derivatives, over the volume of the gain layer. The
treatment will therefore retain the time-dependence of the problem.

In this paper the discussion will be restricted to an idealized, 
spherically symmetric situation and possible convective mixing will be assumed 
to lead to efficient homogenization of the unstable layer. Certainly,
this is not a good assumption for the convective overturn that takes place
in the region between gain radius and shock front, where prominent, large-scale 
inhomogeneities develop (Herant et al.~1994, Burrows et al.~1995,
Janka \& M\"uller 1996). Bethe (1995) has made attempts to discuss the physical
implications of the simultaneous presence of low-entropy downstreams and 
high-entropy rising bubbles. For this purpose he introduced free parameters, 
e.g., to quantify the fraction of neutrinos that hits the cold downflows
and is effective for their heating, or to account
for the part of the matter that is added to the neutron star instead
of being pushed outward in the expanding bubbles. This procedure is not really 
satisfactory and will not be copied here. Instead, 
an admittedly simplified and idealized spherical situation will be considered
to highlight the conditions needed for shock revival and to develop
a qualitative understanding of the influence of different effects.
One-dimensional analysis can help developing a better understanding of the 
delayed explosion mechanism, because simulations in spherical symmetry
have produced successful explosions (Wilson 1985; Wilson et al.~1986;
Janka \& M\"uller 1995, 1996). 
Thus they have demonstrated that convection behind the shock is not an
indispensable requirement for an explosion, although it may be an essential
(Herant et al.~1994, Burrows et al.~1995, Janka \& M\"uller 1996) --- 
yet not necessarily sufficient
(Janka \& M\"uller 1996, Mezzacappa et al.~1998b, Lichtenstadt et al.~1999) ---
ingredient to obtain explosions, or to raise the explosion energy in cases
which fail or nearly fail in spherical symmetry.


\section{Basic equations and assumptions}
\label{sec:basic}

The hydrodynamic equations are considered in Eulerian form for spherical
symmetry with source terms for Newtonian gravity and neutrino energy and 
momentum exchange with the stellar medium. The equations of continuity,
momentum, and energy are:
\begin{eqnarray}
{\partial \rho \over \partial t}+{1\over r^2}{\partial \over \partial r}
(r^2\rho v)&=& 0 \ ,\label{eq:2}\\
{\partial (\rho v)\over \partial t}+{1\over r^2}{\partial \over \partial r}
(r^2\rho v^2)&=& -{\partial P\over\partial r} - \rho{\partial \Phi \over
\partial r}\ ,\label{eq:3}\\
{\partial e\over \partial t}+{1\over r^2}{\partial \over \partial r}
\eck{r^2v(e+P)}&=& -\rho v {\partial \Phi \over \partial r} + Q_{\nu}\ .
\label{eq:4}
\end{eqnarray}
Here $r$, $v$, $\rho$, $P$, $t$ are radius, fluid velocity, density, pressure, 
and time, respectively, and $e$ is defined as the sum of internal energy 
density, $\varepsilon$, and kinetic energy density of the gas:
\begin{equation}
e\,=\, {1\over 2}\rho v^2 + \varepsilon \ .
\label{eq:5}
\end{equation}
The term $Q_{\nu}$ denotes the rate of energy gain or loss per unit 
volume by neutrino heating and cooling. $\Phi(r)$ is an effective 
potential which contains contributions from the gravitational potential
and from the momentum transfer to the stellar gas by neutrinos.
Neglecting self-gravity of the gas in the region between neutrinosphere
and supernova shock, it can be written as
\begin{equation}
\Phi\,=\,-\,{G{\widetilde M}\over r}\,=\,-\,{G\over r}\rund{M-{\ave{\kappa_{\mathrm{t}}}
L_{\nu}\over 4\pi Gc\rho}}\ .
\label{eq:6}
\end{equation}
Here $G$ is the gravitational constant, $c$ the speed of light,
$M$ the mass inside $R_{\nu}$ and ${\widetilde M}$ means an effective mass
that includes the momentum transfer term and is defined by the term in brackets 
on the right side of Eq.~(\ref{eq:6}). 
When self-gravity is disregarded, the mass of the gas between $R_{\nu}$ and
$R_{\mathrm{s}}$ must be negligible compared to the neutron star mass $M$, i.e.,
\begin{equation}
\Delta M\,=\,\int_{R_{\nu}}^{R_{\mathrm{s}}}{\mathrm{d}}r\,4\pi r^2\rho(r) \ll M\ .
\label{eq:7}
\end{equation}

In Eq.~(\ref{eq:6}) $L_{\nu} = \sum_{\nu_i}L_{\nu_i}$ is the total neutrino 
luminosity and $\ave{\kappa_{\mathrm{t}}}$ the mean total opacity
calculated as an average of the total opacities of 
neutrinos $\nu_i$ and antineutrinos $\bar\nu_i$ of all flavors according to
\begin{equation}
\ave{\kappa_{\mathrm{t}}}L_{\nu}\,\equiv\,
\sum_{\nu_i}\kappa_{{\mathrm{t}},\nu_i}L_{\nu_i} +
\sum_{\bar\nu_i}\kappa_{{\mathrm{t}},\bar\nu_i}L_{\bar\nu_i}\ .
\label{eq:8}
\end{equation}
The total opacity $\kappa_{{\mathrm{t}},\nu_i}$ of neutrino $\nu_i$ is
considered to be averaged over the spectrum of the corresponding energy flux.
Note that in this paper, opacities are defined as inverse mean free paths and
are thus measured in units of 1/cm.
Equations~(\ref{eq:3}), (\ref{eq:4}), and (\ref{eq:6})
imply that the momentum transfer rate from neutrinos to the stellar gas is
written as $\ave{\kappa_{\mathrm{t}}}L_{\nu}/(4\pi c\,r^2)$ with 
$L_{\nu}$ and $\ave{\kappa_{\mathrm{t}}}/\rho$
not depending on $r$. This is approximately fulfilled in the optically thin regime for 
neutrinos, i.e., exterior to the neutrinosphere where the neutrino luminosities 
and spectra are roughly constant. Yet it is not exactly true, because the 
concept of ``the'' neutrinosphere is fuzzy and neutrino emission and absorption
continue even outside the neutrinosphere. In addition, the opacity depends on the
composition which varies with the radius. During all of the post-bounce
evolution, however, the typical total neutrino luminosity is only a few per 
cent of the Eddington luminosity,
\begin{equation}
L_{{\mathrm{Edd}},\nu}\,\equiv\,{4\pi GMc\rho \over \ave{\kappa_{\mathrm{t}}}}\ .
\label{eq:9}
\end{equation}
Therefore the neutrino source terms for momentum in Eq.~(\ref{eq:3}) and for kinetic 
energy in Eq.~(\ref{eq:4}), which are carried by the potential $\Phi$, are always 
small and the approximate treatment following below is justified.

Neutrinos transfer momentum to the stellar medium
by neutral-current scatterings off neutrons and protons. The corresponding
transport opacity for these scattering processes is
\begin{equation}
\kappa_{\mathrm{sc}}\,\approx\,{5\alpha^2+1\over 24}
{\sigma_0 \ave{\epsilon^2_{\nu}}\over (m_ec^2)^2}\,{\rho\over m_{\mathrm{u}}}\,
(Y_n+Y_p)\ .
\label{eq:10}
\end{equation}
Here $m_{\mathrm{u}}\approx 1.66\times 10^{-24}\,$g is the atomic mass unit,
$m_ec^2 = 0.511\,$MeV the rest-mass energy of the electron,
$\sigma_0 = 1.76\times 10^{-44}\,$cm$^2$, and
$Y_n = n_n/n_{\mathrm{b}}$ and $Y_p = n_p/n_{\mathrm{b}}$ are the number
fractions of free
neutrons and protons, i.e., their particle densities normalized to the
number density $n_{\mathrm{b}}$ of nucleons.
A minor difference between the neutrino-proton and the neutrino-neutron
scattering cross section due to different vector coupling constants
is ignored, and also the axial-vector couplings are assumed to be
the same and to be equal to the charged-current axial-vector coupling
constant in vacuum, $\alpha = -1.26$.
Additional scattering reactions with electrons and positrons can be 
neglected because of their much smaller cross sections, and neutrino scattering
off nuclei is unimportant because the post-bounce medium exterior to the 
neutrinosphere is nearly completely disintegrated into free nucleons.

In case of $\nu_e$ and $\bar\nu_e$ also the charged-current
absorptions on neutrons and protons, respectively, need to be taken
into account due to their large cross sections. The absorption opacity 
is
\begin{equation}
\kappa_{\mathrm{a}}\,\approx\,{3\alpha^2+1\over 4}
{\sigma_0\ave{\epsilon^2_{\nu}}\over (m_ec^2)^2}\,{\rho\over m_{\mathrm{u}}}\,
\left\{\matrix{Y_n \cr Y_p \cr}\right\}\ .
\label{eq:11}
\end{equation}
In Eqs.~(\ref{eq:10}) and (\ref{eq:11}) the recoil of the nucleon and
phase space blocking effects for the fermions are neglected, which is very good
at the conditions considered in this paper. In both neutral-current and
charged-current processes only the leading terms depending on the 
squared neutrino energy, $\epsilon_{\nu}^2$, are taken into account. 
Averaging over the spectrum of the neutrino energy flux yields the 
factor $\ave{\epsilon_{\nu}^2}$, for which Eq.~(\ref{eq:25}) provides a 
suitable definition, if minor differences between the spectra of neutrino 
energy density and flux density are disregarded.

The total opacity 
includes the contributions from scattering and absorption and is
given as $\kappa_{{\mathrm{t}},\nu_i} = \kappa_{{\mathrm{sc}},\nu_i}+
\kappa_{{\mathrm{a}},\nu_i}$. With typical values $L_{\nu_e} = 
L_{\bar\nu_e} = L_{\nu_x} = {1\over 6}L_{\nu}$ (with $\nu_x \in 
\lbrace \nu_{\mu},\,\bar\nu_{\mu},\,\nu_{\tau},\,\bar\nu_{\tau}\rbrace $
and $L_{\nu_x}$ being the luminosity of each individual type of $\nu_x$),
$\ave{\epsilon_{\bar\nu_e}^2}\approx 2 \ave{\epsilon_{\nu_e}^2}$
and $\ave{\epsilon_{\bar\nu_x}^2}\approx 4 \ave{\epsilon_{\nu_e}^2}$,
the total opacity averaged for all neutrinos and antineutrinos
can be estimated from Eq.~(\ref{eq:8}) as
\begin{eqnarray}
\ave{\kappa_{\mathrm{t}}} &\approx& {113\alpha^2+25\over 144}
{\sigma_0\ave{\epsilon^2_{\nu_e}}\over (m_ec^2)^2}\,{\rho\over m_{\mathrm{u}}}\,
(Y_n+Y_p)\nonumber \\
&\approx& 1.9\times 10^{-7}\rho_{10}\,
\rund{{kT_{\nu_e}\over 4\,{\mathrm{MeV}}}}^{\! 2}\ \ \eck{{1\over {\mathrm{cm}}}}
\ .
\label{eq:12}
\end{eqnarray}
For deriving the first expression, the factor $Y_n+2Y_p$ in the absorption term
was replaced by $Y_n+Y_p$. This is a reasonably good approximation because
$Y_p \la Y_n$ between the neutrinosphere and the shock, and the corresponding 
change of the absorption opacity of electron antineutrinos implies only a minor 
error in the total opacity.
In the second equation use was made of $Y_n+Y_p \approx 1$.
If the neutrino flux spectrum has Fermi-Dirac shape with vanishing degeneracy,
the neutrino temperature $T_{\nu}$ is related to the mean squared neutrino
energy by $\ave{\epsilon^2_{\nu}}\approx 21\,(kT_{\nu})^2$.
$k$ is the Boltzmann constant and $\rho_{10}$ the density measured in
10$^{10}\,$g/cm$^3$. For a total neutrino luminosity 
$L_{\nu} = 10^{53}\,$erg/s,
neutrino momentum transfer reduces ${\widetilde M}$ in Eq.~(\ref{eq:6})
relative to $M$ by about $3.8\times 10^{-2}$M$_{\odot}$, which is indeed a 
small correction.

\section{The characteristic radii}
\label{sec:radii}

The neutrinospheric radius $R_{\nu}$, the gain radius $R_{\mathrm{g}}$ and
the transition radius of the EoS properties, $R_{\mathrm{eos}}$, 
will be formally defined below. They are characteristic of the atmospheric
structure in the postshock region, which determines, together with the
infall region ahead of the shock, the shock radius $R_{\mathrm{s}}$ and the
shock velocity $U_{\mathrm{s}} \equiv \dot R_{\mathrm{s}}$.

\subsection{The neutrinosphere}
\label{sec:neutrinosphere}

The neutrinosphere relevant for the discussion in the following sections
is the ``energy-sphere'', where neutrinos decouple energetically
from the stellar background. It usually does not coincide with the 
sphere of last scattering, the
so-called ``transport-sphere'', outside of which the neutrino distribution
becomes strongly forward peaked (for a detailed discussion, see Janka 1995).   
Only inside their energy-sphere neutrinos can be considered to be roughly
in thermodynamic equilibrium with the stellar medium.
Besides neutrino-nucleon scattering, which is important for all
neutrinos, electron neutrinos $\nu_e$ and electron antineutrinos $\bar\nu_e$
interact via frequent charged-current absorption and emission reactions with 
nucleons, whereas muon and tau neutrinos and antineutrinos do not. Therefore the
energy-spheres of electron neutrinos and antineutrinos are typically 
located farther out in the star at larger radii than those of muon and 
tau neutrinos.

The energy deposition in the gain region, however, is clearly dominated 
by $\nu_e$ and $\bar\nu_e$. For this reason one can concentrate on their 
transport properties and neglect muon and tau neutrinos and antineutrinos in
the discussion. Scattering off nucleons acts on all neutrinos equally.
The charged-current absorption reactions of $\nu_e$ and $\bar\nu_e$ on neutrons 
and protons, respectively, yield an even larger contribution to the total opacity.
The opacities of $\nu_e$ and $\bar\nu_e$ are nearly equal, because 
$\bar\nu_e$ absorption and emission [Eq.~(\ref{eq:18})] is similarly 
frequent as $\nu_e$ absorption and emission [Eq.~(\ref{eq:17})]
as long as positrons are abundant, i.e., the stellar 
atmosphere is hot and electrons are not very degenerate.
Therefore the transport-spheres
and energy-spheres of electron neutrinos and antineutrinos are all close 
together and it is justified to consider only one, ``the'', neutrinosphere
at radius $R_{\nu}$. Of course, the real situation is more complex and there
is no definite radius interior to which neutrinos are in
equilibrium at the local thermodynamical conditions and diffuse, and exterior
to which they are decoupled from the background and stream freely. The
transition between these two limits is continuous and in case of neutrinos,
whose reaction rates are strongly energy dependent, it is also a function
of the neutrino energy.

The spectral temperature of electron neutrinos will be taken equal to the gas
temperature at the assumed neutrinosphere, $kT_{\nu_e} = kT(R_{\nu})$. Detailed
simulations of neutrino transport show that electron antineutrinos
have somewhat more energetic spectra. A typical result (e.g., Bruenn 1993, Janka 1991a)
is $kT_{\bar\nu_e}\approx 1.5\,kT_{\nu_e}$, which will be used below. The fact that
$\nu_e$ and $\bar\nu_e$ spectra are found to be different in detailed models
is an indication that the picture drawn above is overly simplified. Nevertheless
it is sufficiently accurate for the analysis in this paper. Note that in general the 
neutrino luminosity $L_{\nu}$ can not be related to the neutrinospheric temperature
by the Stefan-Boltzmann law for blackbody emission of a sphere with radius $R_{\nu}$,
$L_{\nu} = \pi R_{\nu}^2\,{7\over 8}ac (kT_{\nu})^4$. This formula is frequently
taken for the combined luminosity of neutrinos plus antineutrinos, assuming their 
chemical potentials to be zero. However, the effective temperature $kT_{\mathrm{eff}}$,
which should be used in the Stefan-Boltzmann law, is typically not equal to the
spectral temperature $kT_{\nu}$ (for a discussion, see Janka 1995).
Transport simulations show that due to non-equilibrium effects the difference can 
be quite significant. For this
reason two parameters, $T_{\nu}$ and $L_{\nu}$, will be retained here to describe
the spectrum and the luminosity of the neutrinos emitted from the neutrinosphere. 
Moreover,
the radius of the neutrinosphere will be considered as the position in the star
where the mean value of the cosine of the neutrino propagation angle relative to
the radial direction has a value of 0.25 
[see Eq.~(\ref{eq:26}) and Janka 1991a,b, 1995].

Keeping in mind the simplifications associated with the concept of the
neutrinosphere, the radius $R_{\nu}$ can be defined by the requirement that the
effective optical depth to energy exchange for neutrinos with average energy is
\begin{equation}
\tau_{\mathrm{eff}}\,=\,\int_{R_{\nu}}^\infty{\mathrm{d}}r\,\kappa_{\mathrm{eff}}(r)
\,=\,{2\over 3}
\label{eq:13}
\end{equation}
(Suzuki 1989). The effective opacity $\kappa_{\mathrm{eff}}$ in case of $\nu_e$ 
and $\bar\nu_e$ is defined from
the scattering opacity $\kappa_{\mathrm{sc}}$ and the
absorption opacity $\kappa_{\mathrm{a}}$ as
\begin{equation}
\kappa_{\mathrm{eff}}\,=\,\sqrt{\kappa_{\mathrm{a}}\rund{
\kappa_{\mathrm{a}}+\kappa_{\mathrm{sc}}}}
\label{eq:14}
\end{equation}
(Rybicki \& Lightman 1979, Shapiro \& Teukolsky 1983, Suzuki 1989).
Using Eqs.~(\ref{eq:10}) and (\ref{eq:11}) one obtains for 
the effective opacity, again averaged over the spectrum of the energy flux
which is supposed to have Fermi-Dirac shape with zero degeneracy,
in case of electron neutrinos the expression
\begin{eqnarray}
\kappa_{{\mathrm{eff}},\nu_e} &=& 1.62\,
{\sigma_0\ave{\epsilon_{\nu_e}^2}\over (m_ec^2)^2}\,{\rho\over m_{\mathrm{u}}}\,
Y_n\,\sqrt{1+0.21\,{Y_p\over Y_n}}
\\
&=& 2.2\times 10^{-7}\rho_{10}\,
\rund{{kT_{\nu_e}\over 4\,{\mathrm{MeV}}}}^{\! 2}Y_n\,\sqrt{1+0.21\,{Y_p\over Y_n}}
\ ,\nonumber
\label{eq:15}
\end{eqnarray}
and in case of electron antineutrinos the analogue result with
$Y_n\sqrt{1+0.21\,Y_p/Y_n}$ being replaced by $Y_p\sqrt{1+0.21\,Y_n/Y_p}$.
Since nuclei are nearly completely dissociated into free nucleons and
$Y_n$ is larger than $Y_p$, i.e., $Y_n \sim 0.8$ and $Y_p\sim 0.2$, but 
$kT_{\nu_e}$ is usually somewhat lower than $kT_{\bar\nu_e}$, 
i.e., $kT_{\nu_e}\approx 4\,$MeV and
$kT_{\bar\nu_e}\approx 6\,$MeV, we verify the above statement that the 
effective opacities of $\nu_e$ and $\bar\nu_e$ are approximately equal. 
Assuming equal luminosities, $L_{\nu_e} = L_{\bar\nu_e}$, a suitable average
value for the effective opacity therefore is
\begin{eqnarray}
\ave{\kappa_{\mathrm{eff}}} &=& {1\over 2}\,\kappa_{{\mathrm{eff}},\nu_e}
+ {1\over 2}\,\kappa_{{\mathrm{eff}},\bar\nu_e}
\nonumber\\
&\approx& 1.5\times 10^{-7}\rho_{10}\,
\rund{{kT_{\nu_e}\over 4\,{\mathrm{MeV}}}}^{\! 2} , 
\label{eq:16}
\end{eqnarray}
where the composition dependent term in the weighted average has been
approximated by $\sqrt{2}$. Knowing the density profile $\rho(r)$, the
density $\rho_{\nu}$ at the neutrinosphere can be determined by using 
Eq.~(\ref{eq:16}) in Eq.~(\ref{eq:13}).

\subsection{The gain radius}
\label{sec:gainradius}

Heating and cooling of the gas outside the neutrinosphere mainly
proceed via the charged-current absorption and emission processes
of $\nu_e$ and $\bar\nu_e$ (Bethe \& Wilson 1985; Bethe 1993, 1995, 1997):
\begin{eqnarray}
\nu_e + n &\longleftrightarrow& p + e^- \ , \label{eq:17}\\
\bar\nu_e + p &\longleftrightarrow& n + e^+ \ .\label{eq:18}
\end{eqnarray}
To leading order in the particle energies, the cross sections for neutrino
and electron/positron absorption, respectively, are
\begin{eqnarray}
\sigma_{{\mathrm{a}},\nu_e}&\approx&\sigma_{{\mathrm{a}},\bar\nu_e}\,\approx\,
{3\alpha^2+1\over 4}\,\sigma_0\,\rund{{\epsilon_{\nu}\over m_ec^2}}^{\! 2}\ ,
\label{eq:19}\\
\sigma_{{\mathrm{a}},e^-} &\approx&\sigma_{{\mathrm{a}},e^+} \,\approx\,
{3\alpha^2+1\over 8}\,\sigma_0\,\rund{{\epsilon_e\over m_ec^2}}^{\! 2}\ .  
\label{eq:20}
\end{eqnarray}
At the considered densities and temperatures, fermion phase space blocking
and dense-medium effects can be safely ignored, and electrons are 
relativistic ($kT \ga m_ec^2$). The heating
rate $Q^+_{\nu_i}$ of the stellar medium by neutrinos $\nu_i$ is given by
\begin{equation}
Q^+_{\nu_i} \,=\, {3\alpha^2+1\over 4}\,{\sigma_0 c\, n_j\over (m_ec^2)^2}
\int\limits_0^\infty{\mathrm{d}}\epsilon_{\nu}\int\limits_{-1}^{+1}{\mathrm{d}}\mu\,
{{\mathrm{d}}^2 n_{\nu_i}\over {\mathrm{d}}\epsilon_{\nu}{\mathrm{d}}\mu}\,
\epsilon_{\nu}^3 \ ,
\label{eq:21}
\end{equation}
where $n_j$ is the number density of the target nucleons ($j = p,\,n$),
and ${\mathrm{d}}^2 n_{\nu_i}/({\mathrm{d}}\epsilon_{\nu}{\mathrm{d}}\mu)$ 
the neutrino distribution,
\begin{equation}
{{\mathrm{d}}^2 n_{\nu_i}\over {\mathrm{d}}\epsilon_{\nu}{\mathrm{d}}\mu}\,
\,=\, {2\pi\over (hc)^3}\,f_{\nu_i}(\epsilon_{\nu},\mu)\,
\epsilon_{\nu}^2 \ ,
\label{eq:22}
\end{equation}
with $f_{\nu_i}(\epsilon_{\nu},\mu)$ being the neutrino phase space occupation 
function at some radius $r$, which depends on the neutrino energy 
$\epsilon_{\nu}$ and the cosine of the angle of neutrino propagation 
relative to the radial direction, $\mu = \cos\theta$. In Eq.~(\ref{eq:22})
the factor $h$ in the denominator is Planck's constant. Introducing
Eq.~(\ref{eq:22}) into Eq.~(\ref{eq:21}) and performing 
the phase space integration over all energies and angles yields
\begin{equation}
Q^+_{\nu_i} \,=\, {3\alpha^2+1\over 4}\,\sigma_0\,n_j\,
{\ave{\epsilon_{\nu_i}^2}\over (m_ec^2)^2}
{L_{\nu_i}\over 4\pi r^2 \ave{\mu_{\nu_i}}}\ .
\label{eq:23}
\end{equation}
Here the neutrino luminosity $L_{\nu_i}$, the average squared neutrino 
energy $\ave{\epsilon_{\nu_i}^2}$, and the mean value of the cosine of
the propagation angle, $\ave{\mu_{\nu_i}}$, are calculated from the 
neutrino phase space occupation function $f_{\nu_i}(\epsilon_{\nu},\mu)$
by
\begin{eqnarray}
L_{\nu_i} &=& 4\pi r^2c\,{2\pi \over (hc)^3}
\,\int\limits_0^\infty{\mathrm{d}}\epsilon_{\nu}\int\limits_{-1}^{+1}{\mathrm{d}}\mu\,
\mu\,\epsilon_{\nu}^3\,f_{\nu_i}(\epsilon_{\nu},\mu)\ ,\label{eq:24}\\
\ave{\epsilon_{\nu_i}^2} &=& 
\int\limits_0^\infty {\mathrm{d}}\epsilon_{\nu}\int\limits_{-1}^{+1}{\mathrm{d}}\mu\,
\epsilon_{\nu}^5\,f_{\nu_i}
\rund{
\int\limits_0^\infty {\mathrm{d}}\epsilon_{\nu}\int\limits_{-1}^{+1}{\mathrm{d}}\mu\,
\epsilon_{\nu}^3\,f_{\nu_i}}^{\!\! -1}\!\! ,\label{eq:25}\\ 
\ave{\mu_{\nu_i}} &=&
\int\limits_0^\infty {\mathrm{d}}\epsilon_{\nu}\int\limits_{-1}^{+1}{\mathrm{d}}\mu\,
\mu\,\epsilon_{\nu}^3\,f_{\nu_i}
\rund{
\int\limits_0^\infty {\mathrm{d}}\epsilon_{\nu}\int\limits_{-1}^{+1}{\mathrm{d}}\mu\,
\epsilon_{\nu}^3\,f_{\nu_i}}^{\!\! -1}\!\! .\label{eq:26}
\end{eqnarray}
The quantity $\ave{\mu_{\nu}}$ is also called flux factor and can be
understood as the ratio of the neutrino energy flux, $L_{\nu}/(4\pi r^2)$,
to the neutrino energy density times $c$. Typically,
it is close to 0.25 near the neutrinosphere of $\nu_e$ and $\bar\nu_e$ and
approaches unity when the neutrino distribution get more and more forward
peaked in the limit of free streaming with increasing distance from the
neutrinosphere (Janka 1991b, 1992, 1995).
The total heating rate $Q^+_{\nu}$ is the sum of the contributions from $\nu_e$
and $\bar\nu_e$: 
\begin{equation}
Q^+_{\nu} \,=\, Q^+_{\nu_e}+Q^+_{\bar\nu_e}\ .
\label{eq:27}
\end{equation}
To derive a simple expression, one can
again assume that $L_{\nu_e} \approx L_{\bar\nu_e}$,
$\ave{\epsilon_{\bar\nu_e}^2}\approx 2\ave{\epsilon_{\nu_e}^2}$, and 
that the $\nu_e$ spectrum has
Fermi-Dirac shape with zero degeneracy, i.e., $\ave{\epsilon_{\nu_e}^2}\approx
21(kT_{\nu_e})^2$. In addition, the equality
$\ave{\mu_{\nu_e}} = \ave{\mu_{\bar\nu_e}} \equiv \ave{\mu_{\nu}}$ is reasonably well
fulfilled because the opacities of electron neutrinos and antineutrinos are very
similar and therefore the neutrinospheres of both of them are nearly at 
the same radius. Putting everything together, the heating rate per unit volume
is derived as
\begin{eqnarray}
Q^+_{\nu} 
&=& {3\alpha^2+1\over 4}\,{\sigma_0\ave{\epsilon_{\nu_e}^2}\over (m_ec^2)^2}
\,{\rho\over m_{\mathrm{u}}}\,{L_{\nu_e}\over 4\pi r^2\ave{\mu_{\nu}}}\,
(Y_n + 2Y_p) \nonumber\\
&\approx& 160\,\,{\rho\over m_{\mathrm{u}}}\,{L_{{\nu_e},52}\over r_7^2\ave{\mu_{\nu}}}
\rund{{kT_{\nu_e}\over 4\,{\mathrm{MeV}}}}^{\! 2}\ \ \eck{{{\mathrm{MeV}}\over 
{\mathrm{s}}}}\ .
\label{eq:28}
\end{eqnarray}
The numerical factor gives the rate in MeV per baryon, $r_7$ is the radius in
$10^7\,$cm, and $L_{{\nu_e},52}$ the $\nu_e$ luminosity normalized to 
10$^{52}\,$erg/s. In the layers where most of the heating and cooling between
neutrinosphere and shock take place, nuclei are nearly fully dissociated into
free nucleons (Bethe 1993, 1995, 1997; Thompson 2000) and $Y_p < Y_n$, therefore 
using $Y_n + 2Y_p\approx 1$ in the last expression is a reasonable approximation.

The cooling rate of the stellar gas by emission of $\nu_e$ and $\bar\nu_e$
is calculated as
\begin{equation}
Q^-_{\nu}\,=\, {3\alpha^2\!\! +\!\! 1 \over 8}\,{\sigma_0\, c \over (m_ec^2)^2}
\int\limits_0^\infty\! {\mathrm{d}}\epsilon\,\epsilon^3\! \rund{\!
n_p {{\mathrm{d}}n_{e^-}\over {\mathrm{d}}\epsilon} +
n_n {{\mathrm{d}}n_{e^+}\over {\mathrm{d}}\epsilon}\! } ,
\label{eq:29}
\end{equation}
where use was made of Eq.~(\ref{eq:20}), and the distributions of 
relativistic electrons and positrons are given by 
\begin{equation}
{{\mathrm{d}}n_{e^\pm}\over {\mathrm{d}}\epsilon} \,=\, {8\pi \over (hc)^3}\,
{\epsilon^2 \over 1 + \exp (\epsilon/kT - \eta_{e^\pm})} \ .
\label{eq:30}
\end{equation}
$T(r)$ is the local gas temperature and $\eta_{e^\pm}$ the degeneracy parameter
of electrons or positrons, defined as the ratio of the chemical potential to the 
temperature. A factor of 2 was taken into account as the statistical weight 
for positive and negative spin states. Inserting Eq.~(\ref{eq:30}) into 
Eq.~(\ref{eq:29}) one gets for the cooling rate per unit volume
\begin{eqnarray}
Q^-_{\nu} &=& (3\alpha^2\! +\! 1)\,
{\pi\,\, \sigma_0\, c\,\, (kT)^6 \over (hc)^3 (m_ec^2)^2}\,
{\rho \over m_{\mathrm{u}}} \nonumber \\
& &\phantom{(3\alpha^2\! + +\! 1)\pi\pi\,\sigma_0\, c\,\, T^6\,}
\times
\eck{Y_p{\cal F}_5(\eta_e) + Y_n{\cal F}_5(-\eta_e)} \nonumber \\
&\approx& 145\,\,{\rho\over m_{\mathrm{u}}}\,
\rund{{kT\over 2\,{\mathrm{MeV}}}}^{\! 6} \ \ \eck{{{\mathrm{MeV}}\over 
{\mathrm{s}}}} \ ,
\label{eq:31}
\end{eqnarray}
where the numerical factor is the rate in MeV per nucleon when 
$Y_n+Y_p\approx 1$ and the equilibrium relation
$\eta_{e^-} = -\eta_{e^+}\equiv \eta_e$ with $\eta_e\approx 0$ are used. 
The latter approximation is good in the shock-heated layers because the
electron fraction $Y_e = n_e/n_{\mathrm{b}}$ and thus the electron degeneracy 
is rather low
and $e^\pm$ pairs are abundant. ${\cal F}_5(\eta)$ is the Fermi integral for
relativistic particles,
\begin{equation}
{\cal F}_j(\eta) \,=\, \int_0^\infty{\mathrm{d}}x\, 
{x^j\over 1 + \exp(x-\eta)} \ ,
\label{eq:32}
\end{equation}
with ${\cal F}_5(0)\approx 118$. (Useful formulae for sums and
differences of these Fermi integrals can be found in Bludman \& Van Riper, 1978,
and simple approximations in Takahashi et al.~1978.)

Heating balances cooling at the gain radius, i.e., the gain radius $R_{\mathrm{g}}$
has to fulfill the condition $Q^+_{\nu} = Q^-_{\nu}$ by definition.
With Eqs.~(\ref{eq:28}) and (\ref{eq:31}) one obtains the following relation:
\begin{equation}
R_{{\mathrm{g}},7}\rund{{kT_{\mathrm{g}}\over 2\,{\mathrm{MeV}}}}^{\! 3} 
\,\approx\,1.05\,\,
\sqrt{{L_{\nu_e,52}\over \ave{\mu_{\nu}}_{\mathrm{g}}}}
\rund{{kT_{\nu_e}\over 4\,{\mathrm{MeV}}}}
\ .
\label{eq:33}
\end{equation}
$R_{{\mathrm{g}},7}$ is the gain radius in units of $10^7\,$cm and 
$T_{\mathrm{g}} = T(R_{\mathrm{g}})$ the temperature at the gain radius.
Depending on the position of the gain radius, 
$\ave{\mu_{\nu}}_{\mathrm{g}}$ is a 
factor somewhere between 0.25 (value at the neutrinosphere) and unity
(limit for $r\to \infty$).

\subsection{The EoS transition radius}
\label{sec:eostransition}

It is interesting to consider the conditions for which the pressure is
dominated by nonrelativistic nucleons or radiation plus
relativistic $e^\pm$ pairs ($kT \ga 0.5\,$MeV).
In the first case $P \approx P_{\mathrm{b}} = kT\rho/m_{\mathrm{u}}$, 
if nuclei are fully dissociated into free nucleons. In the latter case
$P \approx P_{\mathrm{r}} = P_{e^\pm}+P_{\gamma}\approx{11\over 12}a_{\gamma}(kT)^4$,
when $\eta_e\approx 0$ is again assumed for the electron degeneracy and the
constant is 
$a_{\gamma} = 8\pi^5/\eck{15(hc)^3} \approx 8.56\times 10^{31}\,$MeV$^{-3}$cm$^{-3}$. 
Setting $P_{\mathrm{b}}$ equal to $P_{\mathrm{r}}$ gives
\begin{equation}
{(kT)^3\over \rho} \,=\, {12\over 11}\,\,{1\over m_{\mathrm{u}}a_{\gamma}}
\label{eq:34}
\end{equation}
or, using the temperature 
$kT\approx kT_{\nu_e}\approx 4\,$MeV (compare Fig.~\ref{fig:2}),
\begin{equation}
\rund{{kT\over 4\,{\mathrm{MeV}}}}^{\! 3}\rho_{10}^{-1} \,\cong\, 1.2 \ .
\label{eq:35}
\end{equation}
This means that the transition from the baryon-dominated to the 
radiation-dominated regime occurs at a density significantly below that
of the neutrinosphere. The latter is typically above $10^{11}\,$g/cm$^3$. 
When the electron degeneracy is negligibly small, the contributions of
relativistic and nonrelativistic gas components
to the pressure are equal for a value of the radiation entropy per nucleon of
$s_{\mathrm{r}} = s_{e^\pm} + s_{\gamma}
\approx (\varepsilon_{\mathrm{r}} + P_{\mathrm{r}})/(kT\rho/m_{\mathrm{u}}) 
= 4P_{\mathrm{r}}/P_{\mathrm{b}} = 4$. 
Since the energy density of relativistic particles is
$\varepsilon_{\mathrm{r}} = 3P_{\mathrm{r}}$, 
whereas $\varepsilon_{\mathrm{b}}=3P_{\mathrm{b}}/2$
for nonrelativistic particles, the relativistic electrons-positron pairs 
and photons dominate the energy density at such conditions. The main 
contribution to the entropy, however, then still comes from nucleons and nuclei
(cf.\ Fig.\ 8 in Woosley et al.\ 1986).

\subsection{The shock radius and infall region}
\label{sec:shockradius}

Conservation of the mass flow, momentum flow and energy flow across the 
discontinuity of the shock front is expressed by the three 
Rankine-Hugoniot conditions
\begin{eqnarray}
\rho_{\mathrm{p}}u_{\mathrm{p}} &=& \rho_{\mathrm{s}}u_{\mathrm{s}} \ , 
\label{eq:36}\\
P_{\mathrm{p}} + \rho_{\mathrm{p}}u_{\mathrm{p}}^2 &=&
P_{\mathrm{s}} + \rho_{\mathrm{s}}u_{\mathrm{s}}^2 \ , 
\label{eq:37}\\
{1\over 2}\, u_{\mathrm{p}}^2 + w_{\mathrm{p}} - 
q_{\mathrm{d}} &=&
{1\over 2}\, u_{\mathrm{s}}^2 + w_{\mathrm{s}} \ ,
\label{eq:38}
\end{eqnarray}
where the indices p and s denote quantities just ahead and behind the shock,
respectively (see Fig.~\ref{fig:2}), 
$w = (\varepsilon + P)/\rho$ is the enthalpy per unit mass,
$q_{\mathrm{d}}$ the nuclear binding energy per unit mass absorbed by 
photodisintegration of nuclei within the shock front,
and $u = v - U_{\mathrm{s}}$ the fluid velocity 
relative to the shock when $U_{\mathrm{s}} = \dot R_{\mathrm{s}}$ is the
shock velocity and $v$ the gas velocity relative to the center of the star.
Note that in the infall region $v$ has negative sign.

With the definition $\beta \equiv \rho_{\mathrm{s}}/\rho_{\mathrm{p}}$, 
Eq.~(\ref{eq:36}) gives $u_{\mathrm{s}} = u_{\mathrm{p}}/\beta$, which can
be used to eliminate $u_{\mathrm{s}}$ from Eq.~(\ref{eq:37}). 
For a strong shock, i.e., $P_{\mathrm{s}}\gg P_{\mathrm{p}}$, this yields
\begin{equation}
P_{\mathrm{s}}\,\approx\, \rund{1-{1\over \beta}}
\rho_{\mathrm{p}}\rund{v_{\mathrm{p}}-U_{\mathrm{s}}}^2 \ .
\label{eq:39}
\end{equation}
Combining Eqs.~(\ref{eq:36})--(\ref{eq:38}) one further finds
\begin{equation}
w_{\mathrm{s}}-w_{\mathrm{p}} \,=\, {1\over 2}\rund{P_{\mathrm{s}}-P_{\mathrm{p}}}
\rund{{1\over \rho_{\mathrm{s}}} + {1\over \rho_{\mathrm{p}}}} - q_{\mathrm{d}}\ .
\label{eq:40}
\end{equation}
With $P_{\mathrm{p}}\ll P_{\mathrm{s}}$, $w_{\mathrm{p}}\ll w_{\mathrm{s}}$
and $w_{\mathrm{s}} \approx 4P_{\mathrm{s}}/\rho_{\mathrm{s}}$ for the
radiation-dominated gas in the postshock region, Eq.~(\ref{eq:40}) 
can be rewritten as
\begin{equation}
{\rho_{\mathrm{s}}\over \rho_{\mathrm{p}}}\,\approx\,7 + 
{2\,q_{\mathrm{d}}\,\rho_{\mathrm{s}} \over P_{\mathrm{s}}}\,\approx\,
{7 \over 4 - 3\sqrt{1 + 14\,q_{\mathrm{d}}/(9\,u_{\mathrm{p}}^2)}}\ ,
\label{eq:41}
\end{equation}
where in the second transformation Eq.~(\ref{eq:39}) was used to replace
$P_{\mathrm{s}}/\rho_{\mathrm{s}}$. This shows that for a relativistic gas
the density jump in a strong shock is a factor of 7. Energy consumed by
photodissociation of nuclei increases the density contrast between preshock
and postshock region (Thompson 2000).
In a more general treatment, retaining $w_{\mathrm{p}}$ and taking into account
the (subdominant)
contributions from nonrelativistic nucleons to the gas pressure behind the
shock (but still using $P_{\mathrm{p}} \ll \rho_{\mathrm{p}}u_{\mathrm{p}}^2$
in the infall region) one also derives the right hand side of
Eq.~(\ref{eq:41}), now with the expression
\begin{equation}
q^\ast \,\equiv\, q_{\mathrm{d}}-w_{\mathrm{p}}-{3\over 2\, m_{\mathrm{u}}}\,
\sum_iY_i\,kT_{\mathrm{s}}
\label{eq:42}
\end{equation}
instead of $q_{\mathrm{d}}$ in the denominator. This means that the density
discontinuity is also affected by the preshock enthalpy and the thermal pressure
of nucleons and nuclei behind the shock (the nuclear composition is accounted 
for by the sum of the number fractions, $\sum_iY_i$). 
Considering $q_{\mathrm{d}}$ to be several 
MeV/$m_{\mathrm{u}}$, $kT_{\mathrm{s}}\sim 1\,$MeV, and the preshock medium to be
dominated by relativistic, degenerate electrons in which case  
$w_{\mathrm{p}}\approx \zeta_e Y_e/m_{\mathrm{u}}$ with an electron chemical 
potential $\zeta_e = \eta_e kT$ of a few MeV, one can see that all terms in 
Eq.~(\ref{eq:42}) are of the same order and therefore equally important.

The preshock region is not affected by the postshock conditions. Because
the shock moves supersonically relative to the medium ahead of it, 
sound waves cannot transport information in this direction. The matter 
there falls into the shock with a significant fraction of the 
free-fall velocity,
\begin{equation}
v_{\mathrm{p}}\,=\,-\,\alpha\,\sqrt{{2\,G {\widetilde M}\over R_{\mathrm{s}}}}
\label{eq:43}
\end{equation}
with $\alpha\sim 1/\sqrt{2}$ (Bethe 1990, 1993; Bruenn 1993).
Ahead of the shock free nucleons are absent and therefore $\nu_e$ and $\bar\nu_e$
absorption does not play a role, but neutrinos interact with nuclei by coherent
scatterings. The opacity of the latter reaction scales roughly with $N^2/A$ when $N$
is the neutron number and $A$ the mass number of the nuclei, and the total neutrino 
opacity of the preshock medium turns out to be close to the result of 
Eq.~(\ref{eq:12}). Therefore the momentum transfer by neutrinos was again taken 
into account by using ${\widetilde M}$ instead of $M$ in Eq.~(\ref{eq:43}).
Plugging Eq.~(\ref{eq:43}) into the rate at which mass falls into the shock,
$\dot M = 4\pi R_{\mathrm{s}}^2\rho_{\mathrm{p}}v_{\mathrm{p}}(< 0)$,
gives the density just above the shock:  
\begin{equation}
\rho_{\mathrm{p}}\,=\,-\,{\dot M \over 4\pi\,\alpha\,
\sqrt{2\,G {\widetilde M}}\,R_{\mathrm{s}}^{3/2}} \ .
\label{eq:44}
\end{equation}
On the other hand, if the original presupernova material has a density
distribution $\rho_0(r_0) = H\,r_0^{-3}$ with $H$ being a constant, then
mass conservation yields a density at the footpoint of the shock at time $t$ 
after the start of the collapse of
\begin{equation}
\rho_{\mathrm{p}}\,=\,{2\over 3}\,{H \over \alpha\,\sqrt{2\,G {\widetilde M}}}\,\,
t^{-1}R_{\mathrm{s}}^{-3/2}
\label{eq:45}
\end{equation}
(Bethe 1990, 1993; see also Cooperstein et al.~1984). Comparing 
Eqs.~(\ref{eq:44}) and (\ref{eq:45}) one finds that the rate at which mass
crosses the shock in this case is
\begin{equation}
\dot M\,=\,-\,{8\pi\over 3}\,{H\over t}\ ,
\label{eq:46}
\end{equation}
which depends on the structure of the progenitor star through the constant
$H$ and decreases with time.

\section{Structure of the atmosphere}
\label{sec:atmosphere}

Within the supernova shock, the infalling matter is strongly decelerated 
to a velocity $v_{\mathrm{s}} = (v_{\mathrm{p}}-U_{\mathrm{s}})/\beta + 
U_{\mathrm{s}}$. For a stalled shock, $|v_{\mathrm{s}}| \ll |v_{\mathrm{p}}|$.
Compared to the internal energy and the gravitational energy, the kinetic
energy behind the shock is therefore negligibly small. The gas is
further slowed down as it moves inward and settles onto the nascent neutron
star. Between neutrinosphere and shock front ${\mathrm{d}}v/{\mathrm{d}}t 
\approx 0$ is therefore a good assumption, i.e., the stellar structure is
well approximated by hydrostatic
equilibrium (Chevalier 1989; Bethe 1993, 1995; Fryer et al.~1996). 
Combining Eqs.~(\ref{eq:2}) and (\ref{eq:3}) and using Eq.~(\ref{eq:6}),
the equation of hydrostatic equilibrium is found to be
\begin{equation}
-\,{1\over \rho}\,{\partial P\over \partial r} - {G\,\widetilde{M}\over r^2}
\,=\,0\ .
\label{eq:47}
\end{equation}
In the following, the solutions of this equation in the layers between 
neutrinosphere $R_{\nu}$ and EoS transition radius $R_{\mathrm{eos}}$
and between $R_{\mathrm{eos}}$ and shock position $R_{\mathrm{s}}$ will be 
derived.

\subsection{Hydrostatic equilibrium between $R_{\nu}$ and $R_{\mathrm{eos}}$}
\label{sec:hystat1}

When nonrelativistic baryons dominate the pressure and relativistic electrons
contribute, but positrons and radiation can be ignored because
the electrons are mildly degenerate, the pressure can be expressed as
\begin{eqnarray}
P\,\approx\,P_{\mathrm{b}} + P_{e^-} &=& {\rho\over m_{\mathrm{u}}}\,kT\,
\rund{1 + {Y_e\over 3}\,{{\cal F}_3(\eta_e)\over {\cal F}_2(\eta_e)}}
\nonumber \\
&\equiv& f_{\mathrm{g}}\,{\rho\over m_{\mathrm{u}}}\,kT \ .
\label{eq:48}
\end{eqnarray}
Since $Y_e$ and the electron degeneracy do not vary strongly, the factor 
$f_{\mathrm{g}}$ can be considered as constant.
Between $R_{\nu}$ and $R_{\mathrm{eos}}$ also the temperature is a slowly
changing quantity, $T(r)\approx T_{\nu} \approx T_{\nu_e}$ 
(compare Fig.~\ref{fig:2}). Hydrostatic equilibrium therefore implies
\begin{equation}
\rho(r)\,=\,\rho_{\nu}\,\exp\eck{-\,{G\,\widetilde{M}\,m_{\mathrm{u}}\over
f_{\mathrm{g}}kT_{\nu}R_{\nu}}\,\rund{1-{R_{\nu}\over r}}}\ ,
\label{eq:49}
\end{equation}
where $\rho_{\nu}$ is the density at the neutrinosphere. Near the
neutrinosphere, $r\approx R_{\nu}$, this can be approximated by 
\begin{eqnarray}
\rho(r) &\approx& \rho_{\nu}\,\exp\rund{-\,{x\over h}\,}
\label{eq:50} \\
{\mathrm{with}}&&  x \,\equiv\, r-R_{\nu} \ \ \ {\mathrm{and}}\ \ \
h \,\equiv\, f_{\mathrm{g}}\,{kT_{\nu}\, R_{\nu}^2 \over
G\,\widetilde{M}\,m_{\mathrm{u}}}\ . \nonumber 
\end{eqnarray}
Using typical numbers gives
\begin{equation}
h \,\approx\, 2.9\times 10^4\,f_{\mathrm{g}} R_{\nu,6}^2
\rund{{kT_{\nu}\over 4\,{\mathrm{MeV}}}}\!
\rund{{\widetilde{M}\over {\mathrm{M}}_{\odot}}}^{\!\! -1}
\ \ \eck{{\mathrm{cm}}}\ ,
\label{eq:51}
\end{equation}
where $R_{\nu,6}$ is the radius of the neutrinosphere in units of
$10^6\,$cm.

The density declines exponentially outside the neutrinosphere with a 
scale height $h \ll r$, forming a sharp ``cliff'' 
(Bethe \& Wilson 1985, Bethe 1990, Woosley 1993a). For this
reason the effective optical depth is dominated by the immediate vicinity
of the neutrinosphere. Therefore the integration in Eq.~(\ref{eq:13}) can be 
performed, using Eq.~(\ref{eq:50}) for the density in the effective 
opacity of Eq.~(\ref{eq:16}), to derive the neutrinospheric
density (normalized to $10^{10}\,$g/cm$^3$) as
\begin{equation}
\rho_{\nu, 10}\,\approx\,150\,f_{\mathrm{g}}^{-1}R_{\nu,6}^{-2}
\rund{{\widetilde{M}\over {\mathrm{M}}_{\odot}}}
\!\rund{{kT_{\nu}\over 4\,{\mathrm{MeV}}}}^{\!\! -3}\ .
\label{eq:52}
\end{equation}
This result confirms that the density of the transition from the
baryon-dominated to the radiation-dominated regime
[Eq.~(\ref{eq:35})] is significantly lower than $\rho_{\nu}$.

\subsection{Hydrostatic equilibrium between $R_{\mathrm{eos}}$ and $R_{\mathrm{s}}$}
\label{sec:hystat2}

In the radiation-dominated region a large part of the pressure is due to 
relativistic electron-positron pairs and photons, but also contributions
from nucleons and nuclei with number fractions $Y_i$ might not be 
negligible, therefore
\begin{eqnarray}
P &=& P_{\gamma} + P_{e^\pm} + P_{\mathrm{b}} \nonumber \\
  &=& {a_{\gamma}\over 3}\,(kT)^4
\eck{{11\over 4} + {15\over 2\pi^4}\,\eta_e^2\!\rund{\!\pi^2\!+\!{\eta_e^2\over 2}}
+ {3\,\rho kT\,\sum_i Y_i\over m_{\mathrm{u}}a_{\gamma}(kT)^4} } \nonumber\\
  &\equiv& P_{\mathrm{r}}\rund{1 + {4\over g_{\mathrm{r}}\,s_{\gamma}}}
 \,\equiv\, f_{\mathrm{r}}\,P_{\mathrm{r}} \ ,
\label{eq:53}
\end{eqnarray}
where $P_{\mathrm{r}}$ is the pressure associated with relativistic particles,
\begin{eqnarray}
P_{\mathrm{r}} &=& g_{\mathrm{r}}\,P_{\gamma}\,=\,
{1\over 3}\,a_{\gamma}\,g_{\mathrm{r}}\,(kT)^4 \ ,
\label{eq:54}\\
{\mathrm{with}}\quad
g_{\mathrm{r}} &\equiv& {11\over 4} + {15\over 2\pi^4}\,\eta_e^2
\rund{\pi^2 + {1\over 2}\,\eta_e^2} \ ,
\label{eq:55}
\end{eqnarray}
$s_{\gamma}$ is the entropy per nucleon carried by photons,
$s_{\gamma} = 4a_{\gamma}(kT)^3/(3\rho/m_{\mathrm{u}})$, and 
$\sum_i Y_i \approx 1$ because of the nearly complete disintegration of
nuclei. If both the factor $g_{\mathrm{r}}$ and $s_{\gamma}$ are constant
(which is roughly fulfilled in the radiation-dominated region between 
$R_{\mathrm{eos}}$ and $R_{\mathrm{s}}$ where the electron degeneracy
parameter $\eta_e$ divided by $\pi$ is small, and, as was discussed in 
Sect.~\ref{sec:physics}, convective processes tend to homogenize the 
total entropy and thus also the radiation entropy; see Bethe 1996b)
then also $f_{\mathrm{r}}$ can be considered as constant. In this case the 
pressure is simply proportional to $(kT)^4$, both for the contribution
from nucleons and for the contribution from photons plus electron-positron 
pairs (for a detailed discussion, see Bethe 1993).

This implies that the density $\rho$ is proportional to $T^3$, i.e.,
\begin{equation}
P\,=\, f_{\mathrm{r}}\,g_{\mathrm{r}}\,{a_{\gamma}\over 3}\,(kT)^4
 \,=\, K\,\rho^{4/3}\ .
\label{eq:56}
\end{equation}
Note that Eq.~(\ref{eq:56}) is valid more generally than
for radiation-dominated conditions (Bethe 1996b). Using 
\begin{equation}
n_e \,=\, Y_e\,{\rho\over m_{\mathrm{u}}}\,=\, {8\pi\over 3}\,
{(kT)^3\over (hc)^3}\,\eta_e(\pi^2+\eta_e^2) \ ,
\label{eq:56a}
\end{equation}
the coefficient $K$ can be determined as 
\begin{equation}
K\,=\, f_{\mathrm{r}}g_{\mathrm{r}}{a_{\gamma}\over 3}\,(hc)^4
\rund{{3\over 8\pi}\,{Y_e \over m_{\mathrm{u}}\eta_e(\pi^2 + \eta_e^2)}
}^{\! 4/3} .
\label{eq:56b}
\end{equation}
If $K$ in Eq.~(\ref{eq:56b}) is approximately constant, which is typically
fulfilled in the region between $R_{\mathrm{g}}$ and $R_{\mathrm{s}}$
(Bethe 1996b), Eq.~(\ref{eq:56}) is a useful representation of the 
equation of state.

With Eq.~(\ref{eq:47}), one can now determine the density
distribution between $R_{\mathrm{eos}}$ and $R_{\mathrm{s}}$ in 
hydrostatic equilibrium as
\begin{equation}
\rho(r)\,=\,\eck{\rho_{\mathrm{s}}^{1/3} + {1\over 4}\,
{G\,\widetilde{M}\over K}\rund{{1\over r}-{1\over R_{\mathrm{s}}}}}^3 \ .
\label{eq:57}
\end{equation}
Inserting this in Eq.~(\ref{eq:56}) and setting 
$K = P_{\mathrm{s}}/\rho_{\mathrm{s}}^{4/3}$, the pressure as a function of
radius is obtained,
\begin{equation}
P(r)\,=\,\eck{P_{\mathrm{s}}^{1/4} + {1\over 4}\,
{G\,\widetilde{M}\over K^{3/4}}\rund{{1\over r}-{1\over R_{\mathrm{s}}}}}^4 \ ,
\label{eq:58}
\end{equation}
and $kT(r)$ can also be found from Eq.~(\ref{eq:56}) as
\begin{equation}
kT(r)\,=\, kT_{\mathrm{s}} + {1\over 4}
\rund{{3\over f_{\mathrm{r}}\,g_{\mathrm{r}}\,a_{\gamma}}}^{\! 1/4}
{G\,\widetilde{M}\over K^{3/4}}\rund{{1\over r}-{1\over R_{\mathrm{s}}}}
\label{eq:59}
\end{equation}
when 
$kT_{\mathrm{s}}=\eck{3P_{\mathrm{s}}/(f_{\mathrm{r}}g_{\mathrm{r}}\,a_{\gamma})}^{1/4}$
is used. If the density-pressure relation is more general than
Eq.~(\ref{eq:56}), namely $P = K\rho^\gamma$ with $K$ being constant, hydrostatic 
equilibrium implies
\begin{equation}
\rho(r)\,=\,\eck{\rho_{\mathrm{s}}^{\gamma -1} + {\gamma\! -\! 1\over \gamma}\,
{G\,\widetilde{M}\over K}\rund{{1\over r}-{1\over R_{\mathrm{s}}}}
}^{1/(\gamma -1)} \ ,
\label{eq:59a}
\end{equation}
which replaces Eq.~(\ref{eq:57}).

Instead of the general solutions, Eqs.~(\ref{eq:57})--(\ref{eq:59}),
simple power-laws,
\begin{eqnarray}
\rho(r) &=& \rho_{\mathrm{s}}\rund{{R_{\mathrm{s}}\over r}}^{\! 3}\ ,\ \ 
kT(r)\,=\,kT_{\mathrm{s}}\,{R_{\mathrm{s}}\over r}\ ,\nonumber \\
P(r) &=& P_{\mathrm{s}}\rund{{R_{\mathrm{s}}\over r}}^{\! 4}\ ,
\label{eq:60}
\end{eqnarray}
yield a good approximation for the hydrostatic atmosphere, if 
$K$ fulfills the condition 
$K = G\widetilde{M}/(4 R_{\mathrm{s}}\rho_{\mathrm{s}}^{1/3})$. 
Since $K = P_{\mathrm{s}}/\rho_{\mathrm{s}}^{4/3}$, this is equivalent to
the requirement that 
\begin{equation}
{P_{\mathrm{s}}\over \rho_{\mathrm{s}}}\,=\, 
{G\,\widetilde{M}\over 4\,R_{\mathrm{s}}}\ .
\label{eq:61}
\end{equation}
On the other hand, from Eqs.~(\ref{eq:39}) and (\ref{eq:43}) one gets
\begin{eqnarray}
{P_{\mathrm{s}}\over \rho_{\mathrm{s}}} &\approx&
{\beta-1\over \beta^2}\,{2\alpha^2 G\,{\widetilde M}\over R_{\mathrm{s}}}
\rund{1-{U_{\mathrm{s}}\over v_{\mathrm{p}}}}^{\! 2} \nonumber \\
&\sim& {6\over 49}\,{G\,{\widetilde M}\over R_{\mathrm{s}}}
\rund{1-{U_{\mathrm{s}}\over v_{\mathrm{p}}}}^{\! 2}\ .
\label{eq:62}
\end{eqnarray}
The numerical factor on the right hand side of Eq.~(\ref{eq:62})
was obtained with $\alpha \sim 1/\sqrt{2}$ (Bethe 1990, 1993) and 
$\beta \sim 7$ [Eq.~(\ref{eq:41})]. Equation~(\ref{eq:62}) shows
that the requirement of Eq.~(\ref{eq:61}) is reasonably well, although
for small shock radii not very well, fulfilled. In the following the 
power-laws of 
Eq.~(\ref{eq:60}) will therefore only serve to facilitate analytical 
evaluation, and the use of this approximate solution of hydrostatic 
equilibrium will be avoided where inconsistencies might result.

With Eqs.~(\ref{eq:33}) and (\ref{eq:60}) the gain radius $R_{\mathrm{g}}$ 
and the conditions at the gain radius can be expressed in terms of the
properties at the shock front and the characteristic parameters
($T_{\nu_e}$, $L_{\nu_e}$) of the neutrino emission. Inserting the relation
$kT_{\mathrm{g}} = kT_{\mathrm{s}}(R_{\mathrm{s}}/R_{\mathrm{g}})$ 
into Eq.~(\ref{eq:33}) yields the gain radius (in units of $10^7\,$cm),
\begin{equation}
R_{{\mathrm{g}},7}\,\approx\,0.98\,\,R_{{\mathrm{s}},7}^{3\over 2}
\,(kT_{{\mathrm{s}},2})^{3\over 2}\,(kT_{\nu_e,4})^{-{1\over 2}}
\rund{{L_{\nu_e,52}\over \ave{\mu_{\nu}}_{\mathrm{g}}}}^{\! -{1\over 4}} ,
\label{eq:63}
\end{equation}
and for the temperature at the gain radius one gets
\begin{equation}
kT_{{\mathrm{g}},2}\,\approx\,
1.02\,\,R_{{\mathrm{s}},7}^{-{1\over 2}}\,
(kT_{{\mathrm{s}},2})^{-{1\over 2}}\,(kT_{\nu_e,4})^{1\over 2}
\rund{{L_{\nu_e,52}\over \ave{\mu_{\nu}}_{\mathrm{g}}}}^{\! {1\over 4}} ,
\label{eq:64}
\end{equation}
where $kT_{{\mathrm{g}},2} = kT_{\mathrm{g}}/(2\,{\mathrm{MeV}})$
and $kT_{\nu_e,4}$ is the neutrinospheric temperature and
$kT_{{\mathrm{s}},2}$ the postshock temperature normalized to
4$\,$MeV and 2$\,$MeV, respectively.

The assumptions made in this section to solve the equation of hydrostatic
equilibrium in the layer between $R_{\mathrm{eos}}$ and $R_{\mathrm{s}}$
do not seem to be very restrictive, because two-dimensional as well as
one-dimensional simulations without convection (e.g., Bruenn 1993;
Janka \& M\"uller 1996, Fig.~6; Rampp 2000) yield
density and temperature profiles in the postshock region which are very 
close to power laws with power law indices around 3 and 1, respectively. 
Near $R_{\mathrm{eos}}$ the contributions of relativistic and nonrelativistic
gas components will become equally important. Here the exponentially steep
density decline just outside the neutrinosphere
must change to the power-law behavior behind the shock, and
both of these limiting solutions will not provide a good description.
The exact structure in the intermediate layer between 
$R_{\mathrm{eos}}$ and $R_{\mathrm{g}}$, however, does not play an important
role in the further discussion and therefore a more accurate treatment is not 
necessary.

\section{Heating and cooling}
\label{sec:heatcool}

To discuss energy deposition and emission of neutrinos exterior to the 
neutrinosphere, one starts with the energy equation for $\nu_e$ plus 
$\bar\nu_e$, which is
\begin{equation}
{1\over 4\pi\,r^2}\,{\partial L_{\nu}\over \partial r}\,=\,-\,Q_{\nu}^+
+ Q_{\nu}^- \ ,
\label{eq:65}
\end{equation}
where $L_{\nu} = L_{\nu_e}+L_{\bar\nu_e}$ 
and $Q_{\nu}^+$ and $Q_{\nu}^-$ are the heating and cooling rates of the
stellar medium as given by Eqs.~(\ref{eq:28}) and (\ref{eq:31}), respectively.
In writing Eq.~(\ref{eq:65}), stationarity was assumed for the neutrinos,
which is justified because the neutrino emission of the accreting 
neutrino star changes on a timescale which is typically longer than other
relevant timescales of the discussed problem. 
From Eq.~(\ref{eq:65}) the net effect of heating or cooling
in a layer between radii $r_1$ and $r_2$ can be deduced as
\begin{equation}
\int\limits_{r_1}^{r_2}{\mathrm{d}}r\,4\pi\,r^2\rund{Q_{\nu}^+ - Q_{\nu}^-}
\,=\, L_{\nu}(r_1)\,-\,L_{\nu}(r_2)\ .
\label{eq:68}
\end{equation}
Refering to Eqs.~(\ref{eq:23}) and (\ref{eq:27}), a suitable spectral 
and flavor average for the absorption coefficient of $\nu_e$ and 
$\bar\nu_e$ can be defined as
\begin{equation}
\ave{\kappa_{\mathrm{a}}} \,=\,{Q_{\nu}^+\over L_{\nu}/
(4\pi\,r^2\ave{\mu_{\nu}})} \ ,
\label{eq:66}
\end{equation}
when $\ave{\mu_{\nu_e}} \approx \ave{\mu_{\bar\nu_e}} \equiv \ave{\mu_{\nu}}$
is used. Plugging this into Eq.~(\ref{eq:65}) gives
\begin{equation}
{\partial L_{\nu}\over \partial r}\,=\,-\,\ave{\kappa_{\mathrm{a}}}\,
{L_{\nu}\over \ave{\mu_{\nu}}}\,+\,4\pi\,r^2 Q_{\nu}^- \ .  
\label{eq:67}
\end{equation}
The neutrino luminosity as a function of radius $r \ge r_0$ 
is the general solution of Eq.~(\ref{eq:67}):
\begin{eqnarray}
L_{\nu}(r) &=& \exp\left\lbrace -\int\limits_{r_0}^r {\mathrm{d}}r'\,
{\ave{\kappa_{\mathrm{a}}}\over \ave{\mu_{\nu}}} \right\rbrace\,
L_{\nu}(r_0) \nonumber \\
&+& \int\limits_{r_0}^r {\mathrm{d}}r'\,4\pi\,(r')^2 Q_{\nu}^-
\exp\left\lbrace -\int\limits_{r'}^r {\mathrm{d}}r''\,
{\ave{\kappa_{\mathrm{a}}}\over \ave{\mu_{\nu}}} \right\rbrace \ .
\label{eq:69}
\end{eqnarray}
The first exponential factor represents the absorption damping of the 
luminosity in the shell between $r_0$ and $r$, the
second exponential factor the reabsorption of neutrinos emitted
at $r'$ in the layer enclosed by radii $r'$ and $r$.

\subsection{Heating and cooling between $R_{\nu}$ and $R_{\mathrm{g}}$}
\label{heatcool1}

Here the lower boundary of the considered volume is the neutrinosphere at
radius $r_0 = R_{\nu}$.
Since both $\ave{\kappa_{\mathrm{a}}}\propto \rho(r)$ [cf.\ Eqs.~(\ref{eq:66})
and (\ref{eq:28})] and $Q_{\nu}^-\propto \rho(r)$ [cf.\ Eq.~(\ref{eq:31})]
are steep functions of the radius in the region between $R_{\nu}$ and 
$R_{\mathrm{g}}$, where the density drops exponentially, most of the absorption
and emission occurs in the immediate vicinity of the neutrinosphere.
Therefore the neutrino luminosity at the gain radius, $L_{\nu}(R_{\mathrm{g}})$,
can be approximated by the limit for $r\to \infty$ of Eq.~(\ref{eq:69}), and
the integral $\int_{r'}^r{\mathrm{d}}r''\,\ave{\kappa_{\mathrm{a}}}/\ave{\mu_{\nu}}$
can be replaced by
$\int_{R_{\nu}}^\infty{\mathrm{d}}r\,\ave{\kappa_{\mathrm{a}}}/\ave{\mu_{\nu}}$.
This leads to
\begin{eqnarray}
L_{\nu}(R_{\mathrm{g}}) 
&\approx& \exp\left\lbrace -\int\limits_{R_{\nu}}^\infty {\mathrm{d}}r\,
{\ave{\kappa_{\mathrm{a}}}\over \ave{\mu_{\nu}}} \right\rbrace \times \nonumber\\
&\phantom{\times}&\quad 
\times \eck{L_{\nu}(R_{\nu})\,+\,\int\limits_{R_{\nu}}^\infty{\mathrm{d}}r\,
4\pi\,r^2 Q_{\nu}^-}\, .
\label{eq:70}
\end{eqnarray}
To evaluate the exponential damping factor, $\ave{\kappa_{\mathrm{a}}}$ is
expressed by Eq.~(\ref{eq:66}), making use of Eq.~(\ref{eq:28}) and 
$L_{\nu} = 2\,L_{\nu_e}$. The neutrino spectrum is assumed not to change
outside the neutrinosphere. With Eq.~(\ref{eq:50}) the integral over the
density profile becomes
$\int_{R_{\nu}}^\infty{\mathrm{d}}r\,\rho(r)\approx \rho_{\nu}h$. 
Employing Eqs.~(\ref{eq:51}) and (\ref{eq:52}), one finds
\begin{equation}
\int\limits_{R_{\nu}}^\infty{\mathrm{d}}r\,
{\ave{\kappa_{\mathrm{a}}}\over \ave{\mu_{\nu}}}\,\approx\,{0.42\over 
\widetilde{\ave{\mu_{\nu}}}}\,\equiv\, a \ ,
\label{eq:71}
\end{equation}
where $\widetilde{\ave{\mu_{\nu}}}$ denotes a radial average of
the flux factor $\ave{\mu_{\nu}}$ in the layer between $R_{\nu}$ and
$R_{\mathrm{g}}$.
The energy loss integral is calculated with Eq.~(\ref{eq:31}) where
$T(r)\approx T_{\nu_e}$ is used near the neutrinosphere. With 
$\int_{R_{\nu}}^\infty {\mathrm{d}}r\,r^2\rho(r)\approx \rho_{\nu}h
\eck{h^2+(R_{\nu}+h)^2}\approx \rho_{\nu}R_{\nu}^2h$ [because $h\ll R_{\nu}$,
cf.\ Eq.~(\ref{eq:51})] and Eqs.~(\ref{eq:51}) and (\ref{eq:52}) this leads to:
\begin{equation}
\int\limits_{R_{\nu}}^\infty\! {\mathrm{d}}r\,4\pi\,r^2 Q_{\nu}^- \approx
4.9\times 10^{51} R_{\nu,6}^2 \!
\rund{{kT_{\nu_e}\over 4\,{\mathrm{MeV}}}}^{\! 4}
\eck{{{\mathrm{erg}}\over {\mathrm{s}}}} \equiv b .
\label{eq:71a}
\end{equation}
Eq.~(\ref{eq:70}) now becomes
\begin{equation}
L_{\nu}(R_{\mathrm{g}})\,\approx\,{\mathrm{e}}^{-a} \eck{L_{\nu}(R_{\nu})\,+\,b}\, ,
\label{eq:72}
\end{equation}
and the total energy exchange between $R_{\nu}$ and $R_{\mathrm{g}}$ 
according to Eq.~(\ref{eq:68}) therefore is
\begin{equation}
L_{\nu}(R_{\mathrm{g}})-L_{\nu}(R_{\nu})\equiv L_{\mathrm{acc}}
\approx ({\mathrm{e}}^{-a}\!-1)L_{\nu}(R_{\nu}) + {\mathrm{e}}^{-a}b .
\label{eq:73}
\end{equation}
Since $R_{\mathrm{g}}$ separates the layer of neutrino cooling from the 
one of neutrino heating, the region between $R_{\nu}$ and
$R_{\mathrm{g}}$ must lose energy by neutrino emission.  
Therefore the neutrino luminosity at $R_{\mathrm{g}}$
must be larger than $L_{\nu}(R_{\nu})$, and $L_{\mathrm{acc}}$
represents the luminosity associated with the
accretion of matter through the gain radius onto the surface of
the nascent neutron star. The requirement $L_{\mathrm{acc}} \ge 0$ 
constrains the luminosity of the neutron star core relative to the 
product $R_{\nu}^2(kT_{\nu_e})^4$ by the inequality 
\begin{equation}
L_{\nu}(R_{\nu})\,\le\,b\,\rund{{\mathrm{e}}^a-1}^{-1}\,.
\label{eq:74}
\end{equation}
Provided the core luminosity can be expressed in terms of blackbody emission
of temperature $T_{\nu_e}$,
\begin{eqnarray}
L_{\nu}(R_{\nu}) &\approx& 4\pi R_{\nu}^2\,{c\over 4}\,{7\over 8}\,a_{\gamma}
(kT_{\nu_e})^4  \nonumber \\
&\approx& 2.9\times 10^{51}\,R_{\nu,6}^2\rund{{kT_{\nu_e}
\over 4\,{\mathrm{MeV}}}}^{\! 4}\ \ \eck{{{\mathrm{erg}}\over {\mathrm{s}}}}\,,
\label{eq:75}
\end{eqnarray}
the consistency condition translates into the relation 
\begin{equation}
\widetilde{\ave{\mu_{\nu}}}\,\ga\,0.42\, ,
\label{eq:76}
\end{equation}
which is satisfied above the neutrinosphere in the layer between $R_{\nu}$ and
$R_{\mathrm{g}}$.

\subsection{Heating and cooling between $R_{\mathrm{g}}$ and $R_{\mathrm{s}}$}
\label{heatcool2}

For reasons of simplicity it will be assumed that in the layer bounded by
$R_{\mathrm{g}}$ and $R_{\mathrm{s}}$ nuclei are completely disintegrated 
into free nucleons. Disregarding the occurrence of $\alpha$ particles, in 
particular, is certainly an approximation which becomes invalid when the 
temperature drops below about 1$\,$MeV, i.e., when the shock is at large
radii, typically around 300$\,$km (see Bethe 1993, 1995, 1996a,b,c, 1997). 
The presence of $\alpha$ particles reduces the neutrino heating, because 
electron neutrinos and antineutrinos are absorbed only on nucleons, but energy
released by the recombination of $\alpha$'s during shock expansion supports
the shock 
at a later stage and contributes to the energy budget of the explosion. Since
in the context of this paper we do not attempt to calculate the explosion
energy, but are interested in a qualitative discussion of the revival phase 
of the stalled shock, the recombination of nucleons to $\alpha$ particles 
is probably not a crucial issue.

As will be demonstrated below, the optical depth between $R_{\mathrm{g}}$ and
$R_{\mathrm{s}}$ is small such that $\int_{R_{\mathrm{g}}}^{R_{\mathrm{s}}} 
{\mathrm{d}}r\, \ave{\kappa_{\mathrm{a}}}/\ave{\mu_\nu} \la 0.5$. Therefore
the reabsorption probability of emitted neutrinos is also small and an 
approximation to the solution of Eq.~(\ref{eq:69}) at the shock position is
\begin{equation}
L_{\nu}(R_{\mathrm{s}})\,\approx\,
\rund{\! 1-\! \int\limits_{R_{\mathrm{g}}}^{R_{\mathrm{s}}}
\! {\mathrm{d}}r\,{\ave{\kappa_{\mathrm{a}}}\over \ave{\mu_\nu}} \!} 
\! L_{\nu}(R_{\mathrm{g}})\, + 
\int\limits_{R_{\mathrm{g}}}^{R_{\mathrm{s}}} \! {\mathrm{d}}r\,
4\pi\,r^2 Q_{\nu}^- .
\label{eq:77}
\end{equation}
The net energy deposition is found to be
\begin{eqnarray}
L_{\nu}(R_{\mathrm{g}})-L_{\nu}(R_{\mathrm{s}}) &\approx&
\int\limits_{R_{\mathrm{g}}}^{R_{\mathrm{s}}} \! {\mathrm{d}}r\,
{\ave{\kappa_{\mathrm{a}}}\over \ave{\mu_\nu}}\,\,L_{\nu}(R_{\mathrm{g}})  -
\int\limits_{R_{\mathrm{g}}}^{R_{\mathrm{s}}} \! {\mathrm{d}}r\,
4\pi\,r^2 Q_{\nu}^- \nonumber \\
&\equiv& {\cal H}-{\cal C} \, .
\label{eq:78}
\end{eqnarray}
Since in the layer bounded by $R_{\mathrm{g}}$ and $R_{\mathrm{s}}$ neutrino
heating takes place, $L_{\nu}(R_{\mathrm{g}}) \ge L_{\nu}(R_{\mathrm{s}})$
(this will also be verified below).
Expanding the exponential damping factors only to lowerst
order in the exponent implies a slight overestimation of the energy
input into the stellar medium by neutrinos (because the luminosity entering
at $R_{\mathrm{g}}$ is assumed to decay linearly through the layer), but also
the energy loss by neutrinos is overestimated, because the reabsorption
of emitted neutrinos is not included.

Using $Q_{\nu}^+$ from Eq.~(\ref{eq:28}) in Eq.~(\ref{eq:66}), 
$L_{\nu} = 2L_{\nu_e}$, and the density profile from 
Eq.~(\ref{eq:60}), one finds for the first integral in Eq.~(\ref{eq:78}):
\begin{eqnarray}
{\cal H} \,\approx\,
4.9\times 10^{50}\, {L_{\nu,52}(R_{\mathrm{g}})\over 
\ave{\mu_{\nu}}^\ast} &\!\!\!\!&
(kT_{\nu_e,4})^2\rho_{{\mathrm{s}},9} R_{{\mathrm{s}},7}  \nonumber \\
\times &\!\!\!\! & \eck{\rund{{R_{\mathrm{s}}\over R_{\mathrm{g}}}}^{\! 2}-1}\ \
\eck{{{\mathrm{erg}}\over {\mathrm{s}}}} , 
\label{eq:79}  
\end{eqnarray}
where $kT_{\nu_e}$ was again treated as a constant, $\ave{\mu_\nu}^\ast$
defines an average value of the flux factor in the layer between
$R_{\mathrm{g}}$ and $R_{\mathrm{s}}$, and $\rho_{{\mathrm{s}},9}$ is the
density behind the shock in units of $10^9\,$g$\,$cm$^{-3}$.
The second integral in Eq.~(\ref{eq:78}) is evaluated with $Q_{\nu}^-$
from Eq.~(\ref{eq:31}) and the temperature and density relations from 
Eq.~(\ref{eq:60}):
\begin{eqnarray}
{\cal C} \,\approx\,
2.9\times 10^{50}\, (kT_{{\mathrm{s}},2})^6
&\!\!\!\!& \rho_{{\mathrm{s}},9} R_{{\mathrm{s}},7}^3 \nonumber \\
\times &\!\!\!\! & \eck{\rund{{R_{\mathrm{s}}\over R_{\mathrm{g}}}}^{\! 6}-1}\ \
\eck{{{\mathrm{erg}}\over {\mathrm{s}}}} .
\label{eq:80} 
\end{eqnarray}
Employing the gain condition, Eq.~(\ref{eq:33}), and again making use of 
Eq.~(\ref{eq:60}), one gets
\begin{equation}
\rund{{kT_{\mathrm{s}}\over 2\,{\mathrm{MeV}}}}^{\! 6}\,\approx\,
0.55\,\,{L_{\nu,52}(R_{\mathrm{g}})\over \ave{\mu_\nu}_{\mathrm{g}}
R_{{\mathrm{g}},7}^2} \rund{{kT_{\nu_e}\over 4\,{\mathrm{MeV}}}}^{\! 2}
\rund{{R_{\mathrm{g}}\over R_{\mathrm{s}}}}^{\! 6} ,
\label{eq:81}
\end{equation}
which serves to rewrite Eq.~(\ref{eq:80}) as
\begin{eqnarray}
{\cal C} \,\approx\,
1.6\times 10^{50}&\!\!\!\! & {L_{\nu,52}(R_{\mathrm{g}})\over
\ave{\mu_{\nu}}_{\mathrm{g}}}\,
(kT_{\nu_e,4})^2\rho_{{\mathrm{s}},9} R_{{\mathrm{s}},7}  \nonumber \\
\times &\!\!\!\! & \rund{{R_{\mathrm{s}}\over R_{\mathrm{g}}}}^{\! 2}
\! \eck{1-\rund{{R_{\mathrm{g}}\over R_{\mathrm{s}}}}^{\! 6}}\ \,
\eck{{{\mathrm{erg}}\over {\mathrm{s}}}}\! .
\label{eq:82}
\end{eqnarray}
Combining Eqs.~(\ref{eq:79}) and (\ref{eq:82}) gives the net energy
transfer to the stellar medium in the gain region:
\begin{equation}
{\cal H} - {\cal C} \,\approx\, {\cal H}\! \times \!
\left\lbrace 1 - {1\over 3}\,{\ave{\mu_\nu}^\ast
\over \ave{\mu_\nu}_{\mathrm{g}}}\! \eck{1 \! + \!
\rund{{R_{\mathrm{g}}\over R_{\mathrm{s}}}}^{\! 2} \! + \!
\rund{{R_{\mathrm{g}}\over R_{\mathrm{s}}}}^{\! 4} }\right\rbrace \!.
\label{eq:83}
\end{equation}
Since $\ave{\mu_\nu}^\ast/\ave{\mu_\nu}_{\mathrm{g}}\sim 1$ and 
$R_{\mathrm{s}} > R_{\mathrm{g}}$, typically 
$R_{\mathrm{s}}\sim 2R_{\mathrm{g}}$, we verify that ${\cal H}-{\cal C}>0$
and therefore $L_{\nu}(R_{\mathrm{s}}) < L_{\nu}(R_{\mathrm{g}})$,
as expected for the neutrino heating region. 
For $\beta = \rho_{\mathrm{s}}/\rho_{\mathrm{p}} \approx 7$ [Eq.~(\ref{eq:41})]
and $\alpha = 1/\sqrt{2}$, Eq.~(\ref{eq:44}) yields for the postshock
density
\begin{equation}
\rho_{\mathrm{s}}\,\approx\, 3\times 10^9\,\,{(-\dot M)\over {\mathrm{M}}_\odot
/{\mathrm{s}}}\,\rund{{\widetilde{M}\over {\mathrm{M}}_\odot}}^{\! - 1/2}\!
R_{{\mathrm{s}},7}^{-3/2}\ \eck{{{\mathrm{g}}\over {\mathrm{cm}}^3}}\,.
\label{eq:84}
\end{equation}
Using this and $R_{\mathrm{s}}\sim 2R_{\mathrm{g}}$ in Eq.~(\ref{eq:79}) 
leads to
\begin{eqnarray}
\int\limits_{R_{\mathrm{g}}}^{R_{\mathrm{s}}} {\mathrm{d}}r\,
{\ave{\kappa_{\mathrm{a}}}\over \ave{\mu_{\nu}}} & = &
{{\cal H}\over L_{\nu}(R_{\mathrm{g}})} \nonumber \\
&\sim &
{0.44\over \ave{\mu_\nu}^\ast}\,
{(kT_{\nu_e,4})^2\over R_{{\mathrm{s}},7}^{1/2}}\,
{(-\dot M)\over {\mathrm{M}}_\odot /{\mathrm{s}}}\,
\rund{{\widetilde{M}\over {\mathrm{M}}_\odot}}^{\!-{1\over 2}}\!\!\! .
\label{eq:85}
\end{eqnarray}
For sufficiently small accretion rates $|\dot M|$ this is less than
about 0.5, and the assumption made before Eq.~(\ref{eq:77}) is verified,
i.e., the reabsorption of neutrinos emitted in the gain region can be 
neglected.

\section{Mass accretion onto the neutron star}
\label{sec:massacc}

The shock accretes mass at a rate 
$\dot M \equiv 4\pi R_{\mathrm{s}}^2\rho_{\mathrm{p}}v_{\mathrm{p}}$ as determined 
by the conditions in the core of the progenitor star   
(see Sect.~\ref{sec:shockradius}). In a stationary state, this rate
is equal to the rate at which matter is advected inward from the
shock to the neutrinosphere to be finally added into the neutron
star. The rate at which matter can be absorbed by the neutron star, however,
depends on the efficiency by which neutrinos are able to remove the energy
excess of the infalling material relative to the energy of the strongly bound
matter in the neutron star surface layers. For the
large accretion rates typical of the collapsed stellar core right after
bounce, the density is so high that the infalling matter becomes opaque 
to neutrinos. In this case the efficiency of the energy loss is reduced. 
When the gas is hotter, 
the neutrino opacity increases (because of the energy dependence of the
neutrino cross sections), and the neutrinosphere moves to a larger radius.
Due to this regulatory effect, the neutrinospheric temperature is a rather
inert quantity and, e.g., turns out to be very similar in different 
numerical models.
Therefore it is not a steady-state mass accretion rate which governs the
temperature at the base of the ``atmosphere'' (as for accretion in optically
thin conditions), but the ``surface'' of the nascent neutron star forms
where the temperature is sufficiently high for neutrino opaqueness to set in. 
 
When neutrino cooling is not efficient enough, the advection of matter
through the neutrino cooling region is reduced compared to the accretion 
into the shock, and matter piles up on top of the neutron star. Similarly,
strong neutrino heating in the gain region can reduce the inflow of matter. 
The transition from accretion to an explosion is characterized by 
an inversion of infall to outflow. For this reason the analysis of 
the conditions for shock revival requires the inclusion of this sort of
time-dependence in the discussion. In the simplified model considered here,
the mass accretion rate is allowed to change between $R_{\mathrm{s}}$ and
$R_{\mathrm{g}}$. Matter advected through $R_{\mathrm{g}}$ at a rate 
determined by the efficiency of neutrino cooling is then assumed to be 
added into the neutron star (compare Fig.~\ref{fig:2}).
 
Using Eqs.~(\ref{eq:2}) and (\ref{eq:6}) and the definition 
$\dot M(r) = 4\pi r^2\!\rho v$, Eq.~(\ref{eq:4}) can be rewritten in
the following form:
\begin{eqnarray}
{\partial \over \partial r}\eck{\dot M \rund{{e\! +\! P\over \rho}-
{G\widetilde{M}\over r}}} &=& 4\pi r^2 (Q_{\nu} + Q_{\mathrm{d}})
\nonumber \\ 
& - & 4\pi r^2\!\rho\, {\partial \over \partial t}\rund{{e\over \rho}}
\nonumber \\
& + & \rund{{e\over \rho} - {G\widetilde{M}\over r}}{\partial\dot M
\over \partial r} \ ,
\label{eq:86}
\end{eqnarray}
where $Q_{\nu} = Q_{\nu}^+ - Q_{\nu}^-$ is the net rate (per unit volume) 
of energy transfer between
neutrinos and the stellar medium and $Q_{\mathrm{d}}$ denotes the energy 
consumed or released
by the photodisintegration of nuclei. The latter term has to be 
introduced in the equation when rest-mass contributions from nucleons and 
nuclei are not included in the internal energy density $\varepsilon$
[Eq.~(\ref{eq:5})]. The nuclei present in the accretion flow through
the shock are assumed to be dissociated to free nucleons within the shock
front [cf.\ Eq.~(\ref{eq:38})]. Therefore the rate $Q_{\mathrm{d}}$ in terms
of the (positive) nuclear binding energy per unit mass, $q_{\mathrm{d}}$, is
\begin{equation}
Q_{\mathrm{d}}\,=\,\rho\, v\, q_{\mathrm{d}}\,\delta(r-R_{\mathrm{s}})\,.
\label{eq:87}
\end{equation}
Here $\delta(x)$ is the delta function. For $v < 0$, which is true in
case of accretion, energy is extracted from the stellar medium, i.e., 
$Q_{\mathrm{d}} < 0$. Now integrating Eq.~(\ref{eq:86}) between $R_{\nu}$ and
a radius $r$ that is infinitesimally larger than $R_{\mathrm{s}}$ gives
\begin{eqnarray}
\dot M \eck{{e\! +\! P\over \rho} - {G\widetilde{M}\over r}}_{R_{\mathrm{s}}}
\!\!\! & - & \dot M' \eck{{e\! +\! P\over \rho} - 
{G\widetilde{M}\over r}}_{R_{\nu}} = \nonumber \\
\dot M'\,q_{\mathrm{d}} 
& + &\! \int\limits_{M(R_{\nu})}^{M(R_{\mathrm{s}})}\!\!\! {\mathrm{d}}M \!
\eck{{Q_{\nu}\over \rho}
-{\partial\over \partial t}\rund{{e\over \rho}}} \nonumber \\
& + & \int\limits_{\dot M'}^{\dot M}\, {\mathrm{d}}\dot M \!
\rund{q_{\mathrm{d}} + {e\over \rho} - {G\widetilde{M}\over r}}\,,
\label{eq:88}
\end{eqnarray}
where $\partial M/\partial r = 4\pi r^2\!\rho$ was used. The mass
accretion rate through the shock was defined as
$\dot M = 4\pi R_{\mathrm {s}}^2\rho_{\mathrm{p}}v_{\mathrm{p}}$ 
and the corresponding accretion rate through the neutrinosphere as
$\dot M' \equiv \dot M(R_{\nu}) = 4\pi R_{\nu}^2\rho_{\nu} v_{\nu}$.
The term for the rate of energy consumption
by nuclear dissociation was split into two parts according
to $\dot M(r) = \dot M' + \int_{\dot M'}^{\dot M(r)}{\mathrm{d}}{\dot M}$.

From Eq.~(\ref{eq:88}) an approximation for $\dot M'$ can be derived by
taking into account that $|Q_{\nu}/\rho | \gg \partial (e/\rho)/
\partial t$ in the region between $R_{\nu}$ and $R_{\mathrm{s}}$,
where strong neutrino heating and cooling occurs. 
Moreover, the integrand of the last term on the right hand
side of Eq.~(\ref{eq:88}) is usually small, because $q_{\mathrm{d}}$
corresponds to about 8--9 MeV per nucleon for complete disintegration
of nuclei into free nucleons, $G\widetilde{M}/R_{\mathrm{s}}\sim
14\,(\widetilde{M}/{\mathrm{M}}_{\odot})/R_{{\mathrm{s}},7}$ MeV per 
nucleon, and $e/\rho \approx {1\over 2}v_{\mathrm{p}}^2 \sim {1\over 2}
G\widetilde{M}/R_{\mathrm{s}}$ immediately above the shock, where the 
infall velocity $v_{\mathrm{p}}$ is given by Eq.~(\ref{eq:43}) and the
specific internal energy is typically much smaller than the specific 
kinetic energy. For the same reason, the first term on the left hand
side of Eq.~(\ref{eq:88}) is much smaller than the second term when
$\dot M$ and $\dot M'$ are of the same order.
With all this one gets
\begin{equation}
\dot M'\,\approx\, -\int\limits_{R_{\nu}}^{R_{\mathrm{s}}}{\mathrm{d}}
r\,4\pi r^2\,Q_{\nu}\times \rund{\eck{{e\!+\! P\over \rho}}_{\! R_{\nu}}
\!\!\! - {G\widetilde{M}\over R_{\nu}} + q_{\mathrm{d}}}^{\!\! -1} 
\!\!\! .
\label{eq:89}
\end{equation}
Because of the large gravitational binding energy of matter at the 
neutrinosphere, the term in brackets in Eq.~(\ref{eq:89}) is negative. The
integral adds up the contributions from neutrino cooling between $R_{\nu}$ 
and $R_{\mathrm{g}}$ and from neutrino heating between $R_{\mathrm{g}}$ and 
$R_{\mathrm{s}}$. If cooling is stronger (which is the case in the first
second after bounce), the integral is negative and $\dot M' < 0$, i.e., the
neutron star accretes matter. If neutrino heating dominates, there is
mass outflow, $\dot M' > 0$. 

Such mass loss takes place during the later
phase of the neutrino-cooling evolution of the nascent neutron star, where a
baryonic wind, the so-called neutrino-driven wind, is blown off the 
neutron star surface due to neutrino energy deposition just outside the 
neutrinosphere (Qian \& Woosley 1996). The transition from accretion
to mass outflow and the onset of mass loss can be discussed with
the formulae presented here. A description of the wind regime (where the
fluid velocity $v$ approaches the local speed of sound), however, is beyond
the scope of the present work, because it requires retaining
the velocity gradient in the momentum equation, Eq.~(\ref{eq:3}).
Assuming steady-state conditions, this leads to the well known set of
dynamic wind equations which can also be discussed by analytic means (see
Qian \& Woosley 1996, and references therein). In contrast, the 
toy model developed in this paper does not make use of
steady-state assumptions for the mass flow through the gain layer,
i.e., it is allowed that $\dot M \not= \dot M'$ in general.

The integral in Eq.~(\ref{eq:89}) was evaluated in Sect.~\ref{sec:heatcool}:
\begin{equation}
\int\limits_{R_{\nu}}^{R_{\mathrm{s}}}{\mathrm{d}}r\,4\pi r^2\,Q_{\nu}
\,=\,-\,L_{\mathrm{acc}} + {\cal H} - {\cal C} \,.
\label{eq:89a}
\end{equation}
Equation~(\ref{eq:73}) gives the net energy exchange between neutrinos and
stellar medium in the layer $[R_{\nu},R_{\mathrm{g}}]$, Eq.~(\ref{eq:83})
the corresponding result for the interval $[R_{\mathrm{g}},R_{\mathrm{s}}]$,
when ${\cal H}$ is taken from Eq.~(\ref{eq:79}) and the neutrino luminosity
$L_{\nu}(R_{\mathrm{g}})$ from Eq.~(\ref{eq:72}) with $a$ and $b$ provided by
Eqs.~(\ref{eq:71}) and (\ref{eq:71a}), respectively. Plugging in 
numbers representative for the early post-bounce evolution, 
$L_{\nu}\approx 5\times 10^{52}$ erg$\,$s$^{-1}$, $R_{\nu} \approx 50$ km,
$\widetilde{M}\approx 1$ M$_{\odot}$,
$a\sim 1$, one finds $L_{\nu}(R_{\mathrm{g}})\approx 6.3\times 10^{52}$
erg$\,$s$^{-1}$ and $\int_{R_{\nu}}^{R_{\mathrm{g}}}{\mathrm{d}}r\,4\pi r^2
Q_{\nu} = -L_{\mathrm{acc}} = L_{\nu}(R_{\nu})-L_{\nu}(R_{\mathrm{g}}) 
\approx -1.3\times
10^{52}$ erg$\,$s$^{-1}$, and using $R_{\mathrm{s}} \approx 2R_{\mathrm{g}} \approx
200$ km, $\ave{\mu_{\nu}}^\ast \sim 1$, $\ave{\mu_{\nu}}_{\mathrm{g}}\sim 0.75$,
yields $\int_{R_{\mathrm{g}}}^{R_{\mathrm{s}}}{\mathrm{d}}r\,4\pi r^2Q_{\nu}=
{\cal H} -{\cal C} \approx 7.7\times 10^{51}$ erg$\,$s$^{-1}$.
The gravitational energy at the
neutrinosphere at 50 km is about $-28$ MeV per nucleon, $q_{\mathrm{d}}$ is
roughly 8 MeV per nucleon, and the internal energy plus pressure account for
typically $\sim 10$ MeV per nucleon:
\begin{equation}
\rund{{e+P\over \rho}}_{\! R_{\nu}} \approx \rund{{5\over 2} +
{4\over 3}\,Y_e\,{{\cal F}_3(\eta_e)\over {\cal F}_2(\eta_e)}}
{kT_{\nu}\over m_{\mathrm{u}}}  \,\approx\,
{7\over 2}\,{kT_{\nu}\over m_{\mathrm{u}}} \, ,
\label{eq:89b}
\end{equation}
where $e = \varepsilon$ has been applied because ${1\over 2}\rho v^2 \ll
\varepsilon$ at the neutrinosphere. Therefore the sum of the terms in the
denominator of Eq.~(\ref{eq:89}) can be estimated to be about $-10^{19}$
erg$\,$g$^{-1}$. This leads to a mass accretion rate of the neutron star of
$\dot M'\sim -0.3$ M$_{\odot}\,$s$^{-1}$, a value which is in the range of
the results of detailed numerical simulations and is of the order of the 
mass infall rate on the shock, $\dot M$.

\section{Mass and energy conservation in the gain region}
\label{sec:meconservation}

Mass and energy conservation in the gain region between $R_{\mathrm{g}}$ and 
$R_{\mathrm{s}}$ determine the early postbounce
evolution of the supernova shock. For example, the shock is pushed outward 
when the matter that falls through the shock stays hot and piles up on top 
of the neutron star, forming an extended envelope instead of being 
accreted into the dense core quickly after efficient energy loss in the 
neutrino cooling layer below $R_{\mathrm{g}}$. Similarly, strong neutrino 
heating in the gain region causes an increase of the postshock pressure
and thus drives an expansion of the shock. On the other hand, enhanced 
neutrino emission will extract mass and/or energy from the layer which 
supports the supernova shock. The consequence will be a retraction of the
shock in radius.
These effects need to be accounted for by an appropriate discussion of the 
delayed explosion mechanism. A steady-state picture is certainly not adequate.

\subsection{Mass in the gain region}
\label{sec:mascon}

The mass $\Delta M_{\mathrm{g}}$ in the gain region can be calculated as 
volume integral over the density:
\begin{equation}
\Delta M_{\mathrm{g}} \,=\,\int\limits_{R_{\mathrm{g}}}^{R_{\mathrm{s}}}
{\mathrm{d}}r\,4\pi r^2\,\rho(r)\ .
\label{eq:90}
\end{equation}
with the density $\rho(r)$ given by Eq.~(\ref{eq:57}). Alternatively, 
since the latter equation is the exact
solution for hydrostatic equilibrium, one can use 
$\rho(r) = -r^2({\mathrm{d}}P/{\mathrm{d}}r)/(G\widetilde{M})$ with $P(r)$
from Eq.~(\ref{eq:58}). Defining the coefficients $c_1 \equiv
\rho_{\mathrm{s}}^{1/3}-G\widetilde{M}/(4KR_{\mathrm{s}})$ and $c_2\equiv
G\widetilde{M}/(4K)$, one finds:
\begin{eqnarray}
\Delta M_{\mathrm{g}} \,=\, 4\pi\, \Biggl\lbrack & &\!\!\!\!\!\!
{1\over 3}\,c_1^3\rund{R_{\mathrm{s}}^3-R_{\mathrm{g}}^3} +
{3\over 2}\,c_1^2c_2\rund{R_{\mathrm{s}}^2-R_{\mathrm{g}}^2} + \nonumber \\
& &\!\!\!\!\!\!
3\,c_1c_2^2\rund{R_{\mathrm{s}}-R_{\mathrm{g}}}
+ c_2^3\ln\! \rund{{R_{\mathrm{s}}\over R_{\mathrm{g}}}}\Biggr\rbrack 
\nonumber \\
=\, 4\pi\, \Biggl\lbrack & &\!\!\!\!\!\!
{1\over 3}\rund{R_{\mathrm{s}}^3\rho_{\mathrm{s}}-R_{\mathrm{g}}^3\rho_{\mathrm{g}}}
+ {c_2\over 2}\rund{R_{\mathrm{s}}^2\rho_{\mathrm{s}}^{2\over 3}-
R_{\mathrm{g}}^2\rho_{\mathrm{g}}^{2\over 3}} + \nonumber \\
& &\!\!\!\!\!\!
c_2^2\rund{R_{\mathrm{s}}\rho_{\mathrm{s}}^{1\over 3}-
R_{\mathrm{g}}\rho_{\mathrm{g}}^{1\over 3}}
+ c_2^3\ln\! \rund{{R_{\mathrm{s}}\over R_{\mathrm{g}}}}\Biggr\rbrack 
\,.
\label{eq:91}
\end{eqnarray}
In deriving the second form of Eq.~(\ref{eq:91}), use was made of 
$\rho(r) = (c_1 + c_2/r)^3$. Moreover, with $\rho = (P/K)^{3/4}$ the
density in Eq.~(\ref{eq:91}) can be substituted by the pressure $P$.
Note that the quantities $\rho_{\mathrm{g}} = \rho(R_{\mathrm{g}})$ and
$P_{\mathrm{g}} = P(R_{\mathrm{g}})$ at the gain
radius must be expressed by the exact relations of Eqs.~(\ref{eq:57}) and
(\ref{eq:58}), respectively. They depend on the postshock state of the matter
as do the coefficients $c_1$ and $c_2$.
The gain radius $R_{\mathrm{g}}$ is given by Eq.~(\ref{eq:63}). It is also 
a function of the conditions immediately behind the shock. Writing the
postshock temperature in terms of the postshock pressure via Eq.~(\ref{eq:56}),
$kT_{\mathrm{s}} = 
\eck{3 P_{\mathrm{s}}/(f_{\mathrm{r}}g_{\mathrm{r}}a_{\gamma})}^{1/4}$,
and using Eqs.~(\ref{eq:39}), (\ref{eq:43}), (\ref{eq:44}) and 
(\ref{eq:53})--(\ref{eq:55}) with typical values
$\beta\sim 7$, $\alpha\sim 1/\sqrt{2}$, $s_{\gamma}\sim 4$, and $\eta_e\sim 2$
(the exact values of these parameters are not essential for the discussion and 
affect the result rather insensitively), 
one gets in case of $|U_{\mathrm{s}}| = |\dot R_{\mathrm{s}}| \ll |v_{\mathrm{p}}|$:
\begin{equation}
kT_{\mathrm{s}}\,\approx\,2 \,\, R_{{\mathrm{s}},7}^{-{5\over 8}}
\rund{\!{-\dot M\over {\mathrm{M}}_{\odot}/{\mathrm{s}}}\!}^{\! {1\over 4}}
\rund{{\widetilde{M}\over {\mathrm{M}}_{\odot}}}^{\! {1\over 8}} 
\ \ \eck{{\mathrm{MeV}}}\,.
\label{eq:91a}
\end{equation}
Inserting this in Eq.~(\ref{eq:63}) and using $L_{\nu} = 2L_{\nu_e}$ yields
\begin{eqnarray}
R_{{\mathrm{g}},7}\,\approx\,1.13\,\,R_{{\mathrm{s}},7}^{9\over 16}
\,\, (kT_{\nu_e,4})^{-{1\over 2}}
\!\!\!\!\!\! &\phantom{\times} & \!\!
\rund{{L_{\nu,52}(R_{\mathrm{g}})\over\ave{\mu_{\nu}}_{\mathrm{g}}}
}^{\! -{1\over 4}} \nonumber \\
\!\!\!\!\!\! &\times &\!\!
\rund{\!{-\dot M\over {\mathrm{M}}_{\odot}/{\mathrm{s}}}\!}^{\! {3\over 8}}
\rund{{\widetilde{M}\over {\mathrm{M}}_{\odot}}}^{\! {3\over 16}}\! ,
\label{eq:91b}
\end{eqnarray}
where the neutrino luminosity at the gain radius, $L_{\nu}(R_{\mathrm{g}})$,
is given by Eq.~(\ref{eq:72}). 

Instead of the exact expression of Eq.~(\ref{eq:91}) an approximation for
$\Delta M_{\mathrm{g}}$ is sometimes preferable. Performing the integration of
Eq.~(\ref{eq:90}) with the approximate density profile of Eq.~(\ref{eq:60}),
one finds
\begin{equation}
\Delta M_{\mathrm{g}}\,\approx\,4\pi\,\rho_{\mathrm{s}} R_{\mathrm{s}}^3
\ln\!\rund{{R_{\mathrm{s}}\over R_{\mathrm{g}}}}\,\propto\,
{-\dot M\over \widetilde{M}^{1/2}}\,R_{\mathrm{s}}^{3/2}
\ln\!\rund{{R_{\mathrm{s}}\over R_{\mathrm{g}}}} .
\label{eq:91c}
\end{equation}
Here $\rho_{\mathrm{s}}$ was written in terms of $R_{\mathrm{s}}$ by
making use of $\rho_{\mathrm{s}} = \beta\rho_{\mathrm{p}}$ and Eq.~(\ref{eq:44}).
Moreover, from Eq.~(\ref{eq:91b}) one can deduce that 
$R_{\mathrm{s}}/R_{\mathrm{g}}\propto R_{\mathrm{s}}^{7/16}$ for 
$|U_{\mathrm{s}}|\ll |v_{\mathrm{p}}|$. An increase of the shock radius 
therefore means that $\Delta M_{\mathrm{g}}$ will also grow.

The rate at which the mass in the gain region changes in time due to a
shift of the upper and lower boundaries of this region but also due to a
variation of the density of the stellar medium, is determined as
the total time derivative of Eq.~(\ref{eq:90}):
\begin{equation}
{{\mathrm{d}}\over {\mathrm{d}}t}(\Delta M_{\mathrm{g}})\,=\,
4\pi R_{\mathrm{s}}^2\rho_{\mathrm{s}}\dot R_{\mathrm{s}} -
4\pi R_{\mathrm{g}}^2\rho_{\mathrm{g}}\dot R_{\mathrm{g}} +
\!\! \int\limits_{R_{\mathrm{g}}}^{R_{\mathrm{s}}}\!{\mathrm{d}}r
4\pi r^2 {\partial \rho\over \partial t} ,
\label{eq:92}
\end{equation}
where $\dot R_{\mathrm{s}}\equiv {\mathrm{d}}R_{\mathrm{s}}/{\mathrm{d}}t
= U_{\mathrm{s}}$ is the shock velocity and 
$\dot R_{\mathrm{g}}\equiv {\mathrm{d}}R_{\mathrm{g}}/{\mathrm{d}}t$
the velocity of the gain radius. When the integration in Eq.~(\ref{eq:92})
is carried out to a radius infinitesimally smaller than $R_{\mathrm{s}}$
with the help of Eq.~(\ref{eq:2}), one obtains
\begin{eqnarray}
{{\mathrm{d}}\over {\mathrm{d}}t}(\Delta M_{\mathrm{g}}) &=&
4\pi R_{\mathrm{s}}^2\rho_{\mathrm{s}}(\dot R_{\mathrm{s}}-v_{\mathrm{s}})
-
4\pi R_{\mathrm{g}}^2\rho_{\mathrm{g}}(\dot R_{\mathrm{g}}-v_{\mathrm{g}})
\nonumber \\
&=& 
4\pi R_{\mathrm{s}}^2\rho_{\mathrm{p}}\dot R_{\mathrm{s}} -
4\pi R_{\mathrm{g}}^2\rho_{\mathrm{g}}\dot R_{\mathrm{g}} - \dot M + \dot M'\ ,
\label{eq:93}
\end{eqnarray}
with $v_{\mathrm{g}}$ and $v_{\mathrm{s}}$ being the velocities of the 
stellar medium at the gain radius and just behind the shock, respectively.
The lower expression was derived by using the shock jump condition for the 
mass flow, Eq.~(\ref{eq:36}), and the definitions 
$\dot M = 4\pi R_{\mathrm{s}}^2\rho_{\mathrm{p}}v_{\mathrm{p}}$ and
$\dot M' = 4\pi R_{\mathrm{g}}^2\rho_{\mathrm{g}}v_{\mathrm{g}}
= 4\pi R_{\nu}^2\rho_{\nu}v_{\nu}$ as introduced in Sect.~\ref{sec:massacc}.
Equation~(\ref{eq:93}) shows that the mass in the gain region can change
because of inflow and outflow of gas but also due to the motion of the
boundaries $R_{\mathrm{g}}$ and $R_{\mathrm{s}}$.
Knowing the initial mass in this layer, $\Delta M_{\mathrm{g}}^0$,
Eq.~(\ref{eq:93}) allows one to calculate the value at later times.

\subsection{Energy in the gain region}
\label{sec:ergcon}

Since the postshock matter is effectively in hydrostatic equilibrium 
(see Sect.~\ref{sec:atmosphere}) the kinetic energy is negligible 
compared to the internal energy and the gravitational potential energy, 
and the total energy in the gain region is therefore given by
\begin{equation}
\Delta E_{\mathrm{g}}\,=\, \int\limits_{R_{\mathrm{g}}}^{R_{\mathrm{s}}}
{\mathrm{d}}r\,4\pi r^2\eck{\varepsilon(r)-{G\widetilde{M}\over r}\,\rho(r)}
\ .
\label{eq:94}
\end{equation}
To evaluate the right hand side, one substitutes $\varepsilon = P/(\Gamma -1)$,
which relates the internal energy density $\varepsilon$ and the pressure $P$
for an ideal gas, with the
adiabatic index $\Gamma$ being typically between $4/3$ and $5/3$, depending
on whether relativistic or nonrelativistic particles, respectively, dominate
the pressure. In addition making use of hydrostatic equilibrium [Eq.~(\ref{eq:47})]
or, alternatively, applying the virial theorem, one finds
\begin{eqnarray}
\Delta E_{\mathrm{g}}\,=\, {4\pi \over 3(\Gamma -1)}\,
\!\!\!\! &(&\!\!\!\! P_{\mathrm{s}}R_{\mathrm{s}}^3 -P_{\mathrm{g}}R_{\mathrm{g}}^3)
\nonumber\\
&-&\! 4\pi\,G\widetilde{M}\,{3\Gamma - 4\over 3(\Gamma-1)}
\int\limits_{R_{\mathrm{g}}}^{R_{\mathrm{s}}}{\mathrm{d}}r\, r\rho(r)\ .
\label{eq:95}
\end{eqnarray}
The second term is the gravitational potential energy times 
$(\Gamma-{4\over 3})/(\Gamma-1)$. An exact expression for the integral
is obtained when Eq.~(\ref{eq:57}) is used for $\rho(r)$:
\begin{eqnarray}
\int\limits_{R_{\mathrm{g}}}^{R_{\mathrm{s}}}\! {\mathrm{d}}r\,r\rho(r)\,=\,
{c_1^3\over 2}\,(R_{\mathrm{s}}^2\! -\! R_{\mathrm{g}}^2) \!\! &+& \!\!
\rund{\! 3c_1^2c_2 + {c_2^3\over R_{\mathrm{s}}R_{\mathrm{g}}}}(R_{\mathrm{s}}
\! -\! R_{\mathrm{g}}) \nonumber \\
&+& \!\! 3\,c_1c_2^2\, \ln\! \rund{{R_{\mathrm{s}}\over R_{\mathrm{g}}}} \,.
\label{eq:96}
\end{eqnarray}
The coefficients $c_1$ and $c_2$ were already defined in Sect.~\ref{sec:mascon}.
For the following discussion an approximation of this integral
is sufficient. It can be derived by employing the approximate power law
profile for the density, Eq.~(\ref{eq:60}):
\begin{equation}
\int\limits_{R_{\mathrm{g}}}^{R_{\mathrm{s}}}\! {\mathrm{d}}r\,r \rho(r)\,\approx\,
\rho_{\mathrm{s}}\,{R_{\mathrm{s}}^2\over R_{\mathrm{g}}}\,
(R_{\mathrm{s}}-R_{\mathrm{g}}) \ .
\label{eq:97}
\end{equation}

The rate at which the total energy in the gain region changes with time 
can be calculated as the time derivative of Eq.~(\ref{eq:94}).
With the definition $l \equiv (\varepsilon + P)/\rho - G\widetilde{M}/r$
one finds
\begin{eqnarray}
{{\mathrm{d}}\over {\mathrm{d}}t}(\Delta E_{\mathrm{g}}) & = &
4\pi R_{\mathrm{s}}^2 \rho_{\mathrm{s}} l_{\mathrm{s}}\dot R_{\mathrm{s}} -
4\pi R_{\mathrm{g}}^2 \rho_{\mathrm{g}} l_{\mathrm{g}}\dot R_{\mathrm{g}} 
\nonumber \\
& - & 4\pi R_{\mathrm{s}}^2 P_{\mathrm{s}} \dot R_{\mathrm{s}} + 
4\pi R_{\mathrm{g}}^2 P_{\mathrm{g}} \dot R_{\mathrm{g}}
\nonumber \\
& + & \int\limits_{R_{\mathrm{g}}}^{R_{\mathrm{s}}}{\mathrm{d}}r\,
4\pi r^2\rund{{\partial \varepsilon\over \partial t} -
{G\widetilde{M}\over r}\,{\partial \rho\over \partial t}} \,,
\label{eq:102}
\end{eqnarray}
where $\dot R_{\mathrm{s}}$ and $\dot R_{\mathrm{g}}$ have the same meaning
as in Eq.~(\ref{eq:92}).
The partial derivatives in the integral can be substituted by 
Eqs.~(\ref{eq:2}) and (\ref{eq:4}). Making additional use of 
Eqs.~(\ref{eq:5}) and (\ref{eq:6}) and of 
${1\over 2}\rho v^2 \ll \varepsilon$ yields
\begin{eqnarray}
{{\mathrm{d}}\over {\mathrm{d}}t}(\Delta E_{\mathrm{g}}) & = &\!
4\pi R_{\mathrm{s}}^2 \rho_{\mathrm{s}} l_{\mathrm{s}}
(\dot R_{\mathrm{s}} - v_{\mathrm{s}}) - 
4\pi R_{\mathrm{g}}^2 \rho_{\mathrm{g}} l_{\mathrm{g}}
(\dot R_{\mathrm{g}} - v_{\mathrm{g}}) 
\nonumber \\
& - &\! 4\pi R_{\mathrm{s}}^2 P_{\mathrm{s}} \dot R_{\mathrm{s}} +
4\pi R_{\mathrm{g}}^2 P_{\mathrm{g}} \dot R_{\mathrm{g}} 
+ \!\! \int\limits_{R_{\mathrm{g}}}^{R_{\mathrm{s}}}\!\! {\mathrm{d}}r
4\pi r^2 Q_{\nu} .
\label{eq:103}
\end{eqnarray}
Now employing the continuity equation for the mass flow across the shock, 
Eq.~(\ref{eq:36}),
and replacing the integral for the energy exchange with neutrinos
between $R_{\mathrm{g}}$ and $R_{\mathrm{s}}$ by ${\cal H}-{\cal C}$ as given
in Eqs.~(\ref{eq:78}) and (\ref{eq:83}), one ends up with
\begin{eqnarray}
{{\mathrm{d}}\over {\mathrm{d}}t}(\Delta E_{\mathrm{g}}) & = &
4\pi R_{\mathrm{s}}^2 \rho_{\mathrm{p}} l_{\mathrm{s}}
(\dot R_{\mathrm{s}} - v_{\mathrm{p}}) -
4\pi R_{\mathrm{g}}^2 \rho_{\mathrm{g}} l_{\mathrm{g}}
(\dot R_{\mathrm{g}} - v_{\mathrm{g}})
\nonumber \\
&\phantom{=}&
 -\,\, 4\pi R_{\mathrm{s}}^2 P_{\mathrm{s}} \dot R_{\mathrm{s}} +
       4\pi R_{\mathrm{g}}^2 P_{\mathrm{g}} \dot R_{\mathrm{g}}
+ {\cal H} - {\cal C} 
\nonumber \\
&=& 4\pi R_{\mathrm{s}}^2 \rho_{\mathrm{p}} l_{\mathrm{s}} \dot R_{\mathrm{s}}
 -  4\pi R_{\mathrm{g}}^2 \rho_{\mathrm{g}} l_{\mathrm{g}} \dot R_{\mathrm{g}}
 -  4\pi R_{\mathrm{s}}^2 P_{\mathrm{s}} \dot R_{\mathrm{s}}
\nonumber\\
&\phantom{=}&
 +\,\, 4\pi R_{\mathrm{g}}^2 P_{\mathrm{g}} \dot R_{\mathrm{g}}
- \dot M\,l_{\mathrm{s}} + \dot M'\,l_{\mathrm{g}} + {\cal H} - {\cal C} .
\label{eq:104}
\end{eqnarray}
The mass accretion rates $\dot M$ and $\dot M'$ account for the inflow 
of matter into the gain region through the shock and for the mass that 
is advected through the gain radius, respectively
[see Sect.~\ref{sec:massacc} and discussion after Eq.~(\ref{eq:93})].
Equation~(\ref{eq:104}) means that the total energy in the gain region 
changes due to active mass motions, $p{\mathrm{d}}V$ work associated
with these mass motions, the movement of the boundaries, and neutrino 
heating.

Making use of $\varepsilon_{\mathrm{s}} = P_{\mathrm{s}}/(\Gamma -1)$,
$\rho_{\mathrm{s}}/\rho_{\mathrm{p}} = \beta$, and Eq.~(\ref{eq:39}),
one finds for
$l_{\mathrm{s}} = (\varepsilon_{\mathrm{s}}+P_{\mathrm{s}})/
\rho_{\mathrm{s}}-G\widetilde{M}/R_{\mathrm{s}}$:
\begin{equation}
l_{\mathrm{s}} \,\approx\, -\eck{1-{\Gamma\over \Gamma\! -\! 1}\,
{\beta\! -\! 1\over \beta^2}
\rund{1-{U_{\mathrm{s}}\over v_{\mathrm{p}}}}^{\! 2}\,}
{G\widetilde{M}\over R_{\mathrm{s}}} \,,
\label{eq:104a}
\end{equation}
where Eq.~(\ref{eq:43}) with $\alpha \approx 1/\sqrt{2}$ was employed for
$v_{\mathrm{p}}^2 \approx
G\widetilde{M}/R_{\mathrm{s}} \approx 1.3\times 10^{19}
R_{{\mathrm{s}},7}^{-1}(\widetilde{M}/{\mathrm{M}}_{\odot})$ erg$\,$g$^{-1}$.
Because of hydrostatic equilibrium a simple relation exists between 
$l_{\mathrm{g}}$ and $l_{\mathrm{s}}$. With $\varepsilon = P/(\Gamma -1)$ and
Eqs.~(\ref{eq:56}) and (\ref{eq:57}) one obtains
\begin{equation}
l_{\mathrm{g}}\,=\,l_{\mathrm{s}} - 
{3\Gamma - 4\over 4(\Gamma-1)}\, {G\widetilde{M}\over R_{\mathrm{s}}}
\rund{{R_{\mathrm{s}}\over R_{\mathrm{g}}}-1} \,.
\label{eq:105}
\end{equation}
Using the more general density-pressure relation
$P = K\rho^\gamma$ instead of Eq.~(\ref{eq:56}), and the corresponding
hydrostatic density profile of Eq.~(\ref{eq:59a}), leads to
\begin{equation}
l_{\mathrm{g}}\,=\,l_{\mathrm{s}} -
\rund{1 - {\Gamma\over \Gamma\! -\! 1}\,{\gamma\! -\!1\over\gamma}}
{G\widetilde{M}\over R_{\mathrm{s}}}
\rund{{R_{\mathrm{s}}\over R_{\mathrm{g}}}-1} \,.
\label{eq:105a}
\end{equation}
For $\Gamma = \gamma$, this gives $l_{\mathrm{g}} = l_{\mathrm{s}}$.

\section{Evolution of shock radius and shock velocity}
\label{sec:application}

The model developed in the preceding sections allows one to study the
behavior of the supernova shock in response to the processes that play 
a role in the collapsed stellar core. The physics between the neutron star
surface and the shock is constrained by the energy influx from the 
neutrinosphere on the one hand and the mass accretion into the shock 
front on the other. Equations~(\ref{eq:91}), (\ref{eq:95})
[in combination with (\ref{eq:96})], and (\ref{eq:39}) determine the shock
radius $R_{\mathrm{s}}$, the shock velocity $U_{\mathrm{s}}$, and the  
postshock pressure $P_{\mathrm{s}}$.
The state of the matter immediately behind the shock
and that at the gain radius are related via Eqs.~(\ref{eq:57})--(\ref{eq:59})
and (\ref{eq:105}), the gain radius $R_{\mathrm{g}}$ is given by 
Eq.~(\ref{eq:63}), the postshock temperature by $kT_{\mathrm{s}} =
\eck{3 P_{\mathrm{s}}/(f_{\mathrm{r}}g_{\mathrm{r}}a_{\gamma})}^{1/4}$
[Eq.~(\ref{eq:56})], and the postshock density as 
$\rho_{\mathrm{s}} = \beta \rho_{\mathrm{p}}$ with $\rho_{\mathrm{p}}$
from Eq.~(\ref{eq:44}).

The mass accretion rate
$\dot M$ into the shock is a fixed parameter of the problem [in Eq.~(\ref{eq:46})
it is expressed in terms of the constant $H$ which is linked to the structure 
of the progenitor star]. The rate of mass advection into the neutron star,
$\dot M'$, can be calculated from Eq.~(\ref{eq:89}). The radius $R_{\nu}$ 
and mass $M$ of the neutron star, the neutrinospheric luminosity $L_{\nu}$,
and the spectral temperature of the emitted electron neutrinos $T_{\nu_e}$
(assumed to be roughly equal to the  temperature $T_{\nu}$ of the stellar gas
at the neutrinosphere) are also input parameters.
The discussion takes into account the effects of neutrino losses in the cooling
region, expressed by Eqs.~(\ref{eq:71})--(\ref{eq:73}), and of neutrino 
heating in the gain region as given by Eqs.~(\ref{eq:78}), (\ref{eq:79}),
and (\ref{eq:83}).

The time dependence of the considered model requires as initial conditions
the values $\Delta M_{\mathrm{g}}^0$ and $\Delta E_{\mathrm{g}}^0$ for the
initial mass and energy in the gain region. This couples the 
subsequent evolution in $\Delta M_{\mathrm{g}}$ and $\Delta E_{\mathrm{g}}$,
which can be followed with Eqs.~(\ref{eq:93}) and (\ref{eq:104}),
respectively, to the situation that exists right after core bounce.
Knowing $\dot M'(t)$ allows one to include also the changes of the neutron
star mass.

\begin{figure}
\epsfxsize=0.99\hsize
\epsffile{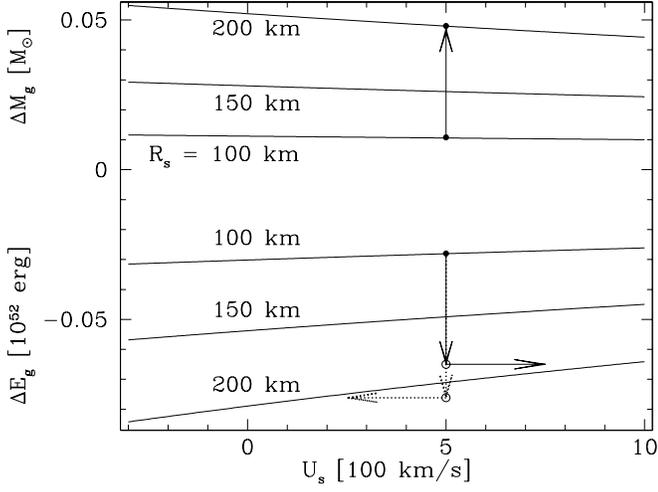}
\caption{
Mass $\Delta M_{\mathrm{g}}$ and energy $\Delta E_{\mathrm{g}}$ in the gain
region as functions of shock velocity $U_{\mathrm{s}}$ for different shock
radii $R_{\mathrm{s}}$. The values of neutron star mass $\widetilde{M}$ and 
mass infall rate $\dot M$ into the shock are fixed.
Two possible cases for a transition from initial
shock radius $R_{\mathrm{s}} = 100$ km to final shock radius
$R_{\mathrm{s}} = 200$ km are indicated. The path marked by solid arrows
corresponds to stable shock expansion, the situation marked by dotted
arrows means a slow-down of the shock.
}
\label{fig:3}
\end{figure}

\subsection{Shock expansion and acceleration}
\label{sec:shockevol}

Combining Eqs.~(\ref{eq:95}) and (\ref{eq:97}) and using Eq.~(\ref{eq:58})
for $P_{\mathrm{g}}$ in terms of $P_{\mathrm{s}}$
with $K^{3/4} = P_{\mathrm{s}}^{3/4}/\rho_{\mathrm{s}}$, one
gets the relation
\begin{eqnarray}
(x-x_0)\,x^3\,\approx\,\rund{\!{R_{\mathrm{g}}\over R_{\mathrm{s}}}\!}^{\! 3}\!
\Biggl\lbrack \,x\! &+& \!\! \rund{\!{R_{\mathrm{s}}\over R_{\mathrm{g}}}- 1\!}
\Biggr\rbrack^4
\nonumber \\
& + & \! 4\,(3\Gamma-4)\rund{\!{R_{\mathrm{s}}\over R_{\mathrm{g}}} - 1\!} x^3 .
\label{eq:98}
\end{eqnarray}
Here $x$ and $x_0$ were defined as
\begin{eqnarray}
x &\equiv& P_{\mathrm{s}}
\rund{{G\widetilde{M}\rho_{\mathrm{s}}\over 4 R_{\mathrm{s}}}}^{\!\! -1}
\propto \rund{1 + U_{\mathrm{s}}\,\sqrt{{R_{\mathrm{s}}\over G\widetilde{M}}}
\,\,}^{\!\! 2}\,,
\label{eq:99} \\ 
x_0 &\equiv& 3(\Gamma-1)\,{\Delta E_{\mathrm{g}}\over 4\pi R_{\mathrm{s}}^3}
\rund{{G\widetilde{M}\rho_{\mathrm{s}}\over 4 R_{\mathrm{s}}}}^{\!\! -1}
\! \propto\,-\,{\Delta E_{\mathrm{g}}\over \dot 
M\sqrt{R_{\mathrm{s}}\widetilde{M}}}\,.
\label{eq:100}
\end{eqnarray}
The proportionality relations can be verified by using Eqs.~(\ref{eq:39}), 
(\ref{eq:43}) (with $\alpha = 1/\sqrt{2}$) and (\ref{eq:44}).

Equation~(\ref{eq:98}) is the key equation to understand the behavior of the
supernova shock under the influence of accretion and neutrino heating.
Typically, $\Delta E_{\mathrm{g}} < 0$ during the shock stagnation phase, 
and therefore $x_0 < 0$. Equation~(\ref{eq:98}) depends on two variables
which constrain the conditions at the shock front, namely on $x > 0$ and
on $y \equiv R_{\mathrm{s}}/R_{\mathrm{g}}$, for which $y \ge 1$ holds.
Fixing the parameters $\dot M$, $\widetilde{M}$ and $R_{\mathrm{s}}$,
one can show that a larger value of $x$ and thus a larger $U_{\mathrm{s}}$
requires that $x_0$ and therefore $\Delta E_{\mathrm{g}}$ is bigger (i.e.,
less negative). Physically, this corresponds to the case where neutrino energy 
deposition leads to a rising postshock pressure $P_{\mathrm{s}}$
[compare Eq.~(\ref{eq:99})], which accelerates the shock front. On the other
hand, if $|U_{\mathrm{s}}|\ll |v_{\mathrm{p}}| \approx 
(G\widetilde{M}/R_{\mathrm{s}})^{1/2}$ the quantity $x$ is essentially constant,
and $y \propto R_{\mathrm{s}}^{7/16}$ [cf. Eq.~(\ref{eq:91b})] is the
variable which reacts to changes of $x_0$. The corresponding discussion is 
more transparent when Eq.~(\ref{eq:98}) is rewritten in the following form:
\begin{equation}
\eck{x\! -\! x_0\! +\! 4(3\Gamma\!-\!4)}x^3y^3 \approx (x+y-1)^4\! + 
4(3\Gamma\!- 4)x^3y^4 .
\label{eq:100a}
\end{equation}
For $x_0 < 0$ this equation has a solution $\hat y$ which shrinks, if
$\Delta E_{\mathrm{g}}$ and thus $x_0$ is larger (i.e., less negative).
Therefore the radius $R_{\mathrm{s}}$ of the shock, which is compatible with the
assumptions, is smaller. Inversely, if $\Delta E_{\mathrm{g}}$ and $x_0$ are
lower (more negative), $R_{\mathrm{s}}$ will be larger. This behavior can
be explained by the observation that $\Delta E_{\mathrm{g}}$ decreases
when matter with negative specific energy is accumulated in the
gain region. Such an accumulation of mass will cause a growth of the
shock radius. It should be noted that for $x$ of order unity (i.e.,
$|U_{\mathrm{s}}|\ll |v_{\mathrm{p}}|$) a solution $\hat y \ge 1$
of Eq.~(\ref{eq:100a}) exists only in case of $x_0 \le 0$. 
A positive value of $\Delta E_{\mathrm{g}}$, on the other hand, is compatible 
only with a sufficiently large shock velocity $U_{\mathrm{s}}$.

The situation is graphically illustrated in Fig.~\ref{fig:3}, where
$\Delta M_{\mathrm{g}}$ from Eq.~(\ref{eq:91}) and $\Delta E_{\mathrm{g}}$
from Eqs.~(\ref{eq:95}) and (\ref{eq:96}) are plotted as functions of 
$U_{\mathrm{s}}$ for
different choices of the shock radius $R_{\mathrm{s}}$ (with parameters:
$\widetilde{M} = 1.25\,$M$_{\odot}$, $\dot M = 0.3\,$M$_{\odot}\,$s$^{-1}$,
$\Gamma = \gamma = 4/3$, $\beta = 7$,
$\alpha = 1/\sqrt{2}$, $g_{\mathrm{r}} = 6.41$, $f_{\mathrm{r}} = 1.16$).
Figure~\ref{fig:3} [or Eq.~(\ref{eq:91c})] 
show that a growth of the mass $\Delta M_{\mathrm{g}}$
in the gain region will cause an increase of $R_{\mathrm{s}}$. This means that
${\mathrm{d}}R_{\mathrm{s}}/{\mathrm{d}}t \ge 0$ can be ensured if
\begin{equation}
{{\mathrm{d}}\over {\mathrm{d}}t}(\Delta M_{\mathrm{g}})\, \ge\, 0\ .
\label{eq:101a}
\end{equation}
This, however, is not a sufficient criterion for a continued outward motion 
of the shock. The dotted arrows in Fig.~\ref{fig:3} indicate a situation where
the decrease of energy in the gain region is so large that the increase of 
the shock radius implies a deceleration of the shock. For moving along the path 
marked by solid arrows, i.e., for obtaining stable shock expansion with 
${\mathrm{d}}U_{\mathrm{s}}/{\mathrm{d}}t \ge 0$, a necessary condition is
\begin{equation}
{{\mathrm{d}}\over {\mathrm{d}}t}(\Delta E_{\mathrm{g}})\, \ge\, 
U_{\mathrm{s}}\eck{{\partial(\Delta E_{\mathrm{g}})\over \partial 
R_{\mathrm{s}}}}_{U_{\mathrm{s}}} \ .
\label{eq:101}
\end{equation}
The right hand side of Eq.~(\ref{eq:101}) cannot be zero in general, because
${\mathrm{d}}(\Delta E_{\mathrm{g}})/{\mathrm{d}}t > 0$ can also be associated
with a shrinkage of $R_{\mathrm{s}}$ if more matter (with negative total 
energy) is lost from the gain region by advection
through the gain radius than is resupplied by gas falling into the shock.
The combined conditions of Eqs.~(\ref{eq:101a}) and (\ref{eq:101}) 
guarantee that $R_{\mathrm{s}}$ and $U_{\mathrm{s}}$ grow at the same time.
Applied to a stalled shock, in which case $U_{\mathrm{s}} = 0$, 
Eq.~(\ref{eq:101a}) together with Eq.~(\ref{eq:101}) can therefore be
considered as ``shock revival criterion'', which states that for an
expansion and acceleration of the shock front to occur, the energy
in the gain region should increase and simultaneously the mass in the 
gain region should not decrease.

\subsection{Shock revival criterion}
\label{sec:criterion}

If the conditions between neutrinosphere and shock vary slowly with time,
$\dot R_{\mathrm{g}} \approx 0$ is a good assumption. Since 
$v_{\mathrm{p}} \neq 0$, Eq.~(\ref{eq:104}) can then be written in the
form
\begin{equation}
{{\mathrm{d}}\over {\mathrm{d}}t}(\Delta E_{\mathrm{g}}) \approx
-\dot M l_{\mathrm{s}} + \dot M\!\! \rund{\! l_{\mathrm{s}}\! -\!
{P_{\mathrm{s}}\over \rho_{\mathrm{p}}}\! }\!\! 
{\dot R_{\mathrm{s}}\over v_{\mathrm{p}}}
+ \dot M' l_{\mathrm{g}} + {\cal H} - {\cal C} .
\label{eq:106}
\end{equation}
Replacing $l_{\mathrm{s}}$ in the bracket on the right hand side of
Eq.~(\ref{eq:106}) by Eq.~(\ref{eq:104a}) and using Eq.~(\ref{eq:39})
for $P_{\mathrm{s}}/\rho_{\mathrm{p}}$, one derives in case
of $|U_{\mathrm{s}}| \ll |v_{\mathrm{p}}|$ the expression
\begin{eqnarray}
{{\mathrm{d}}\over {\mathrm{d}}t}(\Delta E_{\mathrm{g}}) \,\approx\,
-\,\dot M\,l_{\mathrm{s}} \! &-& \!\! \dot M \!\rund{2\!-\! {1\over \beta}\! -\!
{\beta\! -\! 1\over \beta^2}\,{\Gamma\over \Gamma\! -\! 1}}\!
v_{\mathrm{p}}\dot R_{\mathrm{s}}
\nonumber \\
& + &\!\!  \dot M'\,l_{\mathrm{g}} + {\cal H} - {\cal C}\, .
\label{eq:107}
\end{eqnarray}
This equation is correct to first order in 
$|\dot R_{\mathrm{s}}/v_{\mathrm{p}}|\ll 1$.
From the discussion in Sect.~\ref{sec:ergcon} follows that for an
outward acceleration of a stalled shock
($\dot R_{\mathrm{s}} = U_{\mathrm{s}} = 0$), a necessary condition is
[see Eq.~(\ref{eq:101})]:
\begin{equation}
{{\mathrm{d}}\over {\mathrm{d}}t}(\Delta E_{\mathrm{g}}) \,=\,
-\,\dot M\,l_{\mathrm{s}} + \dot M'\,l_{\mathrm{g}} + {\cal H} - {\cal C}
\,>\,0 \ .
\label{eq:108}
\end{equation}
Note that neutrino heating (${\cal H}$) and cooling (${\cal C}$) in the 
gain region as well as the mass accretion
rate $\dot M$ through the shock have a direct influence. But also neutrino
losses {\em below} $R_{\mathrm{g}}$ have an effect by determining
$\dot M'$, and a more indirect one by causing additional energy deposition
in the gain region, where the neutrino energy extracted from the
cooling layer is partly reabsorbed.

The terms proportional to
$\dot M = 4\pi R_{\mathrm{s}}^2\rho_{\mathrm{p}}v_{\mathrm{p}}$
account for the so-called ram pressure of the infalling matter, which is
proportional to $\rho_{\mathrm{p}}v_{\mathrm{p}}^2$ and damps shock expansion,
because the accretion of matter through the shock yields a negative 
contribution to the right hand sides of Eqs.~(\ref{eq:106})--(\ref{eq:108}).
A comparison of Eqs.~(\ref{eq:107}) and (\ref{eq:108}) shows that the
onset of shock expansion enhances this and therefore Eq.~(\ref{eq:108})
gives a minimum requirement.

Instead of just an outward acceleration of the shock, a positive postshock
velocity, i.e., $v_{\mathrm{s}} > 0$, may be considered as a stronger
criterion for the possibility of an explosion.
With $\beta = \rho_{\mathrm{s}}/\rho_{\mathrm{p}}$ and Eq.~(\ref{eq:36})
one derives 
$v_{\mathrm{s}} = \beta^{-1}(v_{\mathrm{p}}-U_{\mathrm{s}}) + U_{\mathrm{s}}$,
which means that $v_{\mathrm{s}} > 0$ translates into $U_{\mathrm{s}} >
-\,v_{\mathrm{p}}/(\beta-1)$. Since $\beta \gg 1$ [Eq.~(\ref{eq:41})],
this condition is fulfilled while $|U_{\mathrm{s}}/v_{\mathrm{p}}|\ll 1$ 
still holds. Using this more rigorous criterion will therefore affect the 
details of the discussion, but will not change the picture qualitatively.

${\mathrm{d}}(\Delta E_{\mathrm{g}})/{\mathrm{d}}t > 0$ can be achieved by
strong neutrino heating (${\cal H}$ large), but can
also result if $\dot M'l_{\mathrm{g}} > \dot Ml_{\mathrm{s}}$. For
$l_{\mathrm{g}} = l_{\mathrm{s}}$, which is true when $\Gamma = \gamma$
[see Eq.~(\ref{eq:105a})], this is equivalent to $\dot M' < \dot M$, i.e., when
less mass is accreted through the shock than is lost from the gain
region into the neutron star [note that $l_{\mathrm{s}} < 0$; Eq.~(\ref{eq:104a})]. 
As a consequence, however, the mass between $R_{\mathrm{g}}$ and $R_{\mathrm{s}}$
and therefore the shock radius will decrease, in conflict with the 
demand for shock expansion (see Sect.~\ref{sec:shockevol}). To make
sure the shock expands, also Eq.~(\ref{eq:101a}) has to be fulfilled.
In case of $U_{\mathrm{s}} = 0$, $\dot R_{\mathrm{g}} = 0$, Eq.~(\ref{eq:93})
yields:
\begin{equation}
{{\mathrm{d}}\over {\mathrm{d}}t}(\Delta M_{\mathrm{g}}) \,=\,
-\,\dot M + \dot M' \,\ge\,0 \ \ \Longleftrightarrow\ \ 
\dot M \,\le\,\dot M' \,.
\label{eq:109}
\end{equation}
Both Eqs.~(\ref{eq:108}) and (\ref{eq:109}) constrain the parameters
for which a revival of the stalled shock can occur.

\begin{figure*}[htp!]
$\phantom{.}$
\vskip1.0truecm 
\centerline{
  \epsfxsize=0.48\hsize\epsffile{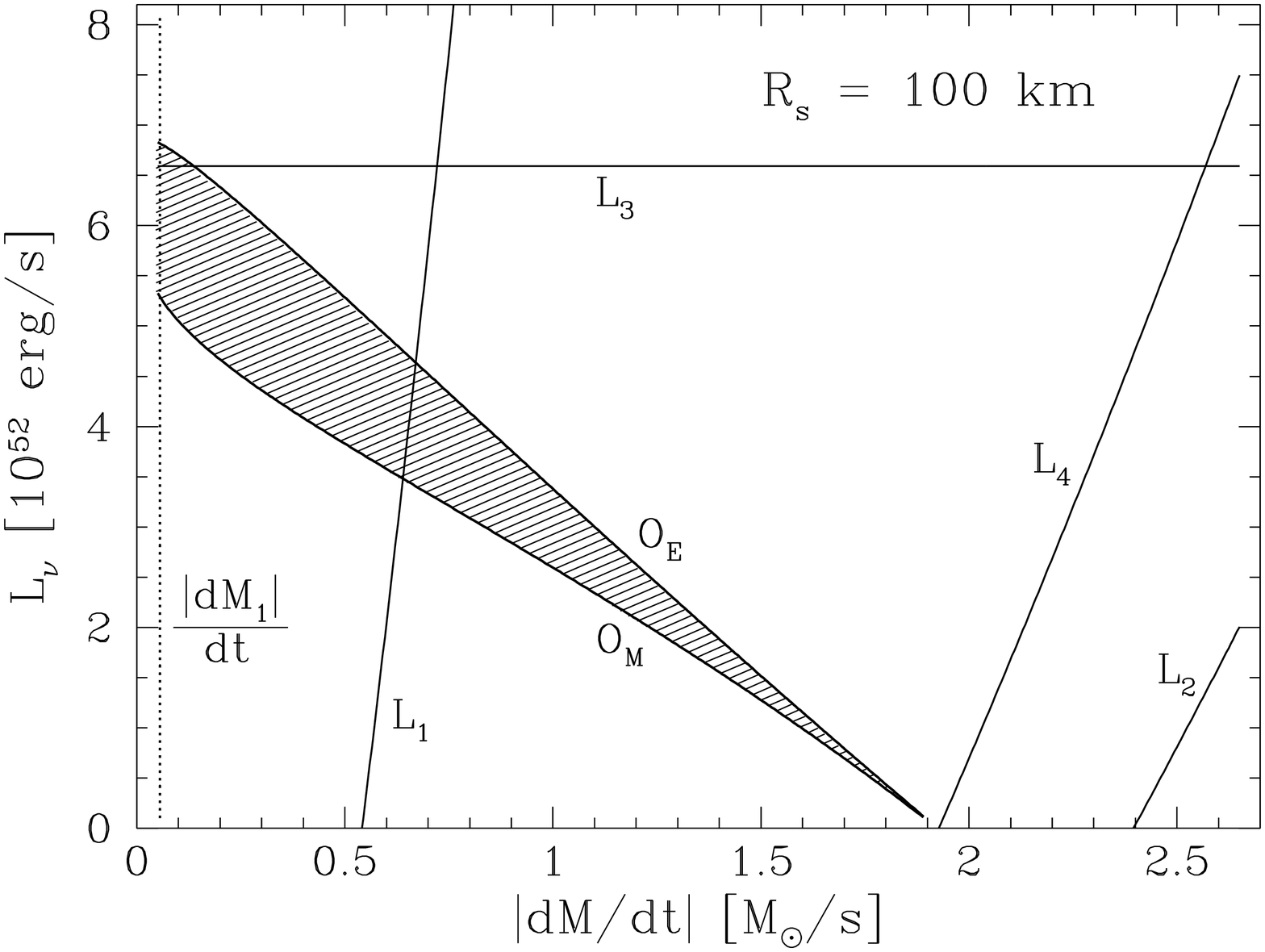}\hfill
  \hskip 0.5truecm
  \epsfxsize=0.48\hsize\epsffile{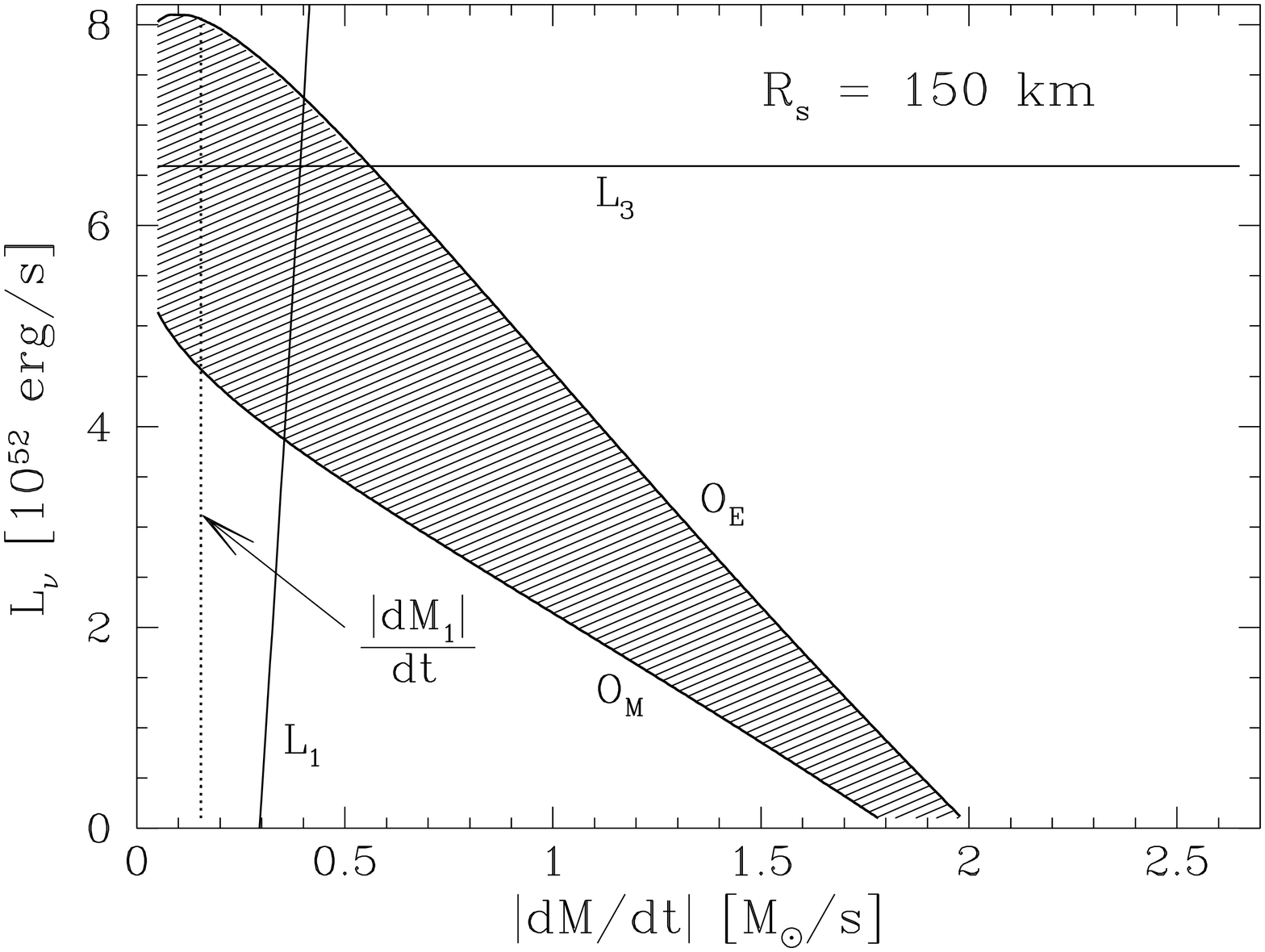}\hfill
  }
\vskip 1.0truecm
\centerline{
  \epsfxsize=0.48\hsize\epsffile{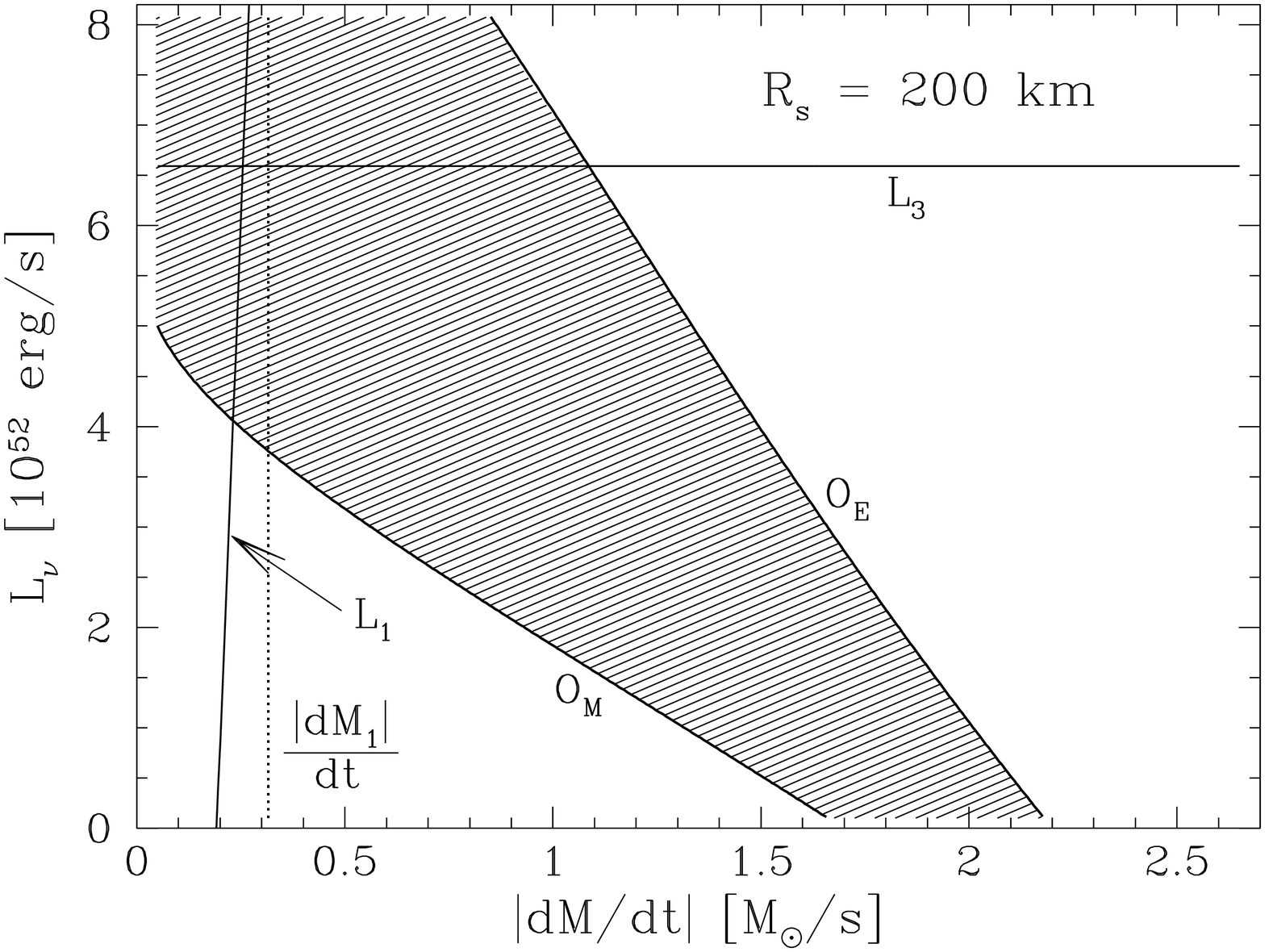}\hfill
  \hskip 0.5truecm
  \epsfxsize=0.48\hsize\epsffile{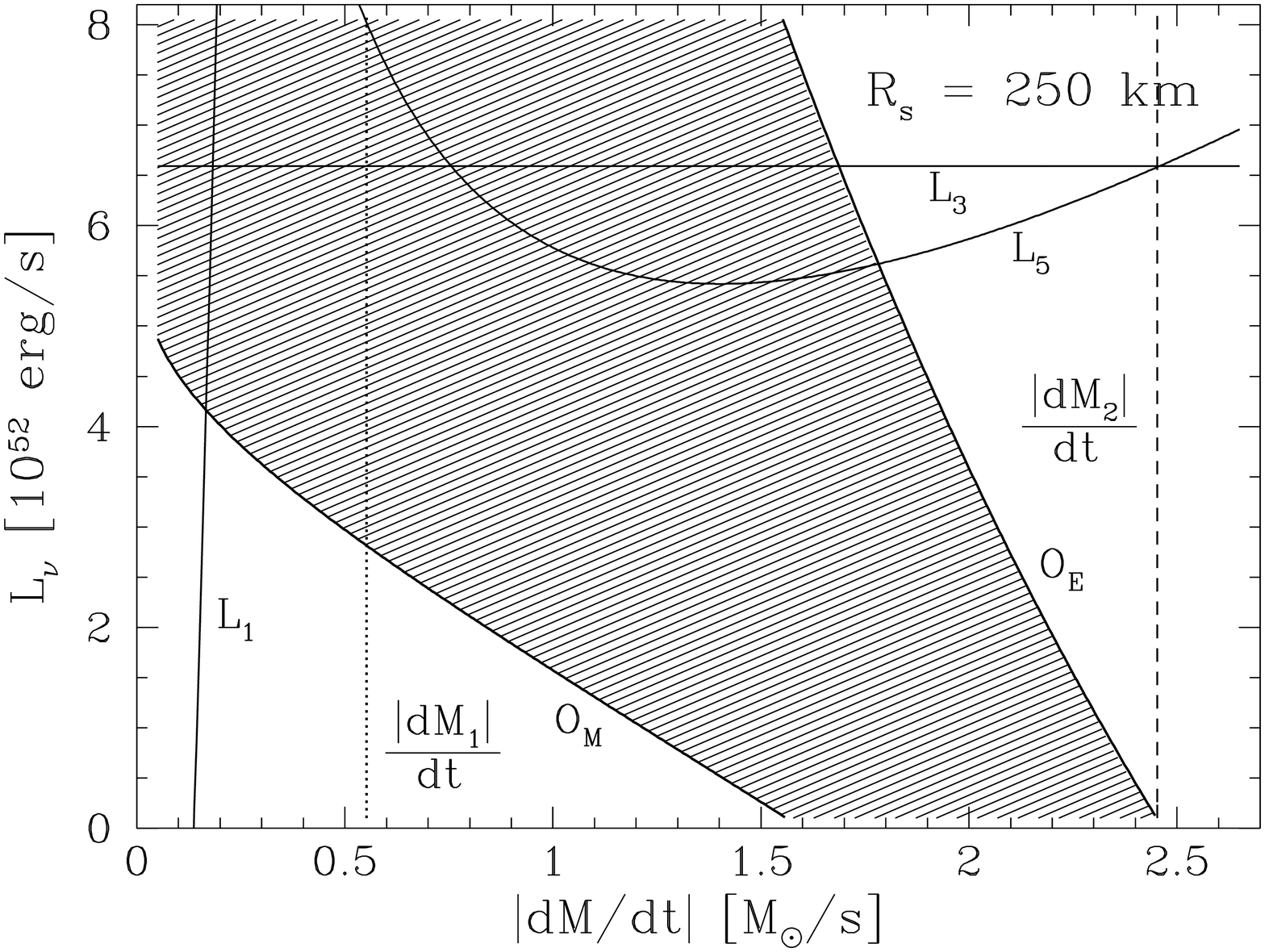}\hfill
  }
\vskip 0.5truecm
\centerline{
\parbox[b]{0.48\hsize}{%
    \epsfxsize=\hsize\epsffile{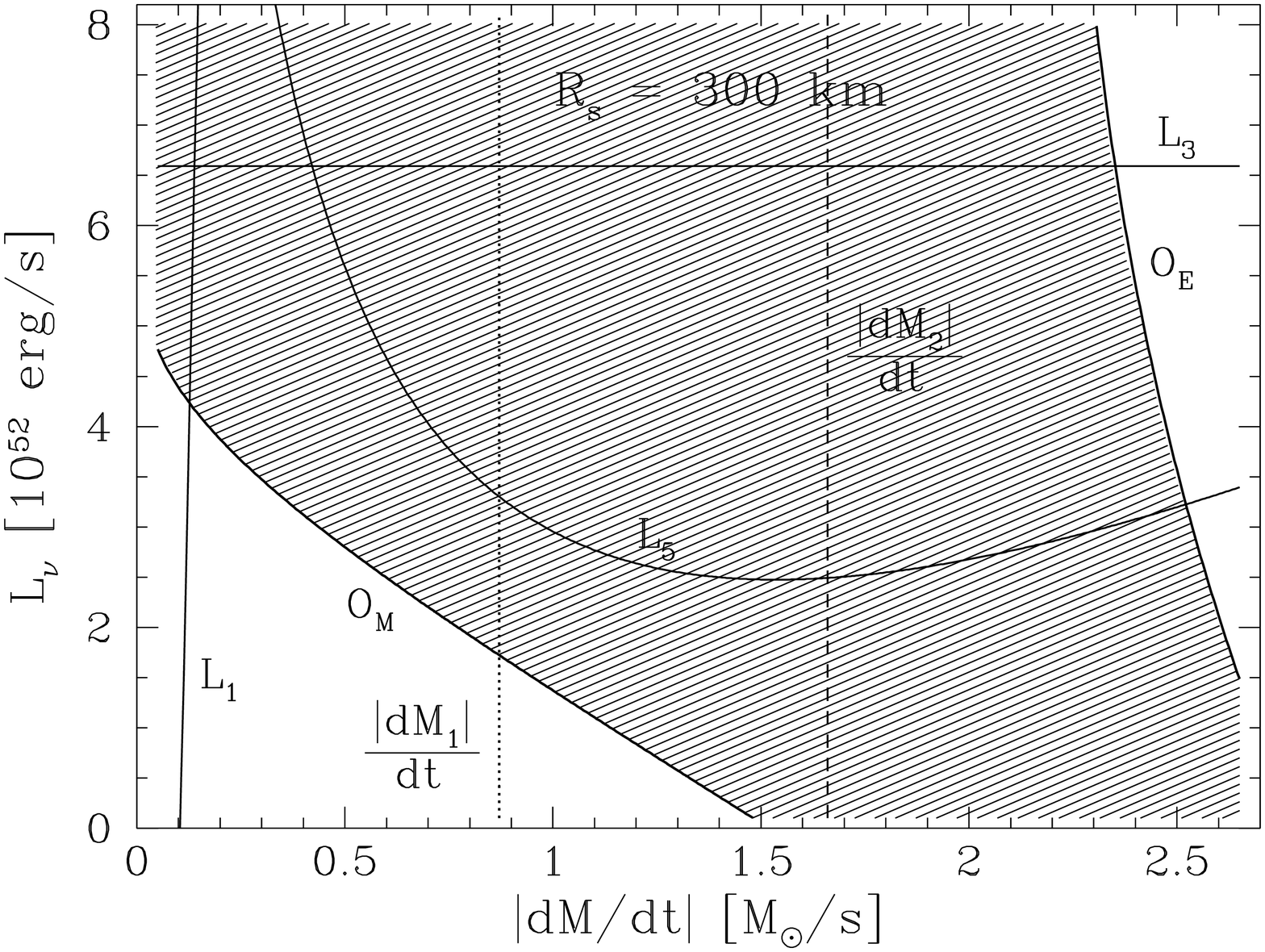}}
  \hfill
  \parbox[b]{0.48\hsize}{%
\caption{Conditions for shock revival by neutrino heating for different
shock stagnation radii $R_{\mathrm{s}}$. The lines labeled with O$_{\mathrm{E}}$
and O$_{\mathrm{M}}$ connect the roots of ${\mathrm{d}}(\Delta E_{\mathrm{g}})/
{\mathrm{d}}t$ and ${\mathrm{d}}(\Delta M_{\mathrm{g}})/{\mathrm{d}}t$,  
respectively, in the plane defined by the mass accretion rate into the shock, 
$\dot M$, and the neutrinospheric luminosity $L_\nu$. The curves with labels $L_i$
($i \in \{1,...,5\}$) and $|{\mathrm{d}}M_j/{\mathrm{d}}t|$ ($j \in \{1,2\}$)
correspond to the constraints (i)--(vi) listed in Sect.~\ref{sec:conditions}.
They represent warning flags indicating that the assumptions of the discussion 
may need to be generalized. The hatched areas mark the regions where the
conditions are favorable for a neutrino-driven
explosion, because Eqs.~(\ref{eq:108}) and (\ref{eq:109}) are both satisfied
such that the supernova shock expands and accelerates. Below the curve 
O$_{\mathrm{M}}$ the rate of mass loss from the gain layer to the neutron star
exceeds the mass accretion rate $\dot M$ and therefore 
${\mathrm{d}}(\Delta M_{\mathrm{g}})/{\mathrm{d}}t$ is negative.
Above the curve O$_{\mathrm{E}}$ the energy deposition by neutrino heating 
cannot compensate for the accumulation of mass with negative
total energy in the gain region and therefore 
${\mathrm{d}}(\Delta E_{\mathrm{g}})/{\mathrm{d}}t$ is negative.
\label{fig:4}}}
}
\end{figure*}

\subsection{Conditions for shock revival}
\label{sec:conditions}

The properties of Eq.~(\ref{eq:108}) together with Eq.~(\ref{eq:109}) 
will now be discussed in more detail. For chosen fixed values of the shock 
stagnation radius, those combinations of mass accretion rate $\dot M$ and
neutrinospheric luminosity $L_{\nu}$ will be determined which allow for
an outward acceleration of the shock front. For these conditions an
explosion driven by neutrino energy deposition may develop.

Assuming $U_{\mathrm{s}} = 0$ the gain radius is given by Eq.~(\ref{eq:91b}).
For the neutrino luminosity $L_{\nu} = 2L_{\nu_e}$ will be taken again.
The accretion rate $\dot M'$ of Eq.~(\ref{eq:89})
can be calculated by using Eqs.~(\ref{eq:89a}) and (\ref{eq:89b}).
Neutrino effects are evaluated from Eqs.~(\ref{eq:71})--(\ref{eq:73}) and 
Eqs.~(\ref{eq:79}) and (\ref{eq:83}) with Eq.~(\ref{eq:84}) for the 
postshock density $\rho_{\mathrm{s}}$.

Several consistency constraints have to be taken into account to make
sure that the assumptions of the analytic model developed in the preceding 
sections are fulfilled:
\begin{itemize} 
\item[(i)] For the gain radius [Eq.~(\ref{eq:91b})]
$R_{\nu}\! +\! h \la R_{\mathrm{g}}\le R_{\mathrm{s}}$ must hold. 
Here $h$ is the scale height 
of the exponential neutron star atmosphere, Eq.~(\ref{eq:51}).
The left inequality constrains the neutrinospheric luminosity to be 
$L_{\nu} \la L_1(\dot M)$, where the limit $L_1$ depends on the accretion 
rate $\dot M$. The right inequality, on the other hand, requires
$L_{\nu} \ge L_2(\dot M)$.  
\item[(ii)] Since the neutrinospheric luminosity $L_{\nu}$ and
temperature $T_{\nu}$ are {\em not} coupled here by the assumption of
blackbody emission, Eq.~(\ref{eq:74}) must be satisfied to have
$L_{\mathrm{acc}} \ge 0$, i.e., to have a cooling layer outside of the
neutrinosphere. This translates into a condition $L_{\nu} \le L_3$.
\item[(iii)] The definition of $R_{\mathrm{g}}$ implies that neutrinos 
transfer energy to the stellar gas for $R_{\mathrm{g}} \le r \le 
R_{\mathrm{s}}$. Therefore Eq.~(\ref{eq:83}) has to fulfill the 
condition ${\cal H}-{\cal C} \ge 0$, corresponding to 
$L_{\nu} \ge L_4(\dot M)$. This constraint is similar to the one
which follows from the requirement that $R_{\mathrm{g}}\le R_{\mathrm{s}}$,
but somewhat stronger, depending on the value of the ratio between
$\ave{\mu_{\nu}}_{\mathrm{g}}$ and $\ave{\mu_{\nu}}^\ast$. 
\item[(iv)] Equation~(\ref{eq:85}) [with ${\cal H}$ taken from 
Eq.~(\ref{eq:79})] has to be less than about 0.5 to justify
the disregard of reabsorption of neutrinos emitted from the gain layer.
This limits the neutrinospheric luminosity to
$L_{\nu} \la L_5(\dot M)$.
\item[(v)] The postshock temperature must be $kT_{\mathrm{s}}\ga 1\,$MeV
because the matter behind the shock is assumed to be completely disintegrated 
into free nucleons, and $\alpha$ particles therefore do not exist.
For this to hold, the absolute value of the mass accretion rate must exceed
some lower limit, $|\dot M| \ga |\dot M_1|$, where $\dot M_1$ depends on the 
shock radius and the effective mass $\widetilde{M}$ of the remnant.
\item[(vi)] Since self-gravity of the gas between neutrinosphere and
shock was neglected, the total mass there must be much smaller than the
mass of the neutron star. This requirement leads to an upper limit for
the rate of mass accretion: $|\dot M| \la |\dot M_2|$. 
\end{itemize}
While conditions (i)--(iii) ensure the logical coherence of the model,
a violation of conditions (iv)--(vi) would just reduce the accuracy of
the discussion. For example, the framework developed in the previous sections
can be generalized such that the (partial) recombination of nucleons to
$\alpha$ particles or heavy nuclei at temperatures below about one MeV
is taken into account. Item (vi) implies
\begin{eqnarray}
\int\limits_{R_{\nu}}^{R_{\mathrm{s}}}\!\!{\mathrm{d}}r\,4\pi r^2\rho(r)
& = & \int\limits_{R_{\nu}}^{R_{\mathrm{eos}}}\!\!{\mathrm{d}}r\,4\pi r^2
\rho(r) + \int\limits_{R_{\mathrm{eos}}}^{R_{\mathrm{s}}}\!\!{\mathrm{d}}r\,
4\pi r^2\rho(r)
\nonumber \\
& \la & \int\limits_{R_{\nu}}^\infty \!\!{\mathrm{d}}r\,4\pi r^2\rho_1(r)
+ \!\!\! \int\limits_{R_{\nu}+h}^{R_{\mathrm{s}}}\!\!\!\!{\mathrm{d}}r\,
4\pi r^2\rho_2(r) \,,
\label{eq:110}
\end{eqnarray}
where $\rho_1(r)\equiv \rho_{\nu}\exp\eck{-(r\!-\!R_{\nu})/h}$ [Eq.~(\ref{eq:50})
with $h$ from Eq.~(\ref{eq:51})] and $\rho_2(r) \equiv \rho_{\mathrm{s}}
(R_{\mathrm{s}}/r)^3$ [Eq.~(\ref{eq:60})]. A reasonable upper bound for this
mass integral is
\begin{equation}
\int\limits_{R_{\nu}}^{R_{\mathrm{s}}} {\mathrm{d}}r\,4\pi r^2\rho(r)
\,\la\, {\widetilde{M}\over 5} \ ,
\label{eq:111}
\end{equation}
which limits the allowed accretion rate according to condition (vi).

The sequence of plots in Fig.~\ref{fig:4} shows the results of 
an evaluation of Eqs.~(\ref{eq:108}) and (\ref{eq:109}) together with 
the constraints (i)--(vi) for different shock stagnation radii:
$R_{\mathrm{s}} = 100, 150, 200, 250$ and 300~km, respectively.
The numerical values chosen for the other parameters were:
$\widetilde{M}= 1.25$~M$_{\odot}$, $R_{\nu} = 50$~km, $kT_{\nu} = 4$~MeV,
$\beta = 7$, $\Gamma = \gamma = {4\over 3}$, $f_{\mathrm{g}} = 1.25$
(corresponding to $\eta_e Y_e \approx 1$ at the neutrinosphere),
$q_{\mathrm{d}} = 8.5\times 10^{18}$ erg$\,$g$^{-1}$, 
$\ave{\mu_{\nu}}^\ast = 0.7$, $\ave{\mu_{\nu}}_{\mathrm{g}} = 0.6$, and 
$\widetilde{\ave{\mu_{\nu}}} = 0.4$.

The roots of ${\mathrm{d}}(\Delta E_{\mathrm{g}})/{\mathrm{d}}t$ and 
${\mathrm{d}}(\Delta M_{\mathrm{g}})/{\mathrm{d}}t$ are represented by the 
lines labeled with O$_{\mathrm{E}}$ and O$_{\mathrm{M}}$, respectively.
These lines separate regions in the
$|\dot M|$-$L_{\nu}$ plane, within which the collapsed stellar core
reacts differently to the mass inflow through the shock and to the 
irradiation by neutrinos emitted from the neutrinosphere
(and from the cooling layer). In this respect
the plots of Fig.~\ref{fig:4} can be considered as ``phase diagrams''
for the post-bounce evolution of the supernova. Within the hatched 
areas both Eqs.~(\ref{eq:108}) and (\ref{eq:109}) are simultaneously 
fulfilled. Additional lines correspond to constraints (i)--(vi). They are
displayed as warning flags that the assumptions of the treatment may 
need to be generalized. Left of the vertical dotted
line, which corresponds to constraint (v), $\alpha$ particles and heavy
nuclei in the
postshock medium would have to be taken into account, and the analysis 
performed here is not very accurate. The vertical dashed line marks the 
boundary right of which Eq.~(\ref{eq:111}) and thus constraint (vi)
is violated.

\begin{figure*}[t]
\centerline{
  \epsfxsize=0.48\hsize\epsffile{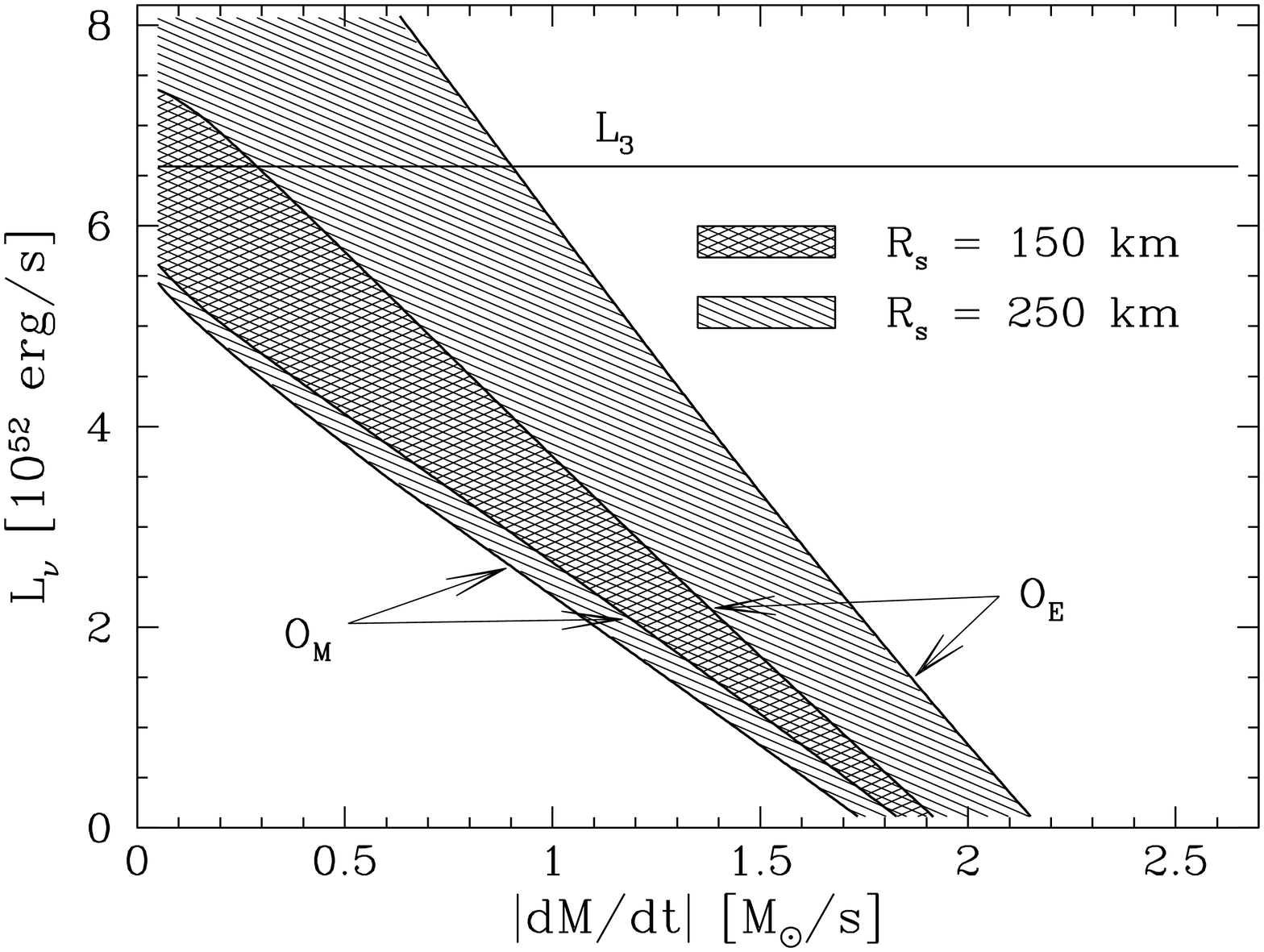}\hfill
  \hskip 0.5truecm
  \epsfxsize=0.48\hsize\epsffile{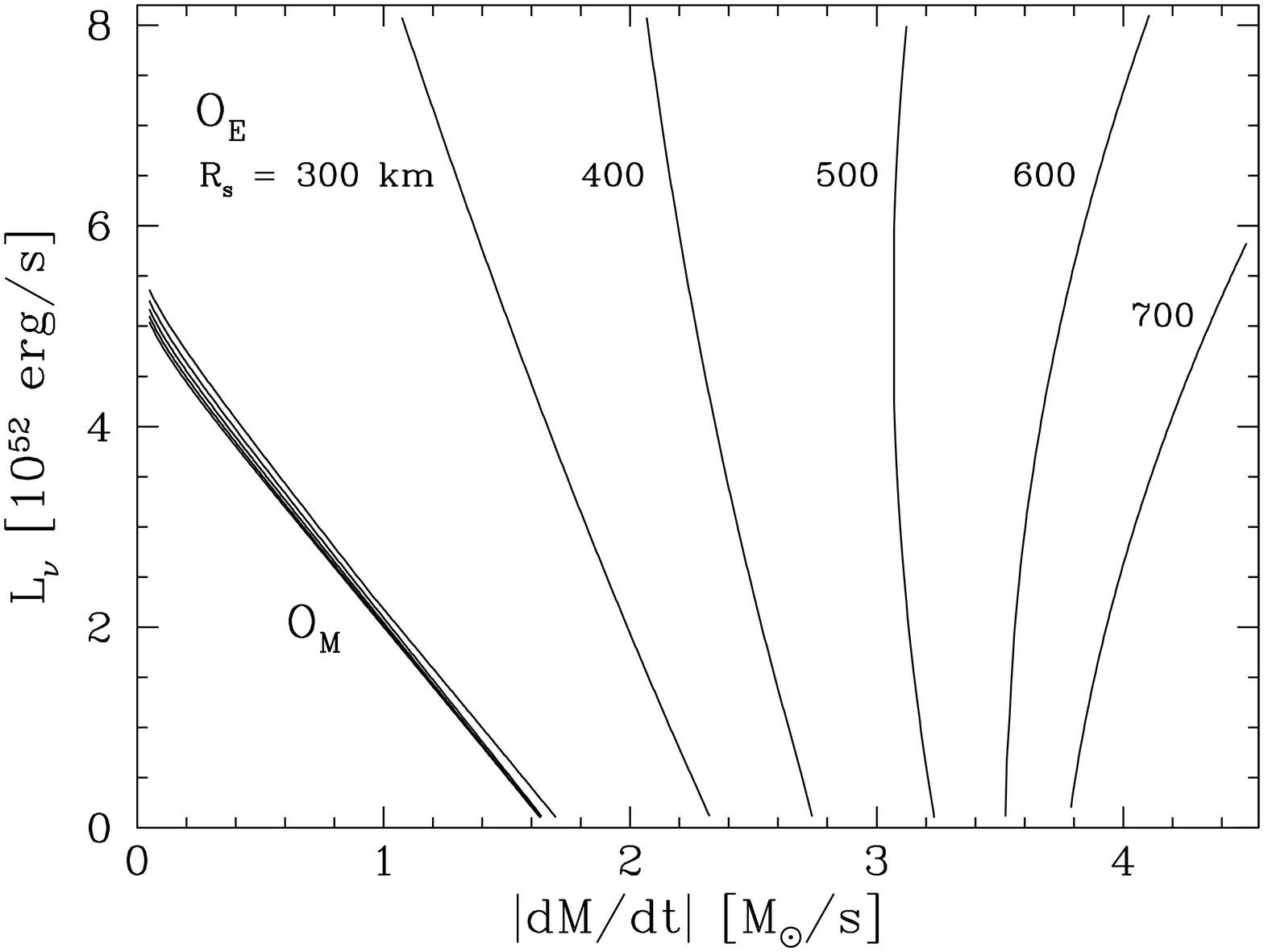}\hfill
  }
\caption{{\em Left:} Conditions for shock revival by neutrino heating for
shock stagnation radii $R_{\mathrm{s}} = 150\,$km (cross-hatched area)
and 250$\,$km (hatched area).
{\em Right:} The lines O$_{\mathrm{M}}$ and O$_{\mathrm{E}}$ which connect
the roots of Eqs.~(\ref{eq:108}) and (\ref{eq:109}), respectively, for
shock radii 300~km, 400~km, 500~km, 600~km, and 700~km.
Different from Fig.~\ref{fig:4}, the gain radius and the integrals for
neutrino heating and cooling in the gain layer were evaluated by using the
exact solution for hydrostatic equilibrium instead of approximate power-law
profiles. In addition, the disappearance of free nucleons and therefore the
quenching of neutrino absorption and emission below a temperature of 1~MeV
were taken into account. Above the line labeled with $L_3$ the accretion
luminosity $L_{\mathrm{acc}}$ [Eq.~(\ref{eq:73})] becomes negative.
The lines corresponding to constraints (v) and (vi) are omitted for reasons
of clarity.
\label{fig:5}  
}
\end{figure*}

Two of the simplifications that entered the analysis for Fig.~\ref{fig:4} 
can be easily removed. On the one hand, the gain radius
$R_{\mathrm{g}}$ and heating and cooling in the gain layer can be
calculated more accurately, when the density and temperature
profiles of Eqs.~(\ref{eq:57}) and (\ref{eq:59}) instead of the power-law 
approximation of the hydrostatic atmosphere [Eq.~(\ref{eq:60})] are used.
In this case $R_{\mathrm{g}}$ must be numerically determined as the root of 
Eq.~(\ref{eq:33}) [with $L_{\nu_e} = {1\over 2}L_{\nu}(R_{\mathrm{g}})$
as given by Eq.~(\ref{eq:72})], 
and the integrals for ${\cal H}$ and ${\cal C}$ in Eq.~(\ref{eq:78})
can also be evaluated numerically. On the other hand, the recombination of free nucleons
to $\alpha$ particles and heavy nuclei at low temperatures can roughly be taken into 
account concerning its effects on the neutrino interaction in the gain region.
Provided that above a certain temperature, say 1~MeV, all nuclei are disintegrated
into free nucleons and below this temperature all nucleons are bound in nuclei,
neutrino absorption and emission reactions will not take place outside of the
corresponding radius $R_{\alpha}$. The latter can also be calculated from the 
temperature profile of Eq.~(\ref{eq:59}). The heating and cooling integrals are
then performed with the upper integration boundary being chosen as the minimum of
$R_{\mathrm{s}}$ and $R_{\alpha}$. The recombination of nucleons to $\alpha$
particles releases a sizable amount of energy, about 7~MeV per nucleon. This 
additional energy source in the gain region was not included in the discussion 
here, because it requires a detailed modelling of the composition history of the 
postshock medium (considering the different degree of disintegration of nuclei
during infall and recombination of nucleons during later expansion in different
volumes of matter).

The results of this more general treatment are displayed in Fig.~\ref{fig:5} 
for shock radii $R_{\mathrm{s}} = 150\,$km and 250$\,$km. The quantitative
changes are significant: Compared to Fig.~\ref{fig:4} the different value
for the ${\cal H}-{\cal C}$ term moves the O$_{\mathrm{M}}$-line slightly upward and the
O$_{\mathrm{E}}$-line more strongly downward. A similar effect is associated with a 
moderate reduction of the shock radius in Fig.~\ref{fig:4}. 
The gain radius obtained by the exact calculation can also 
shrink with growing shock radius, different from the approximate
representation of Eq.~(\ref{eq:91b}). On the other hand, the outer boundary of the
gain layer is defined by the recombination radius $R_{\alpha}$ of $\alpha$ particles
instead of the possibly larger shock radius. Both effects combined, the total 
heating rate in the gain layer is similar. Therefore the qualitative picture remains
unchanged.

The hatched areas in the plots of Figs.~\ref{fig:4} and \ref{fig:5} 
include those combinations of parameters for which the conditions of 
Eqs.~(\ref{eq:108}) and (\ref{eq:109}) are both satisfied and
therefore the initially stagnant shock will expand and will be accelerated.
Below the O$_{\mathrm{M}}$ line neutrino cooling outside of the neutrinosphere
is very efficient and the neutron star swallows matter faster than gas
is resupplied by accretion through the shock. Therefore
${\mathrm{d}}(\Delta M_{\mathrm{g}})/{\mathrm{d}}t$ is negative.
When the mass accretion rate $|\dot M|$ drops below this critical line,
shock expansion can be supported only by an increasing core luminosity,
because a larger value of $L_{\nu}$ 
reduces the neutrino losses from the cooling region. Otherwise advection 
through the gain radius and thus into the neutron star extracts mass 
from the neutrino-heating region and the shock retreats.
Figs.~\ref{fig:4} and \ref{fig:5} show that for given 
rate $|\dot M|$ there is a lower limit of the neutrinospheric luminosity
$L_{\nu}$, which must be exceeded when shock expansion and acceleration
shall occur.

Above the O$_{\mathrm{E}}$ line neutrino heating
[represented by the ${\cal H}-{\cal C}$ term in Eq.~(\ref{eq:108})]
cannot compete with the accumulation of matter with negative total
energy in the gain layer. In this case
${\mathrm{d}}(\Delta E_{\mathrm{g}})/{\mathrm{d}}t$ is negative.
$R_{\mathrm{s}}$ can nevertheless grow 
for such conditions, simply because gas piles up on top of the neutron
star. This pushes the shock farther out, but does not allow positive
postshock velocities to develop. Since the postshock matter is
gravitationally bound ($\Delta E_{\mathrm{g}} < 0$), an explosion, however,
requires sufficiently powerful energy input by neutrinos.
The position of the O$_{\mathrm{E}}$ line shifts with changing shock
radius. For discussing the destiny of the shock this change of the 
overall situation associated with the shock motion therefore has to be 
taken into account. This can be done by solving the equations of the
toy model for time-dependent information about the shock radius and the
shock velocity (see Sect.~\ref{sec:tsolutions}).

The O$_{\mathrm{E}}$ line is very sensitive to the shock position,
whereas the O$_{\mathrm{M}}$ line is only weakly dependent 
(Fig.~\ref{fig:5}). On the one hand,
a high core luminosity $L_{\nu}$ reduces the downward advection of
gas through the gain radius. On the other hand, the neutrino heating 
in the gain layer increases with larger shock radius. Both effects
determine the slopes and positions of the critical lines.
The distance between the O$_{\mathrm{M}}$ and O$_{\mathrm{E}}$ lines in
Figs.~\ref{fig:4} and \ref{fig:5} grows for larger shock radii, and
the hatched area expands. This is caused by an increase of the 
${\cal H}-{\cal C}$ term in Eq.~(\ref{eq:108}).

Acceleration is easier for a shock which has stalled at a large distance
from the center, i.e., the same core luminosity can then ensure favorable
conditions already for a higher value of $|\dot M|$.
Besides stronger neutrino heating in the more extended gain region,
another effect contributes to this. The increase of the postshock pressure
which is necessary to accelerate the standing shock to a positive velocity 
$U_{\mathrm{s}} \ll |v_{\mathrm{p}}|$ is given by
\begin{eqnarray}
\Delta P_{\mathrm{s}}& = &P_{\mathrm{s}}(U_{\mathrm{s}}) - P_{\mathrm{s}}(0)
\nonumber\\
&\cong& -2\rund{\! 1\! -\! {1\over \beta}}
\rho_{\mathrm{p}}v_{\mathrm{p}}U_{\mathrm{s}}\,=\,
-\rund{\! 1\! -\! {1\over \beta}}
{\dot M U_{\mathrm{s}}\over 2\pi\, R_{\mathrm{s}}^2}\ . 
\label{eq:111a}
\end{eqnarray}
This pressure increase is lower when the shock radius $R_{\mathrm{s}}$ is big.

The O$_{\mathrm{M}}$ line defines a critical curve for the shock
evolution, whose slope and position are hardly dependent on the 
shock radius. It can be approximated analytically by solving 
Eq.~(\ref{eq:109}) in case of $\dot M' = \dot M$ for the critical
core luminosity $L_{\nu}^\ast$ as a function of $\dot M$. One derives
\begin{equation}
L_{\nu}^\ast(\dot M)\,\approx \,{\mathrm{e}^{-a}b(1 - 0.1\omega)- 
\dot M(l_{\nu}+q_{\mathrm{d}})
\over 1 - {\mathrm{e}}^{-a}(1-0.1\omega)}\ ,
\label{eq:111b}
\end{equation}
where $l_{\nu}$ is defined as the quantity $l = (e+P)/\rho - G\widetilde{M}/r$
at the neutrinosphere [cf.\ Eqs.~(\ref{eq:89}) and (\ref{eq:89b})], 
$a$ and $b$ were introduced in Eqs.~(\ref{eq:71}) and
(\ref{eq:71a}), respectively, and $\omega$ is given by 
\begin{equation}
\omega\,\equiv\,{(kT_{\nu_e,4})^2\over \ave{\mu_{\nu}}^\ast}\,
{(-\dot M)\over {\mathrm{M}}_{\odot}/{\mathrm{s}}}
\rund{\!{\widetilde{M}\over {\mathrm{M}}_{\odot}}\!}^{\!\! -{1\over 2}}
\!\!\!
R_{\mathrm{s},7}^{-1/2}\eck{\rund{{R_{\mathrm{s}}\over R_{\mathrm{g}}}}^{\!\! 2}
\!\!-\! 1} .
\label{eq:111c}
\end{equation}
To obtain Eq.~(\ref{eq:111b}) use was made of 
Eqs.~(\ref{eq:63}), (\ref{eq:72}), (\ref{eq:73}), (\ref{eq:79}), (\ref{eq:83}),
(\ref{eq:84}), (\ref{eq:89}), (\ref{eq:89a}), and (\ref{eq:91b}).
The term in the curly brackets of
Eq.~(\ref{eq:83}) was assumed to be equal to ${2\over 3}$, and the rather
weak dependence of the gain radius in Eq.~(\ref{eq:111c}) on the neutrino
luminosity was ignored in writing $L_{\nu}^\ast$ in the explicit form of 
Eq.~(\ref{eq:111b}). In the latter equation $\omega$ also depends on the
mass infall rate $\dot M$. For representative shock radii and accretion 
rates, $\omega$ is found to be of order unity: $\omega \sim 1$. With this,
$L_{\nu}^\ast$ becomes a simple linear function of $\dot M$. Inserting
the parameter values used for Figs.~\ref{fig:4} and \ref{fig:5} [listed
after Eq.~(\ref{eq:111})], one ends up with
\begin{equation}
L_{\nu}^\ast(\dot M)\,\approx \,
\rund{5.6 - 3.3\,{(-\dot M)\over {\mathrm{M}}_{\odot}/{\mathrm{s}}}}
\times 10^{52} \,\,{{\mathrm{erg}}\over{\mathrm{s}}}\ .
\label{eq:111d}
\end{equation}
This expression has a root for $\dot M = -1.7$~M$_{\odot}\,{\mathrm{s}}^{-1}$ 
and fits the O$_{\mathrm{M}}$ lines in Figs.~\ref{fig:4} and \ref{fig:5} 
reasonably well.

\begin{figure*}[t]
\centerline{
  \epsfxsize=0.48\hsize\epsffile{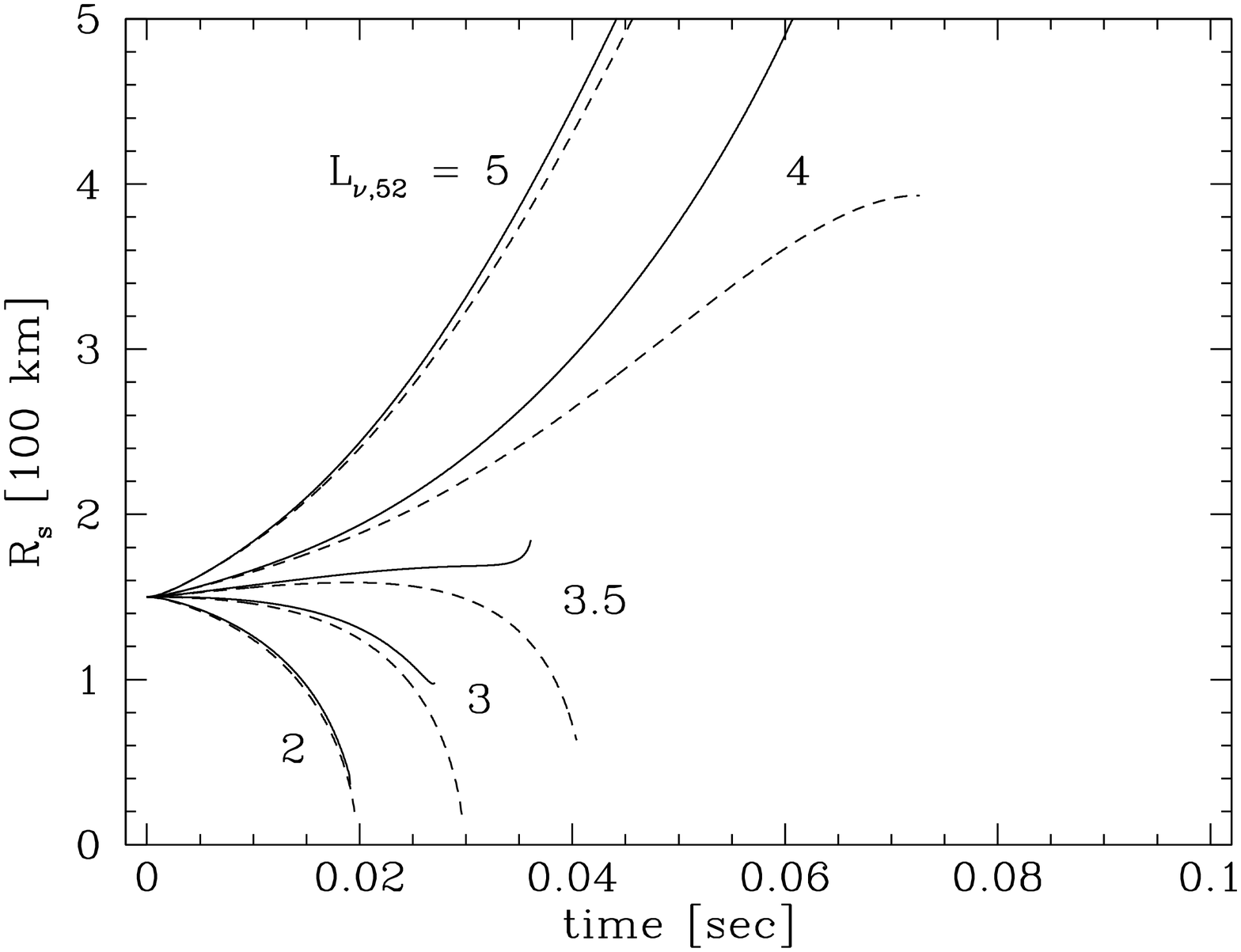}\hfill
  \hskip 0.5truecm
  \epsfxsize=0.48\hsize\epsffile{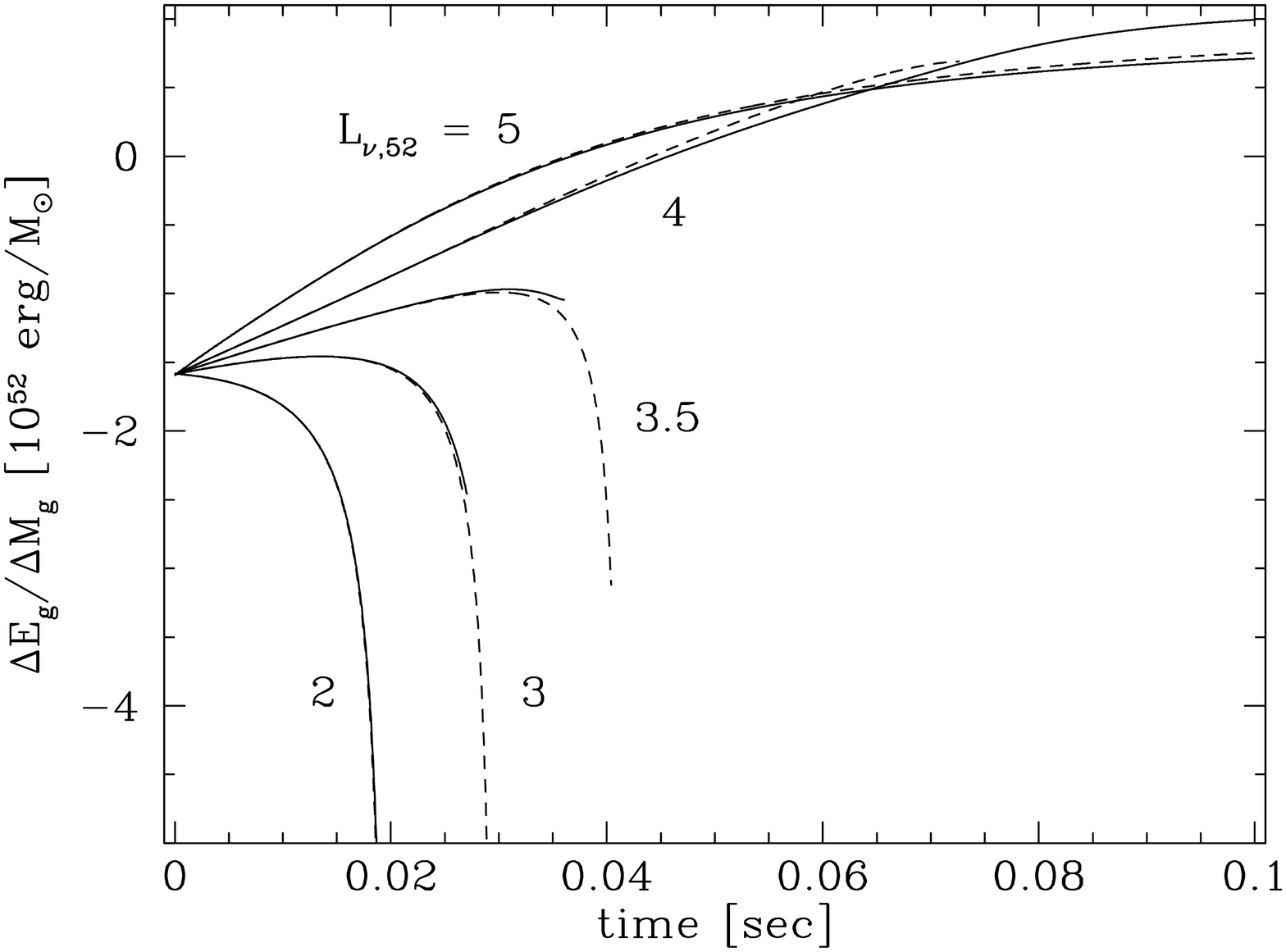}\hfill
  }
\caption{Shock radius (left) and specific energy in the gain layer (right)
as functions of time for different neutrinospheric luminosities 
(measured in units of $10^{52}$~erg$\,$s$^{-1}$) and for an
initial shock stagnation radius of 150~km [with $U_{\mathrm{s}}(t = 0) = 0$].
The structural polytropic index of the gain layer was chosen to be 
$\gamma = 4/3$, the mass accretion rate into the shock $|\dot M| = 
0.3$ M$_{\odot}\,$s$^{-1}$,
the neutron star mass $\widetilde{M} = 1.25\,$M$_{\odot}$, and
the neutrinospheric radius and temperature $R_{\nu} = 50\,$km and
$T_{\nu} = 4\,$MeV, respectively. The gain radius and the integrals for
neutrino heating and cooling in the gain layer were evaluated by using the
exact solution for hydrostatic equilibrium. The disappearance of free 
nucleons and therefore the quenching of neutrino absorption and emission
below a temperature of 1~MeV were also taken into account. The dashed lines
correspond to the case where the density jump in the shock was set to
be large (see text).
\label{fig:6}  
}
\end{figure*}

\subsection{Time-dependent solutions}
\label{sec:tsolutions}

The equations of the toy model developed in this paper can be solved
for the shock radius $R_{\mathrm{s}}(t)$ and the shock velocity 
$U_{\mathrm{s}}(t)$ as functions of time.
For this purpose the mass and energy in the gain layer have to be
evolved according to the conservation laws of 
Eqs.~(\ref{eq:93}) and (\ref{eq:104}). Together
with $U_{\mathrm{s}} = {\mathrm{d}}R_{\mathrm{s}}/{\mathrm{d}}t 
\equiv \Delta R_{\mathrm{s}}/\Delta t$ these equations were
integrated implicitly in time, with the velocity of the gain radius
given by $\dot R_{\mathrm{g}} \equiv \Delta R_{\mathrm{g}}/\Delta t$
for time step $\Delta t$. The
mass in the gain layer, $\Delta M_{\mathrm{g}}$, and the corresponding
total (internal plus gravitational) energy, $\Delta E_{\mathrm{g}}$,
were initially calculated from Eq.~(\ref{eq:91}) and 
Eqs.~(\ref{eq:95}) and (\ref{eq:96}), respectively. The gain radius
$R_{\mathrm{g}}$ and the heating and cooling integrals for the gain
layer were evaluated using the exact solution of hydrostatic equilibrium
[Eqs.~(\ref{eq:57})--(\ref{eq:59})] with the option to chose an 
arbitrary value for the structural polytropic index $\gamma$
[Eq.~(\ref{eq:59a})]. The quenching of neutrino absorption and emission
reactions by the recombination of free nucleons to 
$\alpha$ particles and heavy nuclei below a temperature around 
1~MeV was taken into account. 

The postshock density is related to the preshock density by 
$\rho_{\mathrm{s}} = \beta\rho_{\mathrm{p}}$, and the postshock
pressure is given by Eq.~(\ref{eq:39}). The density contrast $\beta$ as well
as the pressure jump at the shock are affected by the conditions in the
gain layer. The latter is assumed to be in hydrostatic equilibrium
with mass inflow from the infall region and additional gain or loss by mass
exchange with the neutron star. Therefore simultaneous conservation of mass
and energy requires that $\beta$ is allowed to float, just as $P_{\mathrm{s}}$
is a degree of freedom which adjusts in response to the energy input due to
the heating by neutrinos. This means that $\beta$ is also
considered as a variable which the set of equations is solved for.

Although generalization is straightforward, the neutron star mass 
$\widetilde{M}$, the mass accretion rate into the shock, $\dot M$,
and the neutrinospheric parameters 
($L_{\nu}$, $T_{\nu}$, $R_{\nu}$) were kept constant with time for 
reasons of simplicity. Supernova calculations show that during a
transient, but rather short period
of several 10~ms up to about 100~ms after bounce, $L_{\nu}$ and 
$\dot M$ decrease from very high values to a much lower level, and
lateron change only slowly with time (cf., for example, Fig.~2
in Rampp \& Janka 2000). A discussion of the subsequent destiny of 
the supernova shock should not be affected by this variation, because
the shock expansion turns out to occur on a significantly
shorter timescale (see below). The ongoing contraction of the 
neutron star and a corresponding change of the neutrinospheric temperature
and luminosity, however, were found to have considerable influence
(see Janka \& M\"uller 1996). For the exemplary purpose of the 
calculations reported on below, the introduction of additional, 
model-dependent degrees of freedom will nevertheless be abstained from.

\subsubsection{Results for different $L_{\nu}$ and fixed $\dot M$}

The shock radius $R_{\mathrm{s}}(t)$ and the specific energy in the gain layer, 
$\Delta E_{\mathrm{g}}/\Delta M_{\mathrm{g}}$, are shown as functions of time
in Fig.~\ref{fig:6} for different
neutrinospheric luminosities $L_{\nu}$. The structural polytropic exponent
$\gamma$ was set equal to the adiabatic index $\Gamma$ of the equation
of state, both chosen to be ${4\over 3}$. The other parameters of the 
evaluation were $\widetilde{M} = 1.25$~M$_{\odot}$, $R_{\nu} = 50$~km,
$T_{\nu} = 4$~MeV, and $|\dot M| = 0.3$~M$_{\odot}\,$s$^{-1}$. The 
initial shock radius was set to 150~km with $U_{\mathrm{s}}(t = 0) = 0$ and
$\dot R_{\mathrm{g}}(t = 0) = 0$. 

The solid lines display the case
where $\beta$ was allowed to vary in all equations. Only for sufficiently
high neutrinospheric luminosity the shock is able to expand to large radii.
For lower $L_{\nu}$ the specific energy per nucleon in the gain layer begins
to drop again at some stage of the evolution, and continued shock expansion
is not possible, because the postshock pressure is not large enough for
driving the shock out. The sudden positive acceleration of the shock front
towards the end of the solid lines for these unsuccessful cases is a 
mathematical artifact, which occurs in response to the rapid decrease of 
the factor $(1-\beta^{-1})$ in Eq.~(\ref{eq:39}) as $\beta$ falls to 
unphysical values near unity. For a given value of $P_{\mathrm{s}}$, the 
decay of the term $(1-\beta^{-1}) \to 0$ is attempted to be compensated
by a catastrophic increase of the factor $(v_{\mathrm{p}}-U_{\mathrm{s}})^2$
in Eq.~(\ref{eq:39}). In contrast to the solid lines, the dashed curves
were obtained by explicity setting $(1-\beta^{-1}) \approx 1$ in 
Eq.~(\ref{eq:39}). Thus assuming that the density jump in the shock is
large, the shock velocity is solely determined by the value of the 
pressure $P_{\mathrm{s}}$ behind the shock. Therefore the dashed lines 
show the breakdown of the shock expansion more clearly than the solid 
lines. They confirm that shock recession is correlated with a decrease of 
the specific energy in the gain layer.

The properties of the time-dependent solutions for the shock radius agree with 
the discussion of
Sects.~\ref{sec:shockevol}--\ref{sec:conditions}. Keeping $\dot M$ fixed, there
is a threshold value for the core luminosity above which the shock runs out to large
radii and the energy per baryon in the gain layer becomes positive. A high
neutrinospheric luminosity has two favorable effects: On the one hand the
neutrino heating in the gain layer is larger, on the other hand the 
energy loss by neutrino emission in the cooling layer is lower, thus
reducing the mass accretion into the neutron star and the mass loss from the 
gain layer. The case with $L_{\nu} = 4\times 10^{52}$~erg$\,$s$^{-1}$ is near the
borderline between successful shock expansion and failure for the chosen set of 
parameters (compare also Fig.~\ref{fig:5}):
The shock is already very weak when it has reached a radius of 
about 400~km, which is clearly visible from the dashed lines. 

The mass accretion rates, $\dot M'$, of the nascent neutron star, which 
correspond to increasing values of the neutrinospheric luminosity $L_{\nu}$ in
Fig.~\ref{fig:6}, are all negative, with values: $|\dot M'| = 
1.15,\,0.84,\,0.68,\,0.53,$ and 0.21, and the $\nu_e$
plus $\bar\nu_e$ luminosities at the gain radius for these cases are:
$L_{\nu}(R_{\mathrm{g}}) = 5.0,\,5.3,\,5.5,\,5.7,$ and 
$6.0\times 10^{52}$ erg$\,$s$^{-1}$. Because of the contribution from
the accretion luminosity, $L_{\nu}(R_{\mathrm{g}})$ shows much less
variation than the core luminosity $L_{\nu}$, and the neutrino heating 
in the gain layer is also similar. The accretion component is not dominant when
the shock moves out. The breakdown of shock expansion is therefore associated with
a low neutrinospheric luminosity which causes high mass loss from the gain layer, leading
to a decrease of the pressure support behind the shock. At the same time the width 
of the gain layer shrinks, its optical depth drops, and the neutrino energy 
deposition decreases. This leads to a negative feedback and the shock recession
accelerates dramatically.

The optical depth for $\nu_e$ and $\bar\nu_e$ absorption in the gain layer
is given by Eq.~(\ref{eq:85}),
$\tau_{\mathrm{a}} \equiv {\cal H}/L_{\nu}(R_{\mathrm{g}})$. Its value depends on
the particular conditions, the shock position, mass infall rate into the shock,
and the neutrinospheric luminosity, temperature and radius, which influence
the position of the gain radius. For the models shown in
Figs.~\ref{fig:6} and \ref{fig:7} the initial value is between 0.15 and
0.18. In case of shock expansion the optical depth increases for an intermediate
period of time by up to 50 per cent due to the growth of the gain region. 
This improves the conditions for ongoing neutrino heating and
leads to a rapid rise of the energy in the gain layer. The positive feedback also 
causes a sharp bifurcation in the behavior of cases of failing and successful 
shock expansion. Because
neutrinos are not only absorbed, but also reemitted, the net effect of neutrino
energy deposition is lowered somewhat. It scales with
$({\cal H}-{\cal C})/L_{\nu}(R_{\mathrm{g}})$, which has typically only about
half the value of $\tau_{\mathrm{a}}$. When the gain layer expands, the 
temperature decreases and the reemission of neutrinos is reduced.

\subsubsection{Results for different $\dot M$}

When $|\dot M|$ lies below the O$_{\mathrm{M}}$ line of Figs.~\ref{fig:4} and
\ref{fig:5}, the shock expansion is suppressed.
But also high mass infall rates damp the shock expansion, because 
a larger increase of the postshock pressure is needed for shock acceleration
[see Eq.~(\ref{eq:111a})] and the optical depth of the gain layer decreases
(because the gain radius is farther out). For high $|\dot M|$ the shock 
therefore gains speed more slowly.

In case of very high mass infall rates $|\dot M|$ and small shock radii a 
gain layer does not exist. Provided the neutrino luminosity is sufficiently 
large such that $\dot M' > \dot M$ (which is easily fulfilled for high 
$|\dot M|$), the shock is slowly pushed outward by the gas that stays in
the layer between the neutron star and the shock. 
Eventually the postshock temperature will be low enough for a gain layer to
form. With neutrino-heated gas accumulating above the gain radius 
(i.e., $\Delta M_{\mathrm{g}}$ increases) the shock moves even farther out,
but the total energy in
the gain layer decreases because the neutrino heating cannot compensate the
negative binding energy of the growing gas mass. Only when the shock has reached
a sufficiently large radius the situation becomes favorable for an explosion
because then ${\mathrm{d}}(\Delta E_{\mathrm{g}})/{\mathrm{d}}t > 0$
(i.e., the conditions are now left of the corresponding
O$_{\mathrm{E}}$ line in Fig.~\ref{fig:5}).
If this radius is very far out, because $|\dot M|$ is very high,
the energy deposited in the gain layer may not be sufficient to produce
a positive total energy in the gain layer. The gas behind the shock will 
stay bound and an explosion is not possible. 
The critical accretion rate for this to happen depends on the
neutrinospheric parameters ($R_{\nu}$, $T_{\nu}$ and $L_{\nu}$), to some
degree also on the structural polytropic index $\gamma$. For the parameters
and neutrino luminosities considered in the present discussion this
value is found to be around 4~M$_{\odot}\,$s$^{-1}$.

Even for somewhat smaller absolute values of the accretion rate and a 
positive total energy in the gain layer explosions might not occur.
The question, however, whether an outward running supernova shock
will reach the stellar surface and what amount of matter it is able to eject,
requires a global treatment of the problem, including the possible 
energy release by nuclear burning and recombination of nucleons, and including
the energy which will be spent on lifting the stellar mantle and envelope
in the gravitational field of the star. This is far beyond the
limits of the current treatment, which focusses on a discussion of the
conditions that are necessary for reviving a stalled shock and for pushing it 
out to a radius of $\ga 1000$~km by the neutrino heating mechanism.

\subsubsection{Thermodynamic conditions}

\begin{figure*}[t]
\centerline{
  \epsfxsize=0.48\hsize\epsffile{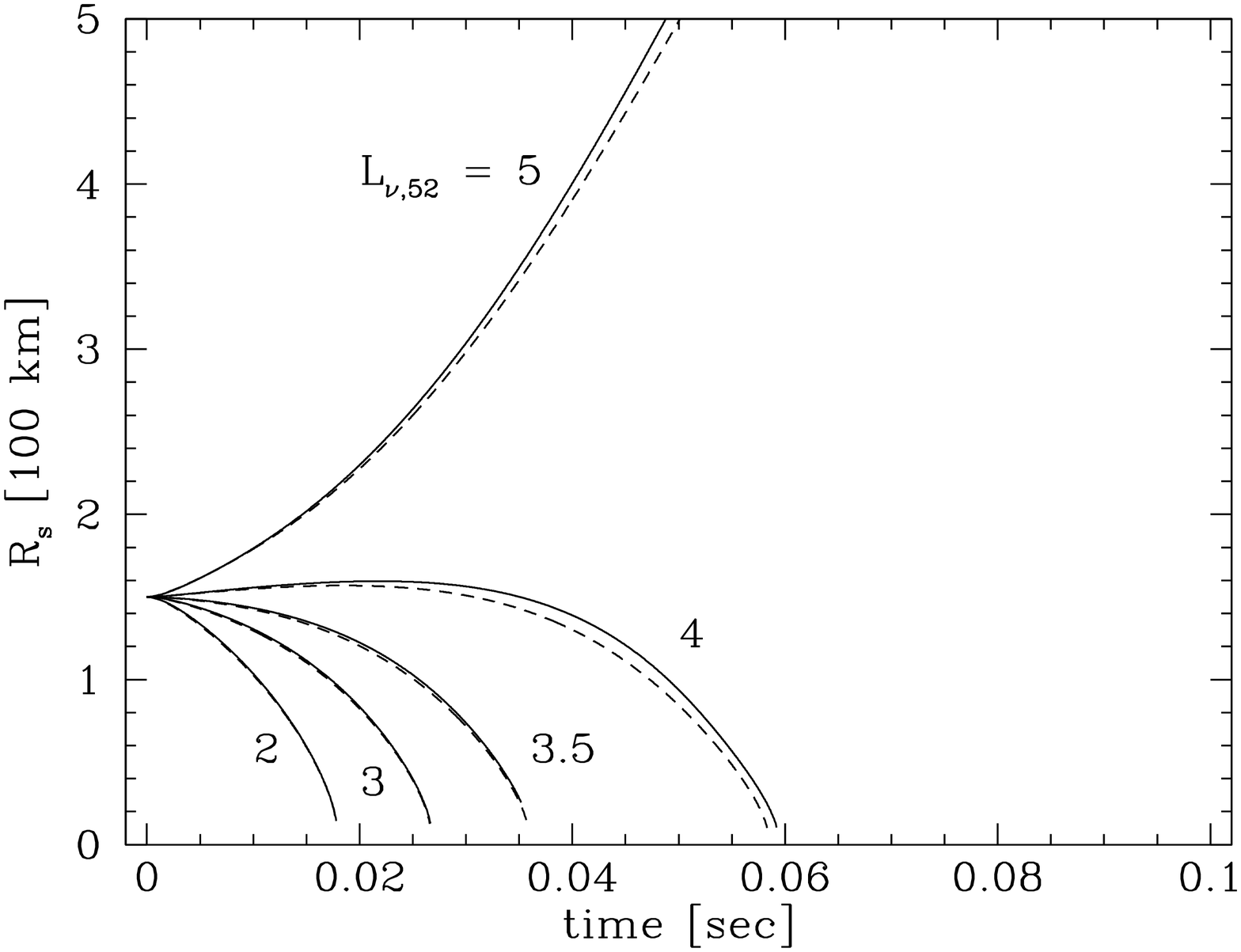}\hfill
  \hskip 0.5truecm
  \epsfxsize=0.48\hsize\epsffile{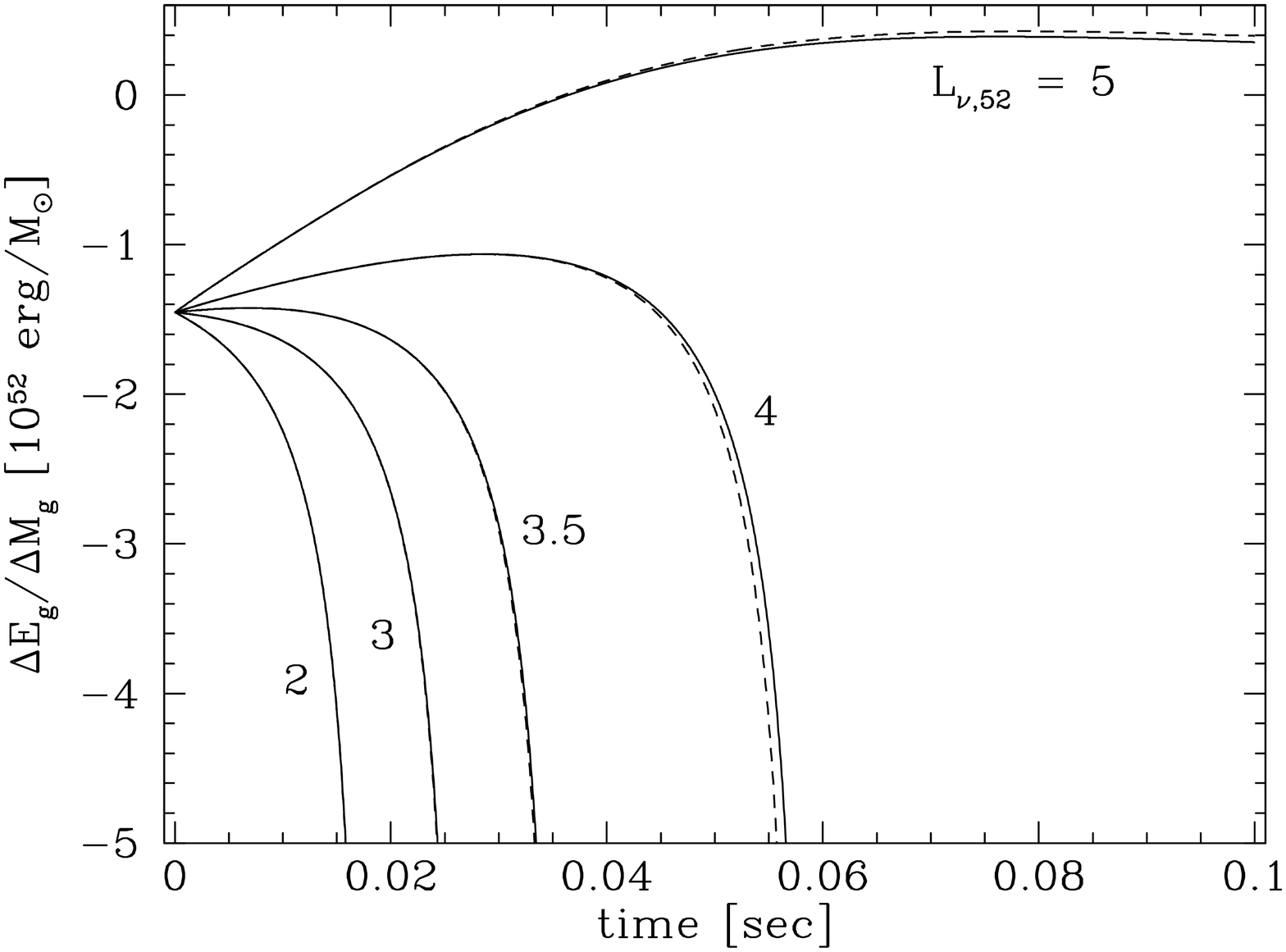}\hfill
  }
\caption{Same as Fig.~\ref{fig:6}, but for a structural polytropic index
$\gamma = 1.45$ in the gain layer. 
\label{fig:7}  
}
\end{figure*}

The entropy per nucleon in the gain layer, where relativistic electrons, 
positrons and photons as well as nonrelativistic nucleons and nuclei
contribute to the pressure, is given by
\begin{equation}
s\, =\, {\varepsilon + P\over kT\,(\rho/m_{\mathrm{u}})} - \sum_i \eta_i Y_i \, ,
\label{eq:112a}
\end{equation}
where the sum runs over all kinds of particles $i$. When nuclei are fully
dissociated, in which case $Y_p = Y_e$ and $Y_n = 1-Y_e$, this sum can 
be written as 
\begin{eqnarray}
\sum_i \eta_i Y_i &=& Y_e(\eta_e + \eta_p - \eta_n) + \eta_n \nonumber \\
&\approx& Y_e\eta_e + Y_e\ln\!\rund{{Y_e\over 1-Y_e}} \nonumber\\
& &\phantom{\approx \eta_e} 
+ \ln\!\rund{1.27\times 10^{-3}\,{(1-Y_e)\rho_9\over (kT)^{3/2}}} \,.
\label{eq:112b}
\end{eqnarray}
Since typically $Y_e\sim 0.2$--0.3 and $\eta_e/\pi \la 1$ around the gain
radius, the combined terms scaling with $Y_e$ are negligibly small.
Using $P = (\Gamma-1)\varepsilon$, 
Eq.~(\ref{eq:112a}) can be evaluated for the models plotted in 
Figs.~\ref{fig:6} and \ref{fig:7}. Near the gain radius
characteristic initial values of the entropy
per nucleon are found to be between 10 and 14, with a contribution from 
relativistic degrees of freedom of $s_{\mathrm{r}}\sim 2$--3.
This is in good agreement with results from detailed hydrodynamical 
models (see, e.g., Rampp \& Janka 2000).
In case of successful shock expansion, the total entropy per nucleon 
increases to values between 25 and 30 towards the end of the computed
evolution. 

Clearly, neither the entropy nor the pressure are 
dominated by radiation and leptons, but baryons play an important role,
at least at the beginning of shock expansion. Nevertheless, the description in 
Sect.~\ref{sec:atmosphere} of the gain layer as being a ``radiation-dominated'' 
region remains justified, although in a generalized sense. While in the region
around the neutrinosphere baryons (and possibly degenerate electrons) yield the 
major contribution to the pressure and internal energy, the importance of 
electron-positron pairs and photons increases at lower densities.
In Sect.~\ref{sec:hystat2} [Eqs.~(\ref{eq:56})--(\ref{eq:56b})] it was argued
that for $r > R_{\mathrm{eos}}$ both the pressure contributions from
relativistic and non-relativistic particles can be written as $P \propto T^4$,
provided that the electron fraction $Y_e$ and the electron degeneracy parameter
$\eta_e$ do not vary strongly. In the gain layer this is fulfilled, because 
the electron degeneracy
is typically small, i.e., $\eta_e \la \pi$, and electron-positron pairs are
abundant (see the detailed discussion by Bethe 1993, 1996b). Indeed, the
hydrodynamical simulation of Rampp \& Janka (2000) shows that $\eta_e$ 
in the gain layer changes only between about 1.5 and 3 during the interesting
phase of the post-bounce evolution. 

Despite of the considerable contribution to the pressure which is provided
by nonrelativistic baryons, also the use of 
$\Gamma = (\partial \ln P/\partial \ln\rho)_s = {4\over 3}$ for the 
adiabatic index of the equation of state in the postshock region is justified, 
although the calculations in this paper are not 
constrained to this specific choice. At the conditions present between the
gain radius and the shock (density between a few $10^8$~g$\,$cm$^{-3}$ and
several $10^{9}$~g$\,$cm$^{-3}$ and temperatures between roughly $\la 1$~MeV and 
$\ga 2$~MeV), a finite mass fraction of $\alpha$ particles is still present
($X_{\alpha}\la 0.5$). 
The disintegration of these $\alpha$'s at nuclear statistical equilibrium
around 1--1.5~MeV and the growing importance of $e^+e^-$ pairs and photons for 
higher temperatures produce $\Gamma$ values between 1.3 and 1.4. This can,
for example, be verified by an inspection of the equation of state of
Lattimer \& Swesty (1991). 

\subsubsection{Steady-state conditions}

Steady-state conditions are realized when $\dot R_{\mathrm{s}} = 
\dot R_{\mathrm{g}} = 0$, which in general requires that 
${\mathrm{d}}(\Delta M_{\mathrm{g}})/{\mathrm{d}}t = 
{\mathrm{d}}(\Delta E_{\mathrm{g}})/{\mathrm{d}}t = 0$. From Eq.~(\ref{eq:109})
one gets $\dot M' = \dot M$, which yields 
\begin{equation}
L_{\mathrm{acc}}\,=\,\dot M\,(l_{\nu} + q_{\mathrm{d}}) +
{\cal H}-{\cal C} \ ,
\label{eq:112}
\end{equation}
when Eqs.~(\ref{eq:89}) and (\ref{eq:89a}) and the 
definition $l = (e+P)/\rho - G\widetilde{M}/r$ are used (note that 
${1\over 2}\rho v^2 \ll \varepsilon$ at the neutrinosphere). 
Neutrino heating and cooling in the gain layer scale with
$L_{\nu}(R_{\mathrm{g}})$ and thus $L_{\nu}(R_{\nu})$ [Eq.~(\ref{eq:72})],
and depend on $R_{\mathrm{s}}$, directly as well as via $R_{\mathrm{g}}$. 
Also $L_{\mathrm{acc}}$ 
depends on the neutrinospheric luminosity $L_{\nu}$ [Eq.~(\ref{eq:73})].
Equation~(\ref{eq:108}), on the other hand, yields
\begin{equation}
\dot M\,=\,{{\cal H}-{\cal C} \over l_{\mathrm{s}}-l_{\mathrm{g}}} 
\label{eq:113}
\end{equation}
for $l_{\mathrm{s}} < l_{\mathrm{g}}$, which is fulfilled if $\Gamma 
< \gamma$. [$\Gamma = \gamma$ requires ${\cal H}-{\cal C} = 0$ and 
therefore $R_{\mathrm{g}} = {\mathrm{min}}(R_{\mathrm{s}}, R_{\alpha})$.] 
From the combined Eqs.~(\ref{eq:112}) and (\ref{eq:113}) the values of 
the shock radius
$R_{\mathrm{s}}$ and the neutrinospheric luminosity $L_{\nu}$ can be 
determined which correspond to steady-state conditions for a given value of 
$\dot M$. The solution for $L_{\nu}$ lies on the critical O$_{\mathrm{M}}$
line displayed in Figs.~\ref{fig:4} and \ref{fig:5}, where $\dot M'$ and   
$\dot M$ are equal. The core luminosity has to satisfy this constraint,
because the neutrinospheric conditions and the temperature in the cooling 
layer above the neutrinosphere are assumed to be regulated by the 
interaction with the high neutrino 
fluxes from the hot, neutrino-opaque neutron star. This inner boundary
condition differs from the one used for discussing ordinary
steady-state accretion onto neutron stars (Chevalier 1989).
There the temperature of the optically thin medium at the base of the 
atmosphere is assumed to adopt a value which ensures that photon or neutrino 
losses carry away the binding energy of the matter which is accreted at a 
given constant rate. This requirement then yields a condition for calculating
the steady-state position of the accretion shock.

\subsection{Convective energy transport}
\label{sec:convection}

The simplified analytic model described in this paper can certainly
not account for the detailed effects associated with convective
overturn in the neutrino-heated layer between gain radius and shock.
This overturn is an intrinsically multi-dimensional
phenomenon where low-entropy downflows and hot, rising bubbles of
neutrino-heated gas coexist in the same region of the star.
Therefore the mixing achieved by the gas motions is not complete
even on a macroscopic scale.
Nevertheless, some consequences and fundamental effects associated
with the existence of convective energy transport in the gain region
can be figured out. 

The described analytic model distinguishes between the adiabatic index
$\Gamma$ of the equation of state in the gain layer, and the structural 
polytropic index $\gamma$.
The gain layer is subject to non-adiabatic
changes, because energy deposition by neutrino heating takes place.
Also, the gain layer is not necessarily isentropic.
Using the developed framework of equations with $P = K\rho^\gamma$,
$\gamma = \Gamma = {\mathrm{const}}$, and $K = {\mathrm{const}}$ 
for the equation of state in the gain region --- which is the default
setting for the analyses in Sect.~\ref{sec:application} ---,
Eq.~(\ref{eq:105a}) yields the same value for the 
total specific energies at $R_{\mathrm{g}}$ and $R_{\mathrm{s}}$.
For a gas with adiabatic index $\Gamma$ this
also means that isentropic conditions are realized. This, therefore,
corresponds to the case where 
convection is very efficient in carrying energy from the gain 
radius, close to which neutrino heating is strongest, to directly behind
the shock. Chosing instead $\gamma > \Gamma$ yields
$l_{\mathrm{g}} > l_{\mathrm{s}}$, a result which is more
characteristic of the situation without convection. Here the energy
deposition by neutrinos establishes negative gradients of the 
entropy and specific energy between gain radius and shock.

Repeating the derivations of 
Sects.~\ref{sec:atmosphere}--\ref{sec:application}
with $\gamma > {4\over 3}$, reveals, on the one hand, that the gain radius
$R_{\mathrm{g}}$ is smaller and therefore the optical depth and the net 
heating in the gain region, ${\cal H}-{\cal C}$, are somewhat
{\em larger} than for the ``standard'' 
case of $\gamma = {4\over 3}$ (because the hydrostatic density and 
temperature profiles are flatter behind the shock).
On the other hand, however, a less efficient
energy transport from the gain radius to the shock has a severe
disadvantage: The gas which is advected inward through $R_{\mathrm{g}}$,
carries away a large fraction of the energy absorbed from neutrinos 
before. In Eq.~(\ref{eq:108}) the term $\dot M' l_{\mathrm{g}}$ yields
a smaller positive or even a negative contribution when $\dot M' < 0$ and
$l_{\mathrm{g}}$ is negative or positive, respectively.
This reduces the net effect of neutrino heating and is harmful
for shock expansion and acceleration.

A comparison of Figs.~\ref{fig:6} and \ref{fig:7} demonstrates the
differences. The time-dependent solutions for shock radius and specific
energy in the gain layer were obtained with $\Gamma = {4\over 3}$ in
both cases, but in Fig.~\ref{fig:7} $\gamma = 1.45$ was chosen 
instead of $\gamma = {4\over 3}$. For 
$\gamma > \Gamma$ the shock expansion is weaker and the specific energy
in the gain layer stays lower. The effect is particularly obvious for
the core neutrino luminosity of $L_{\nu} = 4\times 10^{52}$~erg$\,$s$^{-1}$.
Figure~\ref{fig:6} shows a marginal success for this case, whereas 
in Fig.~\ref{fig:7} the shock expansion fails.

These findings are confirmed by an inspection of the
spherically symmetric simulation of the collapse and post-bounce
evolution of a 15 M$_{\odot}$ progenitor star published recently by
Rampp \& Janka (2000). After an expansion to more than 350 km, the
shock in this model finally recedes to a much smaller radius and 
fails to produce 
an explosion. The shock recession is caused (or accompanied) by a
rapid decrease of the mass in the gain region, because more matter
is flowing through the gain radius than is resupplied by accretion
through the shock. In the hydrodynamical simulation one finds that
$\Delta E_{\mathrm{g}}$ also decreases during this phase, an effect
which should not occur if $l_{\mathrm{g}} = l_{\mathrm{s}} < 0$ [compare
Eq.~(\ref{eq:108})].

This discussion emphasizes the importance of convective energy transport 
between the gain radius and the shock. Postshock convection reduces
the mass loss as well as the energy
loss from the gain region, which are associated with the continuous
inward advection of neutrino-heated gas during the phase of shock
revival. Also an increase of the core luminosity can diminish the
accretion of gas into the neutron star by suppressing
the net neutrino emission from the cooling layer.
Both effects have been demonstrated in numerical simulations to be 
helpful for an explosion. 


\subsection{Limits of the toy model}
\label{sec:limits}

Employing idealized and in many respects simplifying assumptions, a toy
model was developed here on grounds of an approximate solution of the
hydrodynamic equations, Eqs.~(\ref{eq:2})--(\ref{eq:4}).
By reducing the complexity, hopefully without sacrificing 
fundamental properties, the model is intended to help discussing 
the principles and the essence of the neutrino-driven mechanism.
It is, however, not meant to compete with detailed hydrodynamical
simulations, where usually a lot more refinements concerning the 
description of the stellar fluid and of the neutrino transport are
included.
 
The stellar structure outside of the forming neutron star at the center
is considered to consist of three layers: The cooling layer, from where
neutrino loss extracts energy, is bounded by the neutrinosphere on the
one side and the gain radius on the other; the heating layer extends between
gain radius and shock and receives net energy deposition by neutrinos;
in the infall region exterior to the shock, matter of the
progenitor star moves inward at a significant fraction of the free-fall 
velocity. The evolution of the shock depends on the conditions in the
gain layer. Assuming the radial structure of this layer to be given
by hydrostatic equilibrium one can discuss the behavior of the shock in
response to integral properties of the heating region.
The total mass and energy of the gain 
layer change due to inflow and outflow of gas, neutrino heating, and a possible 
shift of the boundaries, and are therefore sensitive to the rate of mass
accretion by the shock on the one hand, and the irradiation by the 
neutrinospheric luminosities on the other.

\subsubsection{Temperature in the cooling layer}

The present work concentrates on a discussion of the phase of shock revival
due to neutrino energy deposition, where the infall of the stellar 
gas is inverted to outflow. This implies that the gas flow through the
gain layer cannot be stationary, i.e., the mass infall rate into the shock,
$\dot M$, is in general different from the mass accretion rate $\dot M'$
by the nascent neutron star. Here $\dot M'$ in dependence of the physical
conditions near the neutron star surface is estimated by assuming that 
the temperature at 
and just outside the neutrinosphere is governed by the interaction of
the stellar medium with the neutrinos streaming out from deeper layers.
This temperature determines the energy loss from the cooling layer around
the neutron star, the efficiency of which then drives
the mass exchange (inflow or outflow) with the gain layer farther out.

This picture is certainly a simplification of the real situation. 
For example, when outward motion of the postshock gas sets in, the 
advection of gas through the gain radius may be quenched as found in 
spherically symmetric simulations. In the described model this effect could be
reproduced if the temperature in the cooling layer would drop. The current set 
of equations, however, does not allow one to calculate this effect because
it does not include how the temperature in the cooling layer depends on the
expansion or compression of the neutron star atmosphere. On the other hand,
in the three-dimensional situation downflows and rising bubbles can coexist when 
convective overturn is present in the gain layer, as suggested by two-dimensional 
hydrodynamical simulations (Herant et
al.\ 1994, Burrows et al.\ 1995, Janka \& M\"uller 1996). In this case
accretion does not need to stop even when the shock is accelerated outward
(Bethe 1993, 1995; however, Burrows et al.\ 1995 see a decrease of
the neutrino luminosity associated with a reduced accretion rate when the
explosion sets in). Convective overturn and thus accretion might in fact
continue until the shock has reached a radius above 1000~km 
(Bethe 1997). It is not easy to estimate the fraction of the gas which 
stays in the gain layer relative to the part which is advected inward to the
neutron star. The ansatz described here may be considered as a
crude attempt to do so.

\subsubsection{Hydrostatic equilibrium}

The stellar structure in the gain layer was calculated as a solution of
the equation of hydrostatic equilibrium. The latter is derived from 
Eq.~(\ref{eq:3}) combined with Eq.~(\ref{eq:2}). When the 
velocity-dependent terms can be neglected, this yields
\begin{equation}
{\partial v\over \partial t} + v\,{\partial v\over \partial r}
\,=\,{{\mathrm{d}} v\over {\mathrm{d}}t}\,=\, 
-\,{1\over\rho}\,{\partial P\over \partial r} -
{\partial \Phi\over \partial r} \,=\, 0 \,.
\label{eq:114}
\end{equation}
Hydrostatic equilibrium holds in the regime where the fluid velocity $v$
is much smaller than the local sound speed $c_{\mathrm{s}}$.
This is well fulfilled when the postshock gas settles inward, and is
also well fulfilled when it starts moving outward behind an
accelerating shock. Of course, omitting
the $v(\partial v/\partial r) = \partial(v^2/2)/\partial r$ 
term from Eq.~(\ref{eq:114}) implies that
the discussed toy model will not reproduce the solutions for a stationary
wind, which has a critical point where $v = c_{\mathrm{s}}$. This
is not a handicap during the shock revival phase and the onset of shock 
expansion, because the gas behind the (slowly propagating) shock front moves
(relative to the shock) with subsonic velocities. It means, however, that
the treatment is not sufficiently general to follow the explosion to 
large radii and high shock velocities, which is the evolutionary 
phase when the expanding medium
around the nascent neutron star forms a neutrino-driven wind.
When the shock is far out and the gain layer has expanded by a large amount,
the assumption of hydrostatic equilibrium therefore becomes inadequate and the 
corresponding structure of the gain layer is not sufficiently accurate any 
longer. Unless the gas velocity approaches $c_{\mathrm{s}}$,
the modification of the stellar density and pressure profiles due to
velocity effects is small. In fact, when the time-dependent calculations
of Sects.~\ref{sec:tsolutions} and \ref{sec:convection} were terminated
at $t = 0.1$ seconds, the
specific kinetic energy of the gas immediately behind the shock was
only a minor fraction of the specific internal energy, at most about
20\%. The integral kinetic energy of the gain layer was
even smaller compared to the total internal energy.

Considering hydrostatic conditions means that the fluid velocity is
assumed not to be relevant for the structure of the neutron star atmosphere. 
In fact, an accretion or outflow velocity field in the gain layer 
was not derived (and was not of direct relevance for the discussion), 
although the toy model employs the mass accretion rates $\dot M$ and 
$\dot M'$. The fluid velocity immediately above the shock is given by the
infall velocity of the gas, and at the gain radius it must be equal to
$v_{\mathrm{g}} = \dot M'/(4\pi R_{\mathrm{g}}^2\rho_{\mathrm{g}})$.
Different rates for the mass accretion into the shock and the mass
flow through the gain radius imply that the mass of the gain layer
can grow or drop due to active mass inflow or outflow (in addition to
the motion of the shock and of the gain radius as the outer and inner
boundaries, respectively, of the considered stellar shell). Therefore
the conditions in the gain layer are non-stationary, i.e.,
$\partial\rho/\partial t = 0$ cannot be true in general. 

Deriving a
velocity profile for the gain layer requires solving Eq.~(\ref{eq:2})
with non-vanishing time derivative of the density and the lower boundary
condition being given by $v(R_{\mathrm{g}}) = v_{\mathrm{g}}$. Doing so, the 
velocity jump at the shock is also influenced by the physical conditions
around the gain radius. This is analogue to the postshock density, which
is sensitive to the mass accumulated in the gain layer, and is similar to the 
postshock pressure, which varies with the integral value of the energy
deposited by neutrinos 
in the gain layer. Mass and energy loss or gain thus affect the 
whole heating layer simultaneously. 
Assuming hydrostatic equilibrium implies that the physical state
of the gas behind the shock front is coupled to the conditions at the gain 
radius, because the sound crossing timescale is considered to be small
compared with all other relevant timescales of the problem. Therefore 
the Rankine-Hugoniot relations for the density jump and the velocity jump at
the shock front cannot be satisfied exactly, which reflects the 
approximative nature of the hydrostatic structure. The violation
of the Rankine-Hugoniot conditions (for specific values of the 
EoS parameters), however, is usually small and the
overall properties of the calculated solutions should be close to 
the true ones, in particular at some distance behind the shock front.

\subsubsection{Equation of state and convection}

The equation of hydrostatic equilibrium in the gain layer was solved 
assuming that gas pressure, density and temperature are related by
$P = K\rho^\gamma = A(kT)^4$ [Eq.~\ref{eq:56})] with $\gamma$, $K$ and
$A$ being constants. As discussed in Sects.~\ref{sec:hystat2} and 
\ref{sec:tsolutions}, 
this is reasonably well fulfilled in the gain layer, where the stellar
gas consists of a mixture of relativistic electrons, positrons and photons 
and nonrelativistic nucleons and nuclei. On the one hand the electron degeneracy
is low ($\eta_e \la \pi$) and both $\eta_e$ and
the electron fraction $Y_e$ do not vary strongly (the variation of $Y_e$
is at most a factor of about two, roughly between 0.25 at the gain radius
and 0.5 at the shock). For these reasons fermion captures on nucleons
were ignored concerning their effect on the $Y_e$ profile in the gain layer.
On the other hand,
heavy nuclei are mostly disintegrated into free nucleons with some admixture 
of $\alpha$ particles. Therefore also the sum of the nuclear number fractions 
$\sum_i Y_i$ is roughly constant; in fact, $\sum_i Y_i \approx 1$ is a good 
approximation. Because there is a finite mass fraction of $\alpha$ particles,
$X_{\alpha} = 4Y_{\alpha} \la 0.5$, nuclear statistical equilibrium yields
an adiabatic index of $\Gamma = 1+P/\varepsilon\approx {4\over 3}$ for the 
equation of state at typical postshock conditions, although nucleons are
responsible for a large fraction of the pressure. None of these assumptions,
however, is rigorously fulfilled in the medium of the gain layer at all
radii and at all times. But making use of these assumptions simplifies 
the discussion considerably because the equation of state can be treated 
analytically. Since the overall properties of the stellar gas are accounted
for, it is very unlikely that the conclusions on the qualitative level of this
paper are affected when more refinements and complications are added. 
This might change details, but should not modify the essence of the toy model.

The energy input to the gain layer by neutrino heating is accounted
for in the model. This energy gain from neutrinos means that the
changes in a fluid element are non-adiabatic. Therefore the structural 
polytropic index $\gamma$ can be different from the adiabatic index $\Gamma$
of the EoS.
Varying $\gamma$ allows one to mimic additional processes which might affect
the evolution and behavior of the gain layer. 
Chosing $\gamma = \Gamma = {4\over 3}$ implies 
that the gain layer is considered to be isentropic, i.e., the energy deposited
by neutrinos is assumed to be efficiently (and instantaneously)
redistributed such that the entropy
is roughly equal everywhere and $l_{\mathrm{s}} = l_{\mathrm{g}}$ holds
[Eq.~(\ref{eq:105a})]. Since neutrino heating is strongest near the gain radius,
this means that energy has to be transported from smaller radii to positions
closer to the shock. Such an effect is realized by the strong postshock
convection seen in multi-dimensional hydrodynamic simulations
(e.g., Herant et al.\ 1994, Burrows et al.\ 1995, Janka \& M\"uller 1996).
Using $\gamma > \Gamma$ a situation is described where more of the 
deposited energy stays near 
the gain radius, corresponding to less efficient energy transport by convection.
The toy model confirms that this has a negative influence on the possibility of
shock expansion.


\subsubsection{Neutrino processes}

Processes different from $\nu_e$ and $\bar\nu_e$ absorption and emission by 
nucleons were not taken into account for the neutrino heating and cooling 
above the neutrinosphere. Both neutrino-electron/positron scattering and 
neutrino-pair annihilation have much smaller reaction cross sections than
the baryonic processes and are less efficient in transferring energy to the 
stellar medium. For this reason they do not play a crucial role for the
explosion mechanism 
(Bethe \& Wilson 1985; Bethe 1990, 1993, 1995, 1997; Cooperstein et al.\ 1987).

Effects due to muon and tau neutrinos and antineutrinos 
($\nu_x$) were completely ignored 
in the discussions of this paper. Because muons and tau leptons cannot be produced
in the low-density medium above the neutrinosphere, $\nu_{\mu}$ and $\nu_{\tau}$
do not interact with nucleons via charged-current reactions and therefore couple 
to the gas less strongly than electron neutrinos and antineutrinos. Energy exchange
by neutral-current scatterings off nucleons contributes in shaping 
their emission spectra near the neutrinosphere (Janka et al.\ 1996, Burrows et 
al.\ 2000) and might also be relevant for the heating in the gain layer.
Although the recoil energy transfer per scattering is reduced by a factor
$\epsilon/(mc^2)$ relative to the absorption of neutrinos with energy $\epsilon$
($m$ is the nucleon mass), the cross sections of both processes are similar
and all flavors of neutrinos and antineutrinos participate in the neutral-current
reactions with neutrons as well as with protons.
Using Eq.~(\ref{eq:10}) for the nucleon scattering opacity and the mean energy
exchange per reaction as given by Tubbs (1979), one can estimate the importance
of nucleon scattering for the energy transfer to the medium relative
to $\nu_e$ and $\bar\nu_e$ absorption as: 
\begin{eqnarray}
{Q_{\nu,{\mathrm{sc}}}^+\over Q_{\nu,{\mathrm{abs}}}^+} &\sim &
{{1+3\alpha^2\over 16}\,5\,(Y_n\!+\!Y_p)
\rund{\ave{\epsilon_{\nu_x}^3} - 6kT\ave{\epsilon_{\nu_x}^2}} \over
{1+3\alpha^2\over 4}\,\ave{\epsilon_{\nu_e}^2}
(Y_n\!+\!2Y_p)\, mc^2 } \nonumber \\ 
&\sim & {15\over 2}\,{(kT_{\nu_x})^2\over (kT_{\nu_e})^2}\,
{kT_{\nu_x}\!-\!kT\over mc^2}\,\sim\, 30\,\,{kT_{\nu_x}\!-\!kT\over mc^2} \,
\label{eq:115}
\end{eqnarray}
where $Q_{\nu,{\mathrm{abs}}}^+$ was taken from Eq.~(\ref{eq:28}) and
$T$ is the gas temperature. $T_{\nu_e}$ and $T_{\nu_x}$ are the 
spectral temperatures of electron neutrinos and heavy-lepton antineutrinos,
respectively, for which $T_{\nu_e}^2 \approx {1\over 2}
T_{\bar\nu_e}^2\approx {1\over 4}T_{\nu_x}^2$ 
was assumed again (compare Sect.~\ref{sec:basic}).
Moreover, the estimate was obtained by using
$L_{\nu_e}\approx L_{\bar\nu_e}\approx L_{\nu_x}$ and $Y_n + 2Y_p\approx 1$. 
The spectral average of the third power of the neutrino energy, 
$\ave{\epsilon_{\nu_x}^3}$, was defined in analogy to Eq.~(\ref{eq:25}).
Since the spectral temperatures and therefore the scattering cross sections
of $\nu_e$ and $\bar\nu_e$ are smaller than those of $\nu_x$,
electron neutrinos plus antineutrinos
were given roughly the same weight as one of the heavy-lepton neutrinos.
For typical values $kT_{\nu_x} \sim 8$~MeV and $kT\sim 2$~MeV one 
therefore derives a relative contribution of scattering processes to
the neutrino heating in the gain layer of about 20\%. 

Apart from this moderate amplification of the heating, 
muon and tau neutrinos have other effects on the shock propagation during the
post-bounce evolution of a supernova. Within the first tens of milliseconds after
shock formation, muon and tau neutrino pairs are produced by $e^\pm$
annihilation in the heated matter immediately behind the shock. In addition to
the disintegration of nuclei and the emission of $\nu_e$ and $\bar\nu_e$,
this extracts energy from the shock-heated layers and weakens the prompt
bounce shock. Somewhat later, between several ten milliseconds and a few hundred
milliseconds after bounce, most of the muon and tau neutrinos come from the
hot mantle layer of accreted material below the neutrinosphere of the
forming neutron star. Since $\nu_{\mu}$ and
$\nu_{\tau}$ pairs now carry away energy which otherwise would be radiated in
electron neutrinos and antineutrinos and would thus be more efficient for
the heating behind the shock, this will have a negative effect on the possibility
of shock rejuvenation. During the following phase of the evolution, when the 
deleptonization of the neutron star advances to deeper layers
and the neutron star enters the Kelvin-Helmholtz cooling stage (Burrows \&
Lattimer 1986), muon and tau neutrinos are mostly produced at higher densities.
Being less strongly coupled to the nuclear medium, they diffuse to the surface 
more rapidly than $\nu_e$ and $\bar\nu_e$. This helps keeping the neutrinospheric
layer hot, where electron neutrinos and antineutrinos take over a larger part
of the energy transport. During this late phase of the evolution, $\nu_{\mu}$ and
$\nu_{\tau}$ might thus even support higher $\nu_e$ and $\bar\nu_e$ fluxes.

\section{Summary and conclusions}
\label{sec:summary}

In this paper an analytic approach was presented which allows one
to discuss the conditions for the revival of a stalled supernova
shock by neutrino heating. The treatment is time-dependent in the
sense that the gas flow is not assumed to be steady and the model
can be used to calculate the shock radius, shock velocity, 
and the properties of the gain layer as functions of time. 

\subsection{Components of the toy model}

The ``atmosphere'' of the collapsed stellar core outside of the neutrinosphere 
is considered to consist of three distinct layers (Sect.~\ref{sec:physics}). 
Between neutrinosphere and gain radius there is a cooling region where neutrino
emission extracts energy. In a heating layer between gain radius and 
shock, $\nu_e$ and $\bar\nu_e$ absorptions on nucleons dominate the
inverse capture reactions of electrons and positrons and deposit energy.
Finally, there is an infall region above the shock, where the gas of the
progenitor star is accelerated to nearly the free-fall velocity. These 
different layers are in contact and exchange mass and energy.

The radial structure of the cooling and heating layers is assumed to be 
described by the conditions of hydrostatic equilibrium, which requires
that the sound travel timescale is smaller than the other relevant timescales 
of the problem. This assumption is reasonably well fulfilled during
the phase when the shock is near stagnation or just starts to gain momentum.
For such conditions the layer behaves like one unit and reacts to changes in 
an infinitesimally short time. In combination with a simple representation
of the equation of state, hydrostatic equilibrium allows one to calculate
the density and pressure profiles analytically (Sect.~\ref{sec:atmosphere}). 
The radius and velocity of the shock then depend on integral properties
of the gain layer, i.e., its total mass and energy.

Changes of the mass and energy integrals are caused by the motion 
of the boundaries or by active mass flow into or out of the gain layer.
In Sect.~\ref{sec:meconservation} conservation
equations for these global quantities were
derived by integrating the equations of hydrodynamics, including the 
terms with time derivatives, over the volume of the gain layer. 
Following these integral quantities
it was then possible to compute the shock position and shock velocity as
well as other important quantities, e.g., the location of the
gain radius, as functions of time. Assuming hydrostatic 
equilibrium thus reduces the mathematical problem to an integration
of a set of ordinary differential equations with time as the independent
variable. Discussing the destiny of the supernova shock therefore means
solving an initial value problem. This expresses the fact that the 
shock evolution depends on the initial conditions, for example,
on the shock stagnation radius and the initial energy in the gain layer,
and is controlled by the cumulative effects of neutrino energy deposition 
and mass accumulation in the gain layer.

In general the gas falling through the shock will not move as a stationary
flow between shock and neutrinosphere: The rate at which mass is 
advected through the gain radius is usually different from the mass
accretion rate by the shock. Steady-state accretion or mass loss are
special cases, which should be limits of the more general situation.

It is not easy to calculate the fraction of the accreted gas which
stays in the gain layer. The neutron star can ``swallow'' matter only
at a finite rate, depending on the efficiency with which the gas
gets rid of the excess energy that prevents its integration into
the neutron star surface. This efficiency is a sensitive function of
the conditions in the cooling layer above the neutrinosphere. Since the
hot, nascent neutron star below the neutrinosphere is a source of 
intense neutrino radiation and these neutrinos interact still frequently
in the cooling region, the temperature there should be close to the
neutrinospheric value. With this assumption and with known radius and
density profile it is possible to calculate the neutrino energy loss from
the cooling layer. This then allows one to derive a rough estimate for
the rate at which gas can be advected through the neutrinosphere into 
the neutron star (Sect.~\ref{sec:massacc}). 

Up to this stage the discussion does not require a detailed solution for
the velocity field of the flow. Since hydrostatic conditions are assumed 
to hold, in which case the kinetic energy is small compared to internal
and gravitational energies, and the time evolution can be discussed
by considering integral quantities, it is sufficient to know the rates
at which mass enters or leaves the gain layer at both boundaries.

The model described here is based on a number of simplifications and
approximations. With the analytic representation of the
equation of state developed in Sect.~\ref{sec:hystat2}, there is no
need to monitor the radial profile and time evolution of the electron 
fraction. In addition to assuming hydrostatic equilibrium this, of course,
limits the accuracy of the radial structure derived for the gain layer.
Other shortcomings are the treatment of the neutron star as a 
point mass, i.e, the gravity of the atmosphere above the neutrinosphere
is neglected, general relativistic effects are ignored, the energy 
release by nucleon recombination in the postshock medium at 
$kT \la 1$--2~MeV is not included, and additional neutrino
heating and cooling by neutrino-electron scattering and neutrino-pair
processes are not considered. Although it may be desirable to include
these effects for a more detailed solution, it seems very unlikely that
more refinements will change the essence of the discussion.

In calculating the neutrino energy deposition in the gain layer the
neutrinospheric luminosity as well as the neutrino emission from the cooling
layer, which is associated with the accretion of gas onto the neutron star, 
are taken into account. The most problematic and probably most serious
weakness of the presented toy model, however, is the overly simplified 
description of the conditions in the cooling layer, which are essential
for estimating both the mass accretion rate and 
the accretion luminosity of the neutron star. The temperature
of the medium around the neutrinosphere is certainly not only determined
by the interaction with the neutrino flow from the neutrinosphere, but
also depends on the processes in contact with the gain layer. This is
currently not included in the toy model. 

Despite of these simplifications the toy model yields interesting insights 
into the interdependence of effects and processes which determine the 
post-bounce evolution of the supernova shock. Thus it may help one
understanding the results of the much more complex hydrodynamic simulations.

\subsection{Results and conclusions}

In particular, the discussion of this paper allows one to draw the following
conclusions:
%
\begin{enumerate}
\item
A criterion for shock revival could be derived in Sects.~\ref{sec:shockevol} 
and \ref{sec:criterion}. It defines the conditions
for which the radius $R_{\mathrm{s}}$ and the velocity 
$U_{\mathrm{s}} = \dot R_{\mathrm{s}}$ of the supernova shock can grow 
simultaneously. This criterion, applied to a stalled shock
($U_{\mathrm{s}} \sim 0$), states that expansion ($\dot R_{\mathrm{s}} > 0$) 
and acceleration ($\dot U_{\mathrm{s}} > 0$) will occur at the same time
when both the mass and the energy in the gain layer increase, i.e., when
${\mathrm{d}}(\Delta M_{\mathrm{g}})/{\mathrm{d}}t > 0$ 
and ${\mathrm{d}}(\Delta E_{\mathrm{g}})/{\mathrm{d}}t > 0$.
\item
For fixed neutrinospheric radius and temperature and given neutron star 
mass, the destiny of a shock at radius $R_{\mathrm{s}}$ depends on the rate
at which the shock accretes matter, $\dot M$, and on the $\nu_e$ plus $\bar\nu_e$ 
luminosity $L_{\nu}$ coming from the neutrinosphere. In the $\dot M$-$L_{\nu}$ 
plane the two inequality relations of the shock revival criterion define two 
critical lines, an upper one and a lower one, 
which enclose the region where favorable conditions for shock expansion and
acceleration occur. Neutrino heating is strong enough there to ensure 
${\mathrm{d}}(\Delta E_{\mathrm{g}})/{\mathrm{d}}t > 0$ for a growing mass
of the gain layer. The existence of a threshold value of the 
core neutrino luminosity (Burrows \& Goshy 1993) for a given  
value of $\dot M$ is confirmed. Besides stronger neutrino heating in the 
gain layer, the main effect of a higher $L_{\nu}$ is
reducing the neutrino emission in the cooling region and thus suppressing
the rate at which mass is advected into the neutron star. 
If $L_{\nu}$ drops below the threshold value,
the gain layer loses more mass than it receives by gas falling
into the shock. Therefore the pressure support for the shock breaks down and
the shock retreats. The same effect can be caused when the mass accretion rate
$\dot M$ drops below a critical limit. This means that not only a low core
luminosity but also a low mass accretion rate by the shock can prevent 
shock expansion.
\item
The efficiency of the neutrino cooling between neutrinosphere and gain
radius is an important factor which contributes to regulating the mass
advection through the gain radius and into the neutron star, and thus
affects also the evolution of the gain layer and of the shock. Although
the description in the toy model is greatly simplified, it demonstrates the
importance of an accurate treatment of the physics, in particular of the
neutrino-matter interactions, in the cooling layer. 
It must be suspected that excessive neutrino emission in the cooling layer,
causing mass and energy loss from the gain layer, and {\em not} insufficient 
neutrino heating in the gain layer, may have been the main reason why 
spherically symmetric simulations ultimately failed to produce explosions,
although the shock had expanded to larger radii, at least for some period
of the post-bounce evolution (see, e.g.,
Bruenn 1993, Bruenn et al.\ 1995, Rampp \& Janka 2000).
\item
The area between the two critical lines in the $\dot M$-$L_{\nu}$ plane
grows for larger shock radii,
because the conditions for neutrino heating in the gain layer improve. 
For parameters $(|\dot M|,L_{\nu})$ above the upper critical line, 
neutrino energy deposition cannot compensate for the inflow of ``negative
total energy'' with the gas that is falling through the shock. Since the
mass in the gain layer increases, the shock nevertheless expands. But only 
after the shock has reached a sufficiently large radius can neutrino heating
raise the total energy in the gain layer. When this radius is very far out, 
the total energy in the gain layer may stay negative, even for high 
neutrino luminosities, because only
a small fraction of the gas in the gain layer experiences favorable 
heating conditions. In this case an explosion cannot occur. For the 
parameters considered in the discussed model, this happens when the 
(absolute value of the) mass accretion rate by the shock is larger than
about 4~M$_{\odot}\,$s$^{-1}$. Also for somewhat lower accretion rates 
and in cases where the energy in the gain layer has become positive, 
explosions might not be possible. The question whether the shock is finally 
able to eject some part of the mantle and envelope of the progenitor star
requires a global treatment of the problem. The toy model developed in this 
paper, however, is suitable to discuss only the phase
of shock revival in dependence of the conditions around the forming 
neutron star. Considering the early evolution of the supernova shock
is in general not sufficient to make predictions about the final   
outcome, in particular in cases where the postshock medium cannot
acquire enough energy to gravitationally unbind the whole mass above the 
gain radius.
\item
The equations of the toy model were integrated for time-dependent solutions
$R_{\mathrm{s}}(t)$ and $U_{\mathrm{s}}(t)$, and characteristic properties 
of the gain layer as functions of time. These solutions confirm the 
conclusions drawn from an evaluation of the shock revival criterion. In 
addition, they yield information about the evolution of the mass
and energy in the gain layer. 
\item
Accounting for the complex effects of multi-dimensional convective overturn
within the simplified discussion of the toy model is not possible. 
Nevertheless, some consequences of convective energy transport in the gain 
layer can be addressed. For a suitable choice of the structural polytropic 
index $\gamma$
of the hydrostatic atmosphere one can describe a situation where the gain layer
is essentially isentropic in contrast to a case where the energy transport by
convection is less efficient and therefore negative gradients of the entropy
and specific energy are present between the gain radius and the shock front.
Without convection such negative gradients must develop 
because neutrino heating is strongest just outside of the gain radius.
Despite of a slightly larger neutrino energy deposition (because
of a smaller gain radius) the absence of ``convection'' 
in the toy model has a negative effect 
on the shock expansion. Because convection redistributes the energy 
deposited by neutrinos mainly near the gain radius to regions closer
to the shock, the energy loss associated with the downward advection of gas
through the gain radius is reduced. Therefore more energy stays in the gain 
layer, in particular at larger radii, the postshock pressure is enhanced,
and the shock is driven out more easily.
\item
Parametric studies with the toy model suggest that successful explosions,
driven by neutrino energy deposition, cannot be very energetic. The total 
energy per unit mass 
(gravitational plus internal energy plus minor kinetic contributions) 
in the expanding gain layer (between gain radius and shock) was observed 
to saturate around 0.1 seconds after shock revival and 
was found to be always limited, even for high core luminosities $L_{\nu}$,
by $\Delta E_{\mathrm{g}}/\Delta M_{\mathrm{g}}\la 10^{52}$~erg$\,$M$_{\odot}^{-1}$,
(corresponding to an energy per nucleon of $\la 5$~MeV). For ``typical'' mass
accretion rates by the shock, $\dot M$, of a few $0.1$~M$_{\odot}\,$s$^{-1}$, the 
integral mass in the gain layer was then between several $10^{-2}$~M$_{\odot}$ and 
(1--$2)\times 10^{-1}$~M$_{\odot}$, the corresponding total energy in the gain 
layer at most around $10^{51}$ erg. For higher accretion rates $\dot M$
the mass in the gain layer was found to be larger, but the energy per nucleon
at the same time lower. The maximum total energies were obtained for intermediate 
values of $\dot M$ (around 1~M$_{\odot}\,$s$^{-1}$)
and high $\nu_e$ plus $\bar\nu_e$ luminosities $L_{\nu}$ (values up to 
$12\times 10^{52}$~erg$\,$s$^{-1}$ were considered), 
but $\Delta E_{\mathrm{g}}$ was less than (2--$3)\times 10^{51}$~erg in all
cases.
%
\end{enumerate}

\subsection{Implications}

The physical mechanism of powering the explosion by neutrino absorption on
nucleons therefore seems to limit the explosion energy to values of at most
a few $10^{51}$~erg. There is no obvious reason why neutrino-driven
explosions could not be less energetic. The upper limit of the explosion energy
is of the order of or a small multiple of the gravitational binding energy of
the gas mass in the neutrino-heating layer around the nascent neutron star.

The reason for this energy limit is a very fundamental one, associated with
the mechanism how the energy for the explosion is delivered, stored and carried
outward. The energy which starts and drives the explosion is mainly transferred
to the stellar gas by electron neutrino and antineutrino absorption on nucleons. 
As soon as the baryons have obtained a sufficiently large mean energy the 
expansion of the heating region sets in and the nucleons move away from the 
central source of the neutrino flux. The specific energy for this to happen
is of the order of the gravitational binding energy of a nucleon. In fact,
because of the inertia of the gain layer and the confinement by the matter falling
into the shock, the nucleons can absorb more energy than that. If this were not
the case, the energy of the expanding layer would be consumed by lifting the 
baryons in the gravitational field of the neutron star, and the kinetic energy 
at infinity could never be large.

The neutrinospheric luminosity $L_{\nu}$ as well as the mass in the gain layer
behind a stalled shock decrease with time, associated with the decreasing rate
$\dot M$ of mass accretion by the shock. This has negative effects
on the possibility of shock revival, as discussed above on grounds of the toy model.
It definitely does not improve the conditions for a strong explosion, because
of the limited energy which a baryon can absorb before it starts moving outward.
With little gas being exposed to neutrino heating the explosion energy will
therefore stay low.
For these reasons ``late'' neutrino-driven explosions appear to be disfavored
compared to delayed ones that develop within the post-bounce period when both
$L_{\nu}$ and $\dot M$ are still high. This is typically the case until about
half a second after core bounce.

Estimating the final explosion energy of the star requires, of course, that the
energy release by nucleon recombination and possible nuclear burning in a fraction 
of the mass of the gain layer are added, and the gravitational binding energy 
of the mantle and envelope material of the progenitor star is subtracted
(see, e.g., Bethe 1990, 1993, Bethe 1996a,c). 
Within the considered toy model these energies cannot be estimated. In
case of a successful neutrino-driven explosion these terms, however, should
not be the dominant ones in the total energy budget.

The total energy of the explosion should also not receive a major contribution
by the energy released during the phase
of the neutrino-driven wind, which succeeds the period of shock revival and early 
shock expansion. Different from the latter phase, the neutrino-driven wind is
characterized by quasi steady-state conditions, with the mass flow rate not 
varying with the radius outside of a narrow region where the mass loss of the 
neutron star is determined.
Baryons interacting with neutrinos near the surface of the
neutron star cannot absorb a particle energy much larger than their gravitational
binding energy before they are driven away from the neutrinosphere. Although 
always positive, the neutrino heating decreases rapidly when the wind accelerates 
outward and the distance from the source of the luminosity increases. Since the
confining effect of mass infall to a shock is absent, the final net energy
of a nucleon moving out with the wind will be even smaller than at earlier times.
In addition, the neutrino luminosity and the neutron star radius shrink
with time. Therefore the mass loss rate during the wind phase will be lower
than right after shock revival. For these reasons the total mass ejected in 
the wind is expected to be less than a few $10^{-2}$~M$_{\odot}$ 
(see Woosley \& Baron 1992, Woosley 1993a, Qian \& Woosley 1996). 

Overcoming the stringent limit on the energy per nucleon that neutrinos can
transfer to the heated matter, requires specific conditions. It could either be
achieved by a sudden, luminous outburst of (energetic) neutrinos,
which builds up on a timescale shorter than the expansion time of the gas around
the neutron star. However, assuming a standard, hydrostatic cooling history 
of the nascent neutron star, there is no theoretical model to support such a 
scenario. Alternatively, the energy of the explosion could be absorbed and 
carried by non-baryonic particles, i.e., electrons and positrons and photons.
In both cases the total
energy is not constrained roughly by the binding energy of the gas in the 
gravitational potential of the neutron star. In fact, neutrino-electron
scattering and neutrino-antineutrino annihilation have been suggested as
important sources of energy for the explosion (Goodman et al.\ 1987, Colgate 1989).
An accurate discussion of the physics of neutrino transport in 
the semi-transparent regime around the neutrinosphere (Janka 1991a,b) 
and a detailed evaluation of the conditions in the heating layer, however,
show that both $\nu e^\pm$ scattering (Bethe \& Wilson 1985; Bethe 1990, 
1993, 1995) and $\nu\bar\nu$ annihilation (Cooperstein et al.\ 1987, Bethe 1997)
are significantly less efficient than $\nu_e$ and $\bar\nu_e$ absorption, and
thus contribute only minor fractions to the explosion energy. 

The situation may be different when the global spherical symmetry is broken, e.g.,
in case of a black hole that accretes gas from a thick disk formed by the 
collapsing matter of a rapidly rotating, massive star. The disk becomes very hot
and loses energy primarily by neutrino emission. Such a scenario was
suggested as source of cosmological gamma-ray bursts and possibly strange,
very energetic supernova explosions (Woosley 1993b, MacFadyen \& Woosley 1999, 
MacFadyen et al.\ 1999). In this case the neutrino luminosities can be higher, 
the region where neutrino pairs annihilate is more compact (which implies that 
the neutrino 
number densities are larger), and the geometry favors more head-on collisions 
between neutrinos. All these effects lead to an enhanced probability of 
$\nu\bar\nu$ annihilation in the close vicinity of the black hole 
(Popham et al.\ 1999).

{\acknowledgements{
It is a pleasure to thank M.~Rampp for comments,
his patience in many discussions, and a comparative evaluation of
his spherically symmetric hydrodynamical simulations.
The author is very grateful to an anonymous referee for thoughtful and
knowledgable comments which helped to improve the manuscript significantly. 
The preparation of Fig.~\ref{fig:2} by Mrs.~H.~Krombach is acknowledged
as well as help by M.~Bartelmann to tame the ``\LaTeX~devil''.
This work was supported by the SFB-375 ``Astroparticle
Physics'' of the Deutsche Forschungsgemeinschaft.
}}


\end{document}